\documentclass[11pt]{article}
\usepackage[french,english]{babel}
\usepackage{amsmath}
\usepackage{amsfonts}
\usepackage{amssymb}
\usepackage{graphicx}   
\usepackage{times}
\usepackage{version}
\def\figskip{\vskip .2cm plus 1mm minus 1mm}
\def\hbar{\not{\hbox{\kern-2.3pt $h$}}}      
\def\psl{\not{\hbox{\kern-2.3pt $p$}}}       
\def\Psl{\not{\hbox{\kern-2.3pt $P$}}}       
\def\ksl{\not{\hbox{\kern-2.3pt $k$}}}       
\def\qsl{\not{\hbox{\kern-2.3pt $q$}}}       
\begin{document}
%
\begin{titlepage}
March 17th 2009 \hfill 
Manuscript ID: EPJC-08-11-017.R1
%
\null\vskip 5.5cm
{\baselineskip 20pt
\begin{center}
{\bf MIXING ANGLES  OF QUARKS AND LEPTONS IN QUANTUM FIELD THEORY 
}
\end{center}
}
\vskip .2cm
\centerline{Q. Duret
     \footnote[1]{LPTHE tour 24-25, 5\raise 3pt \hbox{\tiny \`eme} \'etage,
          UPMC Univ Paris 06, BP 126, 4 place Jussieu,
          F-75252 Paris Cedex 05 (France),\\
Unit\'e Mixte de Recherche UMR 7589 (CNRS / UPMC Univ Paris 06)}
    \footnote[2]{duret@lpthe.jussieu.fr},
           B. Machet
    \footnotemark[1]
     \footnote[3]{machet@lpthe.jussieu.fr}
          \& M.I. Vysotsky 
 \footnote[4]{SSC RF ITEP, lab. 180,
Bolshaya Cheremushkinskaya Ul. 25,
117218 Moscow (Russia)}
 \footnote[5]{vysotsky@itep.ru}
     }
\vskip 1cm
{\bf Abstract:}
Arguments  coming from  Quantum Field Theory are
supplemented with a 1-loop perturbative calculation 
to settle the non-unitarity of mixing matrices linking renormalized mass
eigenstates to bare flavor states for non-degenerate coupled fermions.
We simultaneously diagonalize the kinetic and mass terms and counterterms
in the renormalized Lagrangian.
$SU(2)_L$ gauge invariance  constrains the 
mixing matrix in charged currents of renormalized mass states, for example
the Cabibbo matrix,  to stay unitary.
Leaving aside $CP$ violation, we observe that the  mixing angles
exhibit, within experimental uncertainty, a very simple breaking pattern
of $SU(2)_f$ horizontal symmetry linked to the algebra of weak neutral
currents, the origin of which presumably lies beyond the Standard Model.
It concerns: on one hand, the three quark mixing angles;
on the other hand, a neutrino-like pattern in which $\theta_{23}$ is maximal
 and $\tan (2\theta_{12})=2$. The Cabibbo angle fulfills  the
condition $\tan (2\theta_c)=1/2$ and $\theta_{12}$ for neutrinos
  satisfies accordingly the
``quark-lepton complementarity condition'' $\theta_c + \theta_{12}= \pi/4$.
$\theta_{13} = \pm 5.7\,10^{-3}$ are the only values obtained
for the third neutrino mixing angle that lie within present experimental
bounds.
Flavor symmetries, their breaking by a non-degenerate mass spectrum,
 and their entanglement with the gauge symmetry, are
scrutinized; the special role of flavor rotations as a very mildly broken
symmetry of the Standard Model is outlined.

\bigskip
{\bf PACS:} 11.30.Hv , 11.40.-q , 12.15.Ff , 12.15.Hh , 12.15.Mm, 14.60.Pq 
\vskip 3cm
\vfill

\end{titlepage}
%
\section{Introduction}
\label{section:introduction}

In the Standard Model of electroweak interactions
\cite{GlashowWeinbergSalam},  universality (we think in particular
of gauge neutral currents) is very well verified for mass states,
which are the observed and propagating
states;  non-diagonal transitions (for example $d \leftrightarrow s$
transitions -- see Fig.1 --) 
as well as non-diagonal neutral currents and small violations of universality
are generated at 1-loop by
charged weak currents and the Cabibbo mixing.
This empirical fact is consistent with the gauge Lagrangian for
neutral currents being controlled,  in mass space, by the unit matrix
(this will be  justified later on more precise grounds).
This work, motivated by results of \cite{DuretMachet1} and \cite{DuretMachet2},
which are
summarized below, rests on the fact that, in Quantum Field Theory (QFT)
of non-degenerate coupled systems like fermions, the unit matrix
controlling neutral currents in mass space does not
translate {\em a priori} unchanged when one goes from mass states to
flavor states. We show that neutral gauge currents exhibit, in bare flavor
space, peculiar and regular structures related to flavor transformations
and symmetries.

We have shown in \cite{DuretMachet1}  that, in QFT,
mixing matrices linking bare flavor to renormalized mass eigenstates
for non-degenerate coupled systems should never be parametrized as
unitary. Indeed, assuming that the renormalized ($q^2$ dependent,
effective) quadratic Lagrangian is hermitian at any $q^2$,
different mass eigenstates, which correspond to
different values of $q^2$ (poles of the renormalized propagator),
 belong in general to different orthonormal bases
\footnote{Since, at any {\em given} $q^2$,
the set of eigenstates of the renormalized quadratic Lagrangian form
an orthonormal basis, the mixing matrix with all its elements evaluated at
this $q^2$ is unitary and
the unitarity of the theory is never jeopardized.\label{ftn:unitarity}}
\footnote{Special cases can occur, in which two coupled
states with different masses can be orthogonal: this would be the case
of neutral kaons in a world where they are stable and where
$CP$ symmetry is not violated; the mass eigenstates are then
the orthogonal $K^0_1$ and $K^0_2$ mesons \cite{MaNoVy}.\label{foot:special}};
this is the main property pervading the present work. We  recover
this result in section \ref{section:pert}  from perturbative arguments,
through the introduction of counterterms (that we shall call hereafter
Shabalin's counterterms) canceling,
at 1-loop, on mass-shell $d\leftrightarrow s$ transitions and equivalent
\cite{Shabalin}.

Assuming, for mass states, universality of diagonal neutral currents and absence of
their non-diagonal counterparts, these two properties
 can only be achieved for bare flavor states in two cases 
\footnote{For two generations, one is led to introduce two mixing angles to
parametrize each $2 \times 2$ non-unitary mixing matrix.}:
``Cabibbo-like'' mixing angles (the standard case),
and a set of discrete solutions, unnoticed in the customary approach, including
in particular the so-called maximal mixing $\pi/4 \pm k\pi/2$.
While, for any of these,  one recovers a unitary mixing matrix,
the very small departure from unitarity expected  because of
mass splittings  manifests itself as  tiny violations of the two previous
conditions in the bare mass basis:
 universality gets slightly violated and flavor changing neutral
currents (FCNC's) arise.
We empirically found \cite{DuretMachet2} that these violations obey a
very precise pattern:
in the  neighborhood of a Cabibbo-like solution,
they  become of equal strength for a 
mixing angle extremely close to its measured value
\begin{equation}
\tan (2\theta_c) = \frac12.
\label{eq:cabsol}
\end{equation}
This success was a encouragement to go further in this direction.
We present below the outcome of our investigation 
in the case of three generations of fermions.
The resulting  intricate system of trigonometric equations has been analytically
 solved by
successive approximations, starting from configurations in
which $\theta_{13}$ is vanishing. We will see that this approximation,
obviously inspired by the patterns of mixing angles determined from
experimental measurements, turns out to be a very good one.
Indeed, we show, without exhibiting  all the solutions of our
equations, that the presently observed patterns of quarks as well as
of neutrinos, do fulfill our criterion with a precision smaller than
experimental uncertainty..
While the three angles of the Cabibbo-Kobayashi-Maskawa (CKM)
 solution are ``Cabibbo-like'', the neutrino-like solution
\begin{eqnarray}
\tan (2\theta_{12}) &=& 2\  \Leftrightarrow\  \theta_{12}\  \approx\  31.7^o,\cr
\theta_{23} &=& \frac{\pi}{4},\cr
 \theta_{13} &=& \pm 5.7\, 10^{-3}\ \text{or}\ \theta_{13}=\pm 0.2717
\label{eq:nupre}
\end{eqnarray}
is of a mixed type, where  $\theta_{23}$ is maximal
 while $\theta_{12}$ and $\theta_{13}$ are Cabibbo-like.

Two significant features in these results must be stressed. First,
the values for the third neutrino mixing angle
 $\theta_{13}$ given in (\ref{eq:nupre})  are the only ones
which lie within  present (loose) experimental bounds;
only two solutions satisfy this constraint: a
very small value $\theta_{13} \sim V_{ub} \sim$ a few
$10^{-3}$, and a rather ``large'' one, at the opposite side of the
allowed range (it actually lies slightly beyond present experimental upper
limit). Secondly, our procedure yields in an exact, though quite
simple way, the well-known ``quark-lepton complementarity relation''
\cite{QLC} for 1-2 mixing:
\begin{equation}
\theta_{12} + \theta_c  = \frac{\pi}{4},
\label{eq:qlc}
\end{equation}
where $\theta_{12}$ is the leptonic angle, and $\theta_c$ the
Cabibbo angle for quarks.

The phenomenological results that we obtain for the mixing angles
only depend on the empirical pattern of neutral currents that we uncover
 in bare flavor space,
and  not on the size of the parameter characterizing the
departure of the mixing matrix from unitarity
({\em i.e.}, in practice, the value of the counterterms \cite{Shabalin}).

The latter, that need to be introduced to
cancel unwanted non-diagonal transitions and to restore the standard CKM
phenomenology \cite{DMV}, modify  kinetic
and mass terms of fermions. It turns out that the diagonalization
of the new quadratic Lagrangian (kinetic + mass terms) obtained from the
classical one by their adjunction  requires
non-unitary mixing matrices similar to the ones used in
\cite{DuretMachet1}\cite{DuretMachet2} to connect renormalized mass
eigenstates to bare flavor eigenstates. The difference with respect to
unitary matrices is proportional to Shabalin's kinetic counterterms, and,
thus, depends on wave function renormalization(s). 
Nevertheless, we show, and the $SU(2)_L$ gauge symmetry plays a crucial
role for this, that  the mixing matrix occurring in
charged currents of renormalized mass states,  for example 
the Cabibbo matrix, stays unitary. In this (non-orthonormal) basis,
 the $SU(2)_L$ gauge
 algebra closes on the unit matrix which controls neutral currents
(like it did in the orthonormal basis of bare mass eigenstates).
Mixing angles simply undergo a  renormalization depending on
kinetic counterterms.
By introducing  a  non-unitary renormalization of flavor states,
one can also make unitary the  mixing matrices which connect, in each sector,
the renormalized flavor states to renormalized mass states; the former do
not form either, however, an orthonormal basis.

The above results have been obtained, so far, without connection to
horizontal symmetries; they only rely
on the generalization to three generations of the empirical property
concerning gauge neutral currents in flavor space, that we uncovered in
\cite{DuretMachet1}\cite{DuretMachet2} for two generations of quarks.
This constitutes  a departure from customary approaches,
which rather try to induce some specific form for mass matrices from
suitably guessed horizontal symmetries \cite{Ma}.
So, the last part of this work starts  spanning a bridge between
gauge currents and mass matrices,  investigate which role is eventually
played by  flavor symmetries, and how they are realized in nature.
For the sake of simplicity, we  do this in the case of two
generations only.
A natural horizontal group  arises, which is $SU(2)_f \times U(1)_f$ (or
$U(2)_f$); the expressions of non-trivial parts of 
 gauge neutral currents and of the fermion mass
matrix (that we suppose to be real symmetric) respectively involve
 the $SU(2)_f(\theta)$  generators ${\cal T}_z(\theta)$ and ${\cal
T}_x(\theta)$.
It is a rotated version of the most trivial one (the
generators of which are the Pauli matrices); its orientation depends on the
mixing angle $\theta$.
It is unbroken in the case of mass degeneracy (and
the mixing angle is then arbitrary);  mass splittings alter this situation,
and one  can then find two subgroups leaving respectively invariant
the gauge Lagrangian of neutral currents, or the fermionic mass terms (but
not both).
Mixing angles, associated, as we saw, to specific departure from unity of
the matrix controlling neutral currents in flavor space,
are accordingly also related to a specific pattern of the {\em
breaking} of this $SU(2)_f$
\footnote{That the breaking pattern of some underlying symmetry 
exhibits specific structures is not new since this kind of consideration
is at the origin of mass relations among mesons or baryons in Gell-Mann's
 flavor $SU(3)$ (see for example \cite{GibsonPollard} p.285).}.
We show that 2-dimensional flavor rotations, which are the
transformations generated by the ($\theta$ independent)
generator ${\cal T}_y$,  continuously transform
gauge neutral currents into the mass matrix.

Since introducing a unique constant mass matrix is known to
be problematic in QFT when dealing with coupled systems \cite{Novikov},
we then establish, through the $U(1)_{em}$  Ward identity,
a connection between the photon-fermion-antifermion vertex and the
fermionic self-energy. The same matrix as for other gauge neutral currents
controls, inside the electromagnetic current, the violation of universality
and FCNC's which occur in bare flavor space.
Imposing that both sides of the Ward Identity are invariant by the flavor
transformation that leaves the vertex invariant set constraints on the
self-energy that we propose instead of ``textures'' because they stay,
unlike the latter, invariant by flavor rotations. 

Another important aspect of unitary flavor transformations 
is that, though they may not be  symmetries of the theory (in the sense
that its Lagrangian is not invariant), they should not change the 
``physics'', in particular the Cabibbo  angle occurring in charged
gauge currents.  We show that it is indeed the case,
including its renormalization through the 
counterterms of Shabalin. Among these unitary transformations,
flavor rotations turn out again to be of special interest.
While they do not alter the breaking pattern (flavor group structure)
 of neutral currents in each sector ($(u,c)$ and $(d,s)$),
it is in general not the case for
charged currents unless the rotations in the two sectors are
identical. When it is so, only one of the two mixing angles
(the one of $(u,c)$ {\em or} the one of $(d,s)$) can be turned to zero,
such that the one in the other sector becomes,
as commonly assumed, equal to the Cabibbo
angle.
Flavor rotations appear as a very mildly broken  symmetry of the Standard
Model, in the sense that they only alter the Lagrangian through unphysical
phase shifts and do not modify the ``physics'' (the  Cabibbo
 mixing angle or its leptonic equivalent, masses \ldots).

The paper ends with various remarks and questions. Comparison with previous
works is also done.
The important issue of the alignment of mass and flavor states is
investigated; that it can only occur in one of the two sectors is put in
connection with the group structure of gauge charged currents; the empirical
properties of mixing angles that have been uncovered inside neutral
currents then naturally translate to the physical angles observed in the
former.
Unfortunately, we have in particular not been
able to put the apparent quantization on the $\tan$ of twice the mixing
angles as $n/2, n\in {\mathbb Z}$ in relation with the
$SU(2)_f \times U(1)_f$ flavor group of symmetry that underlies
electroweak physics for two flavors. The connection of the $\tan$ of the
Cabibbo angle with the Golden ratio  \cite{DuretMachet2}\cite{Strumia}
stays a mystery the realm of which probably lies beyond the Standard Model.

\section{Perturbative considerations}
\label{section:pert}

In this section, we show how 1-loop counterterms introduced
by Shabalin \cite{Shabalin} in order to cancel on mass-shell non-diagonal
transitions between quark mass eigenstates entail, that mixing matrices
linking (orthonormal) bare flavor states to renormalized mass states
 are in general non-unitary. This result is obtained by diagonalizing the
whole quadratic (kinetic + mass) renormalized Lagrangian + counterterms.
Kinetic counterterms (wave function renormalization) are shown to drive
this non-unitarity.
Accordingly, renormalized mass states do not form  an orthonormal basis (as
demonstrated in section \ref{section:general} from basic QFT argumentation). 
Neutral currents being controlled in (both bare and renormalized)
mass space, by the unit matrix (which we demonstrate), we exhibit
the non-unit matrix which controls them, at 1-loop,
 in bare flavor space. We also show, by
explicit calculations in the case of two generations, how $SU(2)_L$ gauge
invariance preserves the unitarity of the Cabibbo matrix $\mathfrak C$
 occurring
in charged  currents of renormalized mass eigenstates. It does not write
anymore, however, as the product of the two renormalized mixing matrices
occurring in bare neutral currents.
We also show that, at the price of an additional non-unitary
renormalization of bare flavor states, which then become non-orthonormal,
too,
one can go to  unitary  mixing matrices ${\mathfrak C}_{u,d}$ connecting,
 in each sector, renormalized mass states to renormalized flavor states.
The standard relation
${\mathfrak C} = {\mathfrak C}_u^\dagger {\mathfrak C}_d$
is then restored. 

\subsection{The 1-loop self-energy}
\label{subsection:1loop}

The study of neutral kaons \cite{MaNoVy} has unambiguously shown that,
while flavor eigenstates can be assumed to
 form an orthonormal basis, mass eigenstates
$(K_{Long},K_{Short})$ do not (see footnote \ref{foot:special});
 the corresponding mixing matrix can only be non-unitary.

\vbox{
\begin{center}
\includegraphics[height=4truecm,width=8truecm]{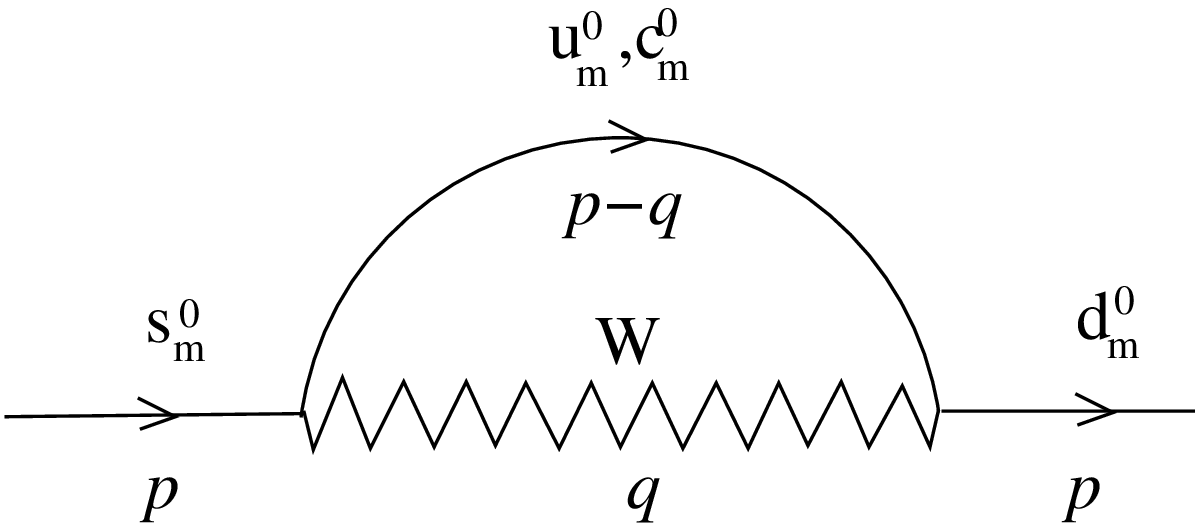}
\figskip
{\em Fig.~1: $s^0_m \to d^0_m$ transitions at 1-loop}
\end{center}
}

The situation could look very similar in the fermionic case, since there
exist, for example, transitions between $s^0_m$ and $d^0_m$
\footnote{$s^0_m$ and $d^0_m$ are the classical mass states obtained after
diagonalization of the classical mass matrix by a bi-unitary
transformation. At the classical level, they form an orthonormal basis; 
however, at 1-loop, non-local  $s^0_m \leftrightarrow d^0_m$ can
 occur.}
, depicted in
Fig.~1. They have the form of a non-diagonal kinetic term (see subsection
\ref{subsub:finite} for renormalization)
\begin{equation}
f_d(p^2, m_u^2, m_c^2, m_W^2)\ \bar d^0_m\, p\!\!/ (1-\gamma^5)\, s^0_m,
\label{eq:kinren}
\end{equation}
in which the function $f_d$ is dimensionless and includes the factors
$g^2 \sin\theta_c \cos\theta_c\; (m_c^2 -m_u^2)$
($\theta_c$ is the classical Cabibbo angle).
One should however also take into consideration the work
\cite{Shabalin}
\footnote{The introduction of these counterterms enabled to show
that, in a left-handed theory, the electric dipole moment of the quarks
vanished up to 2-loops. This resulted in a neutron electric dipole moment
well below experimental limits \cite{Gavela}.}
 which shows how the introduction of  counterterms can
make these transitions vanish for $s^0_m$ {\em or}
 $d^0_m$ on mass-shell
\footnote{Both cannot be of course simultaneously on mass-shell.}.
The following non-diagonal counterterms, which are of two types,
kinetic as well as mass terms,  and with both chiral structures:
\begin{equation}
-A_d\; \bar d^0_m \, p\!\!/(1-\gamma^5)\, s^0_m - B_d\; \bar d^0_m
(1-\gamma^5) s^0_m
-E_d\; \bar d^0_m \, p\!\!/(1+\gamma^5)\, s^0_m - D_d\; \bar d^0_m (1+\gamma^5)
s^0_m,
\label{eq:AC}
\end{equation}
with
\begin{eqnarray}
A_d = \frac{m_d^2\, f_d(p^2 = m_d^2)- m_s^2\, f_d(p^2 = m_s^2)}{m_d^2 - m_s^2},&& 
E_d = \frac{m_s m_d \left(f_d(p^2 = m_d^2) - f_d(p^2 = m_s^2)\right)}{m_d^2 -
m_s^2},\cr 
&& \cr
B_d = -m_s\, E_d,&& D_d= -m_d\, E_d,
\label{eq:BD}
\end{eqnarray}
are easily seen (see Appendix \ref{section:Shab}) to play this role.

The kinetic counterterms for d-type quarks write (the $L$ and $R$ subscripts
meaning respectively, throughout the paper, ``left'' $(1-\gamma^5)$ and
``right'' $(1+\gamma^5)$) 
\begin{equation}
-A_d\; \left(\begin{array}{cc} \bar d^0_{mL} & \bar s^0_{mL} \end{array}\right)
\left(\begin{array}{cc}  & 1 \cr 1 & \cr \end{array}\right)
p\!\!/ \left(\begin{array}{c} d^0_{mL} \cr s^0_{mL} \end{array}\right)
- E_d\; \left(\begin{array}{cc} \bar d^0_{mR} & \bar s^0_{mR}
\end{array}\right)
\left(\begin{array}{cc}  & 1 \cr 1 & \cr \end{array}\right)
p\!\!/ \left(\begin{array}{c}d^0_{mR} \cr s^0_{mR}\end{array}\right),
\end{equation}
and the mass counterterms
\begin{equation}
-\left(\begin{array}{cc} \bar d^0_{mL} & \bar s^0_{mL} \end{array}\right)\left(\begin{array}{cc}  & D_d \cr B_d & \cr
\end{array}\right) \left(\begin{array}{c}d^0_{mR} \cr
s^0_{mR}\end{array}\right)
-\left(\begin{array}{cc} \bar d^0_{mR} & \bar s^0_{mR} \end{array}\right)\left(\begin{array}{cc}  & B_d \cr D_d & \cr
\end{array} \right)\left(\begin{array}{c} d^0_{mL} \cr s^0_{mL} \end{array}\right)
.
\end{equation}
Instead of the customary perturbative treatment of such counterterms in the
bare orthonormal mass basis, order by order in the coupling constant,
 which can be rather cumbersome in this case
\footnote{In particular, when neither $d$ nor $s$ is on mass shell, which
starts occurring at 2-loops, their role does not restrict anymore to the
cancellation of non-diagonal transitions. See also subsection
\ref{subsection:phys}.
}, 
we shall instead consider and re-diagonalize
the effective renormalized Lagrangian at 1-loop 
\begin{eqnarray}
{\cal L} &=& 
\left(\begin{array}{cc} \bar d^0_{mL} & \bar s^0_{mL} \end{array}\right)
\left(\begin{array}{cc} 1 & -A_d \cr -A_d &  1 \cr \end{array}\right)
p\!\!/ \left(\begin{array}{c} d^0_{mL} \cr s^0_{mL} \end{array}\right)
+ \left(\begin{array}{cc} \bar d^0_{mR} & \bar s^0_{mR} \end{array}\right)
\left(\begin{array}{cc} 1 & -E_d \cr -E_d &  1 \cr \end{array}\right)
p\!\!/ \left(\begin{array}{c}d^0_{mR} \cr s^0_{mR}\end{array}\right)\cr
&& -\left(\begin{array}{cc} \bar d^0_{mL} & \bar s^0_{mL} \end{array}\right)\left(\begin{array}{cc} m_d & D_d \cr B_d & m_s\cr
\end{array}\right) \left(\begin{array}{c}d^0_{mR} \cr
s^0_{mR}\end{array}\right)
-\left(\begin{array}{cc} \bar d^0_{mR} & \bar s^0_{mR} \end{array}\right)\left(\begin{array}{cc}m_d  & B_d \cr D_d & m_s\cr
\end{array} \right)\left(\begin{array}{c} d^0_{mL} \cr s^0_{mL} \end{array}\right).
\label{eq:L1ren}
\end{eqnarray}
The advantage of doing so is that a link can then easily be
established with  section  \ref{section:general}  which uses the
general QFT argumentation of \cite{DuretMachet1}\cite{DuretMachet2}
to get similar results.
The diagonalization of the quadratic Lagrangian (\ref{eq:L1ren}) (kinetic +
mass terms) proceeds as follows.

\smallskip

$\bullet$\ Find 2 matrices ${\cal V}_d$ and ${\cal U}_d$ such that,
for the kinetic terms
\begin{equation}
{\cal V}_d^\dagger  \left(\begin{array}{cc} 1 & -A_d \cr -A_d &  1 \cr
\end{array}\right) {\cal V}_d = 1 = 
{\cal U}_d^\dagger  \left(\begin{array}{cc} 1 & -E_d \cr -E_d &  1 \cr
\end{array}\right) {\cal U}_d\; ;
\label{eq:kincon}
\end{equation}
they then rewrite
\begin{equation}
\left(\begin{array}{cc} \bar d^0_{mL} & \bar s^0_{mL} \end{array}\right)
\; p\!\!/ \;({\cal V}_d^\dagger)^{-1} {\cal V}_d^{-1}
\left(\begin{array}{c} d^0_{mL} \cr s^0_{mL} \end{array}\right)
+ \left(\begin{array}{cc} \bar d^0_{mR} & \bar s^0_{mR} \end{array}\right)\; p\!\!/ \;({\cal U}_d^\dagger)^{-1} {\cal
U}_d^{-1} \left(\begin{array}{c}d^0_{mR} \cr s^0_{mR}\end{array}\right),
\end{equation}
which leads to introducing the new states
\begin{equation}
\chi_{dL} = {\cal V}_d^{-1}  \left(\begin{array}{c} d^0_{mL} \cr s^0_{mL} \end{array}\right)
= {\cal V}_d^{-1} {\cal C}_{d0}^{-1}\left(\begin{array}{c}
d^0_{fL} \cr s^0_{fL} \end{array}\right) ,
\ \chi_{dR}=  {\cal U}_d^{-1} \left(\begin{array}{c}d^0_{mR} \cr
s^0_{mR}\end{array}\right)
= {\cal U}_d^{-1} {\cal H}_{d0}^{-1}
\left(\begin{array}{c} d^0_{fR} \cr s^0_{fR} \end{array}\right),
\label{eq:chibasis}
\end{equation}
where ${\cal C}_{d0}$ and ${\cal H}_{d0}$  are the two unitary matrices
by which the classical mass matrix $M_0$ has been diagonalized
into $diag(m_d,m_s)$
\footnote{$diag(m_d,m_s) = {\cal C}_{d0}^\dagger M_0 {\cal H}_{d0}$, where
$M_0$ is the classical mass matrix.  \label{ftn:classM}};
we take them as follows
\footnote{We take a rotation matrix with angle $(-\theta_{dL})$ 
to match the formul{\ae} of \cite{DuretMachet1} \cite{DuretMachet2}.}:
\begin{equation}
{\cal C}_{d0}= {\cal R}(-\theta_{dL}),\quad
{\cal H}_{d0}= {\cal R}(-\theta_{dR}),
\label{eq:Cud0}
\end{equation}
where we have introduced the notation
\begin{equation}
{\cal R}(\theta) =  \left(\begin{array}{rr} \cos\theta & \sin\theta \cr
-\sin\theta & \cos\theta \end{array}\right).
\label{eq:rot}
\end{equation}
Solutions to the conditions (\ref{eq:kincon}) are the {\em non-unitary} matrices
depending respectively of arbitrary angles $\varphi_{Ld}$ and $\varphi_{Rd}$
\footnote{Maximal mixing, for example 
$\displaystyle\frac{1}{\sqrt{2}}\left(\begin{array}{rr} 1 & 1 \cr -1 & 1
\end{array}\right)$, also diagonalizes the kinetic terms, but into
$\left(\begin{array}{cc} 1+ A_d & \cr & 1-A_d \end{array}\right)$, which is
not the canonical form (= the unit matrix).
This accordingly requires two different renormalizations of
the corresponding eigenvectors, which are finally $\sqrt{\frac{1+A_d}{2}}
(d_m^0-s_m^0)$ and  $\sqrt{\frac{1-A_d}{2}} (d_m^0+s_m^0)$. The mixing
matrix connecting  bare mass states to them is 
${\cal V}_d=\frac{1}{\sqrt{2}}\left(\begin{array}{rr}
\frac{1}{\sqrt{1+A_d}} & \frac{1}{\sqrt{1-A_d}}\cr
-\frac{1}{\sqrt{1+A_d}} &\frac{1}{\sqrt{1-A_d}}
\end{array}\right)$,
which is non-unitary (and non normal): it satisfies ${\cal V}_d^\dagger{\cal V}_d
= \left(\begin{array}{cc} \frac{1}{1+A_d} & \cr & \frac{1}{1-A_d}\end{array}
\right)$ and ${\cal V}_d{\cal V}_d^\dagger = \left(\begin{array}{cc}
 1 & A_d \cr A_d & 1 \end{array}\right)$.
This is why we look for general non-unitary ${\cal V}_d$
and ${\cal U}_d$. The  special case outlined here corresponds
 to $\rho_d=0$ and $\varphi_{Ld}= \pi/4$ in (\ref{eq:calV}).
 \label{footnote:maxi}}
and arbitrary parameters $\rho_d$ and $\sigma_d$:

\vbox{
\begin{eqnarray}
{\cal V}_d &\stackrel{A_d\ small}{\approx}& {\cal R}(\varphi_{Ld})
+ A_d \left( \begin{array}{rr}
\displaystyle\frac{\rho_d -1}{2} \sin(\varphi_{Ld})   &   -\displaystyle\frac{\rho_d -1}{2}
\cos(\varphi_{Ld})   \cr
\displaystyle\frac{\rho_d + 1}{2} \cos(\varphi_{Ld})   &   \displaystyle\frac{\rho_d+1}{2}
\sin(\varphi_{Ld})
\end{array}\right)\cr
&& \cr
&=& {\cal R}(\varphi_{Ld}) \Big[ 1 - A_d\big( {\cal T}_z(\varphi_{Ld})
+i\rho_d {\cal
T}_y\big)\Big],\cr
&& \cr
 {\cal U}_d &\stackrel{E_d\ small}{\approx}& {\cal R}(\varphi_{Rd})
+ E_d \left( \begin{array}{rr}
\displaystyle\frac{\sigma_d -1}{2} \sin(\varphi_{Rd})   &   -\displaystyle\frac{\sigma_d -1}{2}
\cos(\varphi_{Rd})   \cr
\frac{\displaystyle\sigma_d + 1}{2} \cos(\varphi_{Rd})   &   \displaystyle\frac{\sigma_d +1}{2}
\sin(\varphi_{Rd})
\end{array}\right)\cr
&& \cr
&=& {\cal R}(\varphi_{Rd}) \Big[ 1 - E_d\big( {\cal T}_z(\varphi_{Rd})
+i\sigma_d {\cal
T}_y\big)\Big],
\label{eq:calV}
\end{eqnarray}
}

where we have introduced the notations
\begin{equation}
{\cal T}_z(\theta) = \frac12 \left(\begin{array}{rr}
\sin 2\theta & -\cos 2\theta \cr -\cos 2\theta & -\sin 2\theta 
\end{array}\right), \quad {\cal T}_y = \frac12 \left(\begin{array}{cc}
 & -i \cr i & \end{array}\right),
\label{eq:Tzy}
\end{equation}
which will be often used in section \ref{section:NCMM}, together with the
${\cal T}_x(\theta)$ generator which closes the corresponding $SU(2)_f$ algebra..

The connection between the flavor states and the $\chi_{L,R}$ states that
diagonalize the kinetic terms goes accordingly through the non-unitary
mixing matrices ${\cal C}_{d0}\, {\cal V}_d$ and ${\cal H}_{d0}\, {\cal U}_d$.
At this stage, one has already made the transition from an {\em orthonormal
bare mass basis} $(d_m^0, s_m^0)$
\footnote{since $(d^0_m,s^0_m)$ is obtained from the bare flavor basis,
 supposed to be orthonormal, by a unitary transformation.\label{foot:nonor}}
 to  {\em non-orthonormal}
 $\chi$ bases; the
next bi-unitary transformation (below) will not change this fact.

$\bullet$\ Express the renormalized mass matrix in the new $\chi_{L,R}$
basis
\begin{equation}
\left(\begin{array}{cc} \bar d^0_{mL} & \bar s^0_{mL} \end{array}\right)\left(\begin{array}{cc} m_d & D_d \cr B_d & m_s\cr
\end{array}\right) \left(\begin{array}{c}d^0_{mR} \cr
s^0_{mR}\end{array}\right) =
\overline{\chi_{dL}}\ {\cal M}\, \chi_{dR}, \ 
{\cal M} = {\cal V}_d^\dagger \left(\begin{array}{cc} m_d
& D_d \cr B_d & m_s\cr
\end{array}\right) {\cal U}_d
\label{eq:calM}
\end{equation}
and diagonalize it by a second bi-unitary transformation
\begin{equation}
V_d^\dagger \,{\cal M}\,
U_d = \left(\begin{array}{cc} \mu_d & \cr
& \mu_s \end{array}\right)
\label{eq:bi2}
\end{equation}
which, since it is bi-unitary, leaves the kinetic terms unchanged.
The new (renormalized) mass eigenstates are accordingly
\begin{equation}
\left(\begin{array}{c} d_{mL} \cr s_{mL} \end{array}\right)
 = V_d^{-1} \chi_{dL}
= V_d^{-1} {\cal V}_d^{-1} {\cal C}_{d0}^{-1} \left(\begin{array}{c}
d^0_{fL} \cr s^0_{fL} \end{array}\right),\  
\left(\begin{array}{c} d_{mR} \cr s_{mR} \end{array}\right)
= U_d^{-1} \chi_{dR}
= U_d^{-1} {\cal U}_d^{-1} {\cal H}_{d0}^{-1} \left(\begin{array}{c}
d^0_{fR} \cr s^0_{fR} \end{array}\right),
\label{eq:xi}
\end{equation}
which correspond to the non-unitary mixing matrices ${\cal C}_{d0} {\cal
V}_d V_d$ and
${\cal H}_{d0}\, {\cal U}_d U_d$ respectively for left-handed and
right-handed fermions. The renormalized mass bases are accordingly
non-orthonormal (see footnote \ref{foot:nonor}).

$\bullet$\ Parametrizing $V_d = {\cal R}(\theta_{2Ld})$,
one uses the arbitrariness of $\varphi_{Ld}$ to choose $\varphi_{Ld} +
\theta_{2Ld} =0$, which  cancels
the influence of the mass counterterms $B_d,D_d$ and gives:
\begin{equation}
{\cal V}_d V_d = \left(\begin{array}{cc} 1 & \displaystyle\frac{1-\rho_d}{2} A_d \cr
\displaystyle\frac{1+\rho_d}{2} A_d & 1 \end{array}\right),
\label{eq:VV}
\end{equation}
The mixing matrix ${\cal C}_d \equiv {\cal C}_{d0} {\cal V}_d V_d$
 connecting the bare flavor states
 to the renormalized mass eigenstates 
\begin{equation}
\left(\begin{array}{c} d^0_{fL} \cr s^0_{fL}\end{array}\right)
= {\cal C}_d \left(\begin{array}{c}
d_{mL} \cr s_{mL} \end{array}\right)
\label{eq:defmix}
\end{equation}
can be written, after some manipulations (commutations)
\begin{equation}
{\cal C}_d =
 \Big[1 - A_d\big({\cal T}_z(\theta_{dL}) +i\rho_d {\cal T}_y\big)  \Big]
{\cal R}(-\theta_{dL})
\quad\Leftrightarrow\quad
{\cal C}_d^{-1} = {\cal R}(\theta_{dL})\Big[
1 + A_d \big( {\cal T}_z(\theta_{dL}) + i\rho_d {\cal T}_y\big)\Big].
\label{eq:Cd1}
\end{equation}
It satisfies in particular
\begin{equation}
({\cal C}_d^{-1})^\dagger {\cal C}_d^{-1} = 1 + 2A_d\, {\cal
T}_z(\theta_{dL}).
\label{eq:Cd2}
\end{equation}
Eq.(\ref{eq:Cd2}) is specially relevant since, once neutral currents are
controlled, as we show later, in the (non-orthonormal)  mass basis
$\xi_{dL}$ by the unit matrix, 
$({\cal C}_d^{-1})^\dagger {\cal C}_d^{-1}$ provides, after introducing
Shabalin's 1-loop counterterms in mass space, the renormalized
1-loop Lagrangian for neutral currents in the bare flavor
basis
\footnote{This is also valid for the electromagnetic current, which is one
among the gauge neutral currents. Up to the electric charge, it is
controlled in mass space  by the unit matrix and by the 
combination $({\cal C}_d^{-1})^\dagger {\cal C}_d^{-1}$ in bare flavor
space.\label{footnote:em}}.

When the system becomes degenerate, the reasons to forbid
 $d^0_m \leftrightarrow s^0_m$ on
mass-shell transitions disappear and Shabalin's counterterms are expected
to vanish. There is then no more need to introduce any non-unitary mixing
matrix. The same result can be reached by the general QFT arguments of
\cite{DuretMachet1} since one can always choose an orthonormal basis of
degenerate mass eigenstates; any connection between them and the
(supposedly orthonormal) bare flavor basis goes then through unitary mixing
matrices.

\subsubsection{Renormalization;  finiteness of the counterterms}
\label{subsub:finite}

The function $f_d$ which appears in (\ref{eq:kinren}), calculated in the
unitary gauge for the $W$ boson and dimensionally
regularized, is proportional to \cite{Shabalin}
\begin{equation}
g^2 \sin\theta_c\cos\theta_c (m_c^2 - m_u^2)\int_0^1 dx \left[
\frac{2x(1-x)}{\Delta(p^2)} + \frac{p^2 x^3(1-x)}{M_W^2 \Delta(p^2)}
+\frac{x+3x^2}{M_W^2 \Delta(p^2)^{2-n/2}}\Gamma(2-n/2)\right].
\label{eq:f}
\end{equation}
$n=4-\epsilon$ is the dimension of space-time,  $\Delta(p^2) = 
(1-x)M_W^2 + x\frac{m_u^2 + m_c^2}{2} -x(1-x)p^2$, and $\Gamma$ is the
Gamma function $\Gamma(\epsilon/2) = 2/\epsilon -\gamma + \ldots$
where $\gamma \approx 0.5772\ldots$ is
the Euler constant.  In particular, it includes a pole $(1/\epsilon)$ term
and finite terms
\begin{equation}
f_d \ni g^2 \sin\theta_c\cos\theta_c (m_c^2 - m_u^2)
\int_0^1 dx \left[\frac{x+3x^2}{M_W^2} (2/\epsilon -\gamma)
+ \ finite(x,p^2, M_W^2, m_c^2,m_u^2)\right];
\label{eq:pole}
\end{equation}
we have decomposed the latter into the one proportional to
the Euler constant, independent of $p^2, m_c^2, m_u^2$, and ``$finite$'',
which depends on them.
The transition corresponding to Fig.~1 gets, after renormalization (for
example in the $MS$ or $\overline{MS}$ schemes), a finite value
\begin{equation}
f_d^R(p^2, m_u^2, m_c^2, m_W^2)\ \bar s^0_m\, p\!\!/ (1-\gamma^5)\, d^0_m.
\label{eq:kinrenR}
\end{equation}
The cancellation of the (now finite) $d_m^0 \leftrightarrow s_m^0$
transitions for $d$ or $s$ on-shell can be obtained by introducing the
finite counterterms $A_d, B_d, C_d, D_d$ given in (\ref{eq:BD}).
The $MS$ and $\overline{MS}$ schemes, which differ by the subtraction of
a constant proportional to $\gamma$ in the integral (\ref{eq:pole}),
lead to different values for $A_d$, but identical values for $B_d, E_d, D_d$.
It is noticeable that  (\ref{eq:BD}), when considered for bare
$f_d$, leads to infinite $A_d$ but to finite $B_d, E_d, D_d$. 
 Likewise, the combination
\begin{equation}
(m_c^2 - m_u^2) A_u - (m_s^2 - m_d^2) A_d,
\label{eq:consA}
\end{equation}
proportional (see (\ref{eq:BD})) to $m_c^2 f_u(p^2 = m_c^2) - m_u^2
f_u (p^2 = m_u^2) - m_s^2 f_d(p^2 = m_s^2) + m_d^2 f_d(p^2 = m_d^2)$,
is finite \footnote{ $f_u$ is defined by a formula analogous to
(\ref{eq:f}), with the
exchange $m_c \leftrightarrow m_s, m_u \leftrightarrow
m_d$.\label{footnote:fu}}.
This property results from the independence of the pole term
in (\ref{eq:f}) on the quark masses, but for the global factors $(m_{c}^2
- m_{u}^2)$ for $f_d$ and $(m_{s}^2 - m_{d}^2)$ for $f_u$.
The finiteness of (\ref{eq:consA}) entails in particular that
$A_u$ and $A_d$ cannot vanish simultaneously
and, thus, that a non-unitary mixing matrix is
always at work in, at least, one of the two fermionic sectors $(u,c
\ldots)$ and $(d,s \ldots)$. Like the $B_{u,d}$, $E_{u,d}$ and $D_{u,d}$
counterterms,
the combination (\ref{eq:consA}), which does not depend on the 
Euler constant $\gamma$,
has the same value in $MS$ and $\overline{MS}$; indeed, the aforementioned
properties of the pole term are shared by
the one proportional to $\gamma$ in (\ref{eq:f}). 

The four bare ``infinite'' functions
$f_d(m_d^2)$, $f_d(m_s^2)$, $f_u(m_u^2)$ and $f_u(m_c^2)$ involved in the
expressions of the counterterms $A_d$, $A_u$, $B_d$ and $B_u$ (and hence
also of $E_d$, $D_d$ and $E_u$, $D_u$) satisfy accordingly
three conditions, resp. $B_d=cst$, $B_u=cst$, (\ref{eq:consA})$=cst$.
The left-over arbitrariness corresponds to the
renormalization prescription for $A_d$
\footnote{or for $A_u$, but the two choices
cannot be independent.}
 which fixes, for example,
$m_s^2 f_d^R(m_s^2) - m_d^2f_d^R(m_d^2)$ (see (\ref{eq:BD})).
It also corresponds to a renormalization prescription for $f_d$.
The most  common choices are  $MS$ and $\overline{MS}$, which lead to
the same values of  $B_{u,d},E_{u,d},D_{u,d}$
and of the combination (\ref{eq:consA}),
 but other  choices are {\em a priori} conceivable,
which are eventually  closer to ``physics''
(see subsection \ref{subsection:cc} and footnote \ref{footnote:scheme},
where we comment about the alignment of mass and flavor states in the
$(u,c)$ sector in connection with flavor rotations),
and which can lead to different values for $B_{u,d},E_{u,d},D_{u,d}$ and
for the combination (\ref{eq:consA}).

A few remarks are due concerning the cancellation of ultraviolet infinities
leading to a finite $W \to q_1 \bar q_2$ amplitude.
That a renormalization of the CKM matrix is mandatory to cancel
infinities between the $(scalar)q_1\bar q_2$ and $Wq_1\bar q_2$ sectors
when mass splittings are present was first shown in \cite{MarcianoSirlin} for
the case of two generations, and then in \cite{DennerSack} in the case of
three. In the present work, which uses the unitary gauge like in the
section 3 of \cite{MarcianoSirlin}, only finite mass renormalization is
needed and  the only infinite
counterterms that occur are the $A_{d,u}$ (kinetic counterterms
corresponding to wave function renormalization).
They become finite by a renormalization of $f_{d,u}$ (see  (\ref{eq:f})),
which also makes finite the $s(c)_m^0 \to d(u)_m^0$  1-loop self-energy
diagrams; both have indeed the same dependence on momentum and chirality.
Showing that ultraviolet divergences cancel between the $(scalar)q_1\bar q_2$
and $Wq_1\bar q_2$ sectors amounts accordingly to showing that this
infinite wave function renormalization is enough to
make the  observable $Wq_1\bar q_2$ vertex finite at 1-loop.
The insertion of non-diagonal self-masses on any of
the external legs of a bare $Wq_1\bar q_2$ vertex gives a vanishing
contribution because one of the two fermions attached to it is always on
mass-shell (Shabalin's counterterms are built up for this). 
So, the looked for cancellations correspond to the
standard property of infinities coming from wave function renormalization
to combine with those arising from the proper vertices
(see for example \cite{MarcianoSirlin}) to make, after a suitable 
charge renormalization, the 1-loop $Wq_1 \bar q_2$ amplitude finite.

\subsubsection{Summary of the perturbative 1-loop procedure}

Since the procedure to go from the bare Lagrangian to the effective
renormalized Lagrangian at 1-loop in flavor space is, though simple, not
completely trivial, we make a brief summary of it below:

* the bare flavor basis can be supposed to be orthonormal;\hfill\break
* the bare mass basis, obtained from the diagonalization of the bare mass
matrix
by (bi)-unitary transformations, is orthonormal, too;\hfill\break
* in this bare mass basis, there appear at 1-loop non-diagonal
transitions, and also flavor changing neutral currents; \hfill\break
* counterterms are introduced in this basis 
 to cancel non-diagonal on mass-shell transitions;\hfill\break
* they alter the matrix of kinetic terms, which, in the same basis,
 is no longer $1$, and the
mass matrix, which is no longer diagonal;\hfill\break
* putting back kinetic terms to the unit matrix requires non-unitary
transformations; the new states $\chi$ so defined do not form anymore an
orthonormal basis;\hfill\break
* the mass matrix, including the newly added counterterms,
 has to be re-expressed in the $\chi$ bases
 and re-diagonalized by a bi-unitary transformation;
this does not change anymore the kinetic terms;\hfill\break
* this last diagonalization defines the renormalized mass states, which are
obtained from the bare flavor states by a product of three matrices, two
being unitary and one non-unitary; they accordingly do not form an orthonormal
basis (the same result is obtained in section \ref{section:general}
from general considerations of QFT).
This is the counterpart of canceling, on mass-shell, through counterterms,
the non-diagonal, non-local  transitions that occurred between orthogonal
bare mass states. 
The situation, after  renormalization, is thus very similar to the
one studied in \cite{MaNoVy} for neutral kaons; \hfill\break
* once the (non-unitary) mixing matrix $\cal C$ linking renormalized
mass states to bare flavor states at 1-loop  has been defined by this
procedure,  we will show in subsection \ref{subsub:gaugeneut} that,
in the renormalized (non-orthonormal)
mass basis, the renormalized Lagrangian at
1-loop for neutral currents is  controlled by the unit matrix.
This entails that the quantity $({\cal C}^{-1})^\dagger {\cal C}^{-1}$ 
determines the
same Lagrangian in the bare flavor basis (it differs from the unit matrix, its
usual expression in the absence of Shabalin's counterterms).

\subsection{Gauge currents and renormalized  mixing matrices}

\subsubsection{$\boldsymbol{SU(2)_L}$ gauge symmetry: how
 the renormalized Cabibbo matrix stays unitary}
\label{subsub:uniCab}

$SU(2)_L$ gauge invariance, through the expression of the covariant
derivatives of the fermionic fields, requires that the same counterterms that
occur for the kinetic terms should also occur inside the gauge couplings.
Let us consider a kinetic fermionic term in its canonical form
$ \overline\Psi\, \overleftrightarrow \partial \Psi \equiv
\frac{1}{2} \big(\overline\Psi \partial \Psi - (\overline{\partial \Psi})\Psi
\big)$, and call $A$ the generic kinetic counterterm. In the kinetic term 
$\partial$ is accordingly replaced with $A \partial$ and, introducing the
covariant $SU(2)_L$ derivative in the two terms of $\overline\Psi\,
\overleftrightarrow \partial \Psi$ yields
$\frac12\overline \Psi A(\partial -ig \vec W.\vec T) \Psi -
\frac12\Big(\overline{ A(\partial -ig \vec W.\vec T) \Psi}\Big) \Psi
= \overline\Psi A\partial \Psi - \frac{ig}{2} \overline\Psi (A\vec T
+ \vec T A).\vec W \Psi$.
Calling
\begin{equation}
A = \left(\begin{array}{ccccc}
1 & -A_u & \vline & & \cr -A_u & 1 & \vline && \cr
\hline
&& \vline & 1 & -A_d  \cr && \vline & -A_d & 1  \end{array}\right)
\end{equation}
the matrix of counterterms,
the Lagrangian in bare mass space must accordingly include
\begin{equation}
\hskip -1cm{\cal L} \in \left(\begin{array}{cccc} \bar u^0_{mL} & \bar c^0_{mL} & \bar
d^0_{mL}  &\bar
s^0_{mL}\end{array}\right)
\left( A p\!\!/ -\frac{ig}{2} (A \vec T + \vec T A). \vec W_\mu) \gamma^u \ldots \right)
\left(\begin{array}{c} u^0_{mL} \cr c^0_{mL} \cr d^0_{mL} \cr s^0_{mL}
\end{array}\right).
\label{eq:Lcharged}
\end{equation}
It is hermitian and involves the (Cabibbo rotated) $SU(2)_L$ generators $\vec T$
\begin{equation}
T^3 = \frac12 
\left(\begin{array}{ccc}
1   & \vline &   \cr  
\hline
& \vline & -1    \end{array}\right),
T^+ = \left(\begin{array}{ccc}
 &   \vline &  {\cal C}_{0} \cr 
\hline
& \vline & \end{array}\right),   
T^- = \left(\begin{array}{ccc}
 &   \vline   &  \cr 
\hline
{\cal C}^\dagger_{0} & \vline &  \end{array}\right);
\label{eq:TTT}
\end{equation}
${\cal C}_0$ is the bare Cabibbo matrix 
\begin{equation}
{\cal C}_0 = {\cal C}_{u0}^\dagger {\cal C}_{d0} = {\cal R}(\theta_c),\quad
\theta_c =\theta_{uL}-\theta_{dL},
\label{eq:bareCab}
\end{equation}
${\cal C}_{d0}$ being the classical unitary mixing matrix in 
$(d,s)$ sector given by (\ref{eq:Cud0}) and ${\cal C}_{u0}$ its equivalent in the
$(u,c)$ sector, with bare mixing angle $(-\theta_{uL})$.

The mixing matrix has become, in the basis of bare mass eigenstates:
\begin{eqnarray}
{\cal C} &=& \frac12 
\left[ \left(\begin{array}{cc} 1 & -A_u \cr -A_u & 1
\end{array}\right) {\cal C}_{0}
+{\cal C}_{0} \left(\begin{array}{cc} 1 & -A_d \cr -A_d & 1
\end{array}\right)\right],
\label{eq:Cab2}
\end{eqnarray}
which is not unitary.
However, going to the final basis of  mass eigenstates
\begin{equation}
\left(\begin{array}{c} u_{mL} \cr c_{mL} \end{array}\right)
 = V_u^{-1} {\cal V}_u^{-1}\left(\begin{array}{c} u^0_{mL} \cr
c^0_{mL}\end{array}\right),\quad
\left(\begin{array}{c}
d_{mL} \cr s_{mL} \end{array}\right)
= V_d^{-1} {\cal V}_d^{-1}\left(\begin{array}{c} d^0_{mL} \cr s^0_{mL}\end{array}
\right),
\label{eq:2V}
\end{equation}
it becomes

\vbox{
\begin{eqnarray}
{\mathfrak C} &=& \frac12 
V_u^\dagger {\cal V}_u^\dagger
\left[ \left(\begin{array}{cc} 1 & -A_u \cr -A_u & 1
\end{array}\right) {\cal C}_{0}
+{\cal C}_{0} \left(\begin{array}{cc} 1 & -A_d \cr -A_d & 1
\end{array}\right)\right]
{\cal V}_d V_d\cr
&=& {\cal C}_u^\dagger {\cal C}_d
-\frac12 ({\cal V}_u V_u)^\dagger
\left[ A_u \left(\begin{array}{cc} & 1 \cr 1 & \end{array}\right){\cal C}_0
+ A_d\, {\cal C}_0  \left(\begin{array}{cc} & 1 \cr 1 & \end{array}\right)
\right]
{\cal V}_d V_d,
\label{eq:C2}
\end{eqnarray}
}

where we have used the expression on the left of (\ref{eq:Cd1}) for ${\cal C}_d$ and its
equivalent for ${\cal C}_u$.
Choosing, as we did before, $\varphi_{Lu} + \theta_{2Lu}=0=\varphi_{Ld} +
\theta_{2Ld}$, and using (\ref{eq:VV}) for ${\cal V}_d V_d$ and its
equivalent for ${\cal V}_u V_u$,  one gets finally:

\vbox{
\begin{eqnarray}
{\mathfrak C} =  \left(\begin{array}{rr}
\cos\theta_c + \displaystyle\frac{\rho_u A_u - \rho_d A_d}{2}\sin\theta_c   &
\sin\theta_c + \displaystyle\frac{\rho_u A_u - \rho_d A_d}{2}\cos\theta_c \cr
-\sin\theta_c + \displaystyle\frac{\rho_u A_u - \rho_d
A_d}{2}\cos\theta_c   &  \cos\theta_c
+ \displaystyle\frac{\rho_u A_u - \rho_d A_d}{2}\sin\theta_c
\end{array}\right)
\approx  {\cal R}\left(\theta_c -\frac{\rho_u A_u -\rho_d
A_d}{2}\right).
\label{eq:Cfin}
\end{eqnarray}
}

So, once  Shabalin's counterterms and  the change of basis have
both been taken into account, the renormalized Cabibbo matrix, which does
not write anymore as the product ${\cal C}_u^\dagger {\cal C}_d$,
becomes again unitary
\footnote{The customary expression ${\cal C}_u^\dagger {\cal C}_d$ for the
CKM matrix is not unitary and should be discarded.
One gets indeed, with straightforward notations,
${\cal C}^\dagger {\cal C} \approx
1 -2 {\cal R}(\theta_{dL})\Big( A_u {\cal T}_z(\theta_{uL})
+ A_d {\cal T}_z(\theta_{dL})\Big) {\cal R}(-\theta_{dL})$}.
This occurs because of $SU(2)_L$ gauge invariance,
and despite the fact that neither ${\cal C}_u$ nor ${\cal C}_d$ is unitary.
 With respect to its classical value, the classical Cabibbo angle
$\theta_c = \theta_{uL} - \theta_{dL}$  gets
renormalized by $\displaystyle\frac{\rho_u A_u -\rho_d A_d}{2}$.

\subsubsection{Neutral currents and the closure of the
$\boldsymbol{SU(2)_L}$ algebra}
\label{subsub:gaugeneut}

Like charged currents, the form of neutral currents
 is determined by  gauge invariance, through the $SU(2)_L$ covariant derivative.
It is given in the bare mass basis by  (\ref{eq:Lcharged}),
which easily translates to the renormalized mass basis since
the latter deduces from the former by the transformations
(\ref{eq:2V}).
 
The procedure is specially simple since the $T^3$
 generator only involves unit matrices in each sector, such that
$V^{-1} {\cal V}^{-1}$ can freely move through it.
It is furthermore easy to check that, in addition to (\ref{eq:kincon}), one has
\begin{equation}
({\cal V}_{u,d}V_{u,d})^\dagger \left(\begin{array}{cc} 1 & -A_{u,d} \cr
-A_{u,d} & 1
\end{array}\right) {\cal V}_{u,d} V_{u,d} =1,
\end{equation}
such that  the 1-loop effective
Lagrangian for neutral currents  gets controlled by the unit matrix
in the renormalized mass basis.

So,  $SU(2)_L$ gauge invariance ensures that 
neutral currents are controlled by the unit matrix: \newline
-  at the classical
level in the basis of bare (orthonormal) mass states;\newline
- in the Lagrangian renormalized at 1-loop  in the basis of
renormalized (non-orthonormal) mass sates.\newline

After Shabalin's   counterterms $A_u \equiv \epsilon_u$ and $A_d
\equiv \epsilon_d$ have been included,
in the renormalized mass bases
the $SU(2)_L$ generators write
\begin{equation}
T^3 = \frac12 
\left(\begin{array}{ccc}
1   & \vline &   \cr  
\hline
& \vline & -1    \end{array}\right),
T^+ = \left(\begin{array}{ccc}
 &   \vline &  {\mathfrak C} \cr 
\hline
& \vline & \end{array}\right),   
T^- = \left(\begin{array}{ccc}
 &   \vline   &  \cr 
\hline
{\mathfrak C}^\dagger & \vline &  \end{array}\right);
\label{eq:TTTrenor}
\end{equation}
of course, the unitarity of $\mathfrak C$ is necessary for its closure
on the unit matrix in the neutral gauge sector.

\subsubsection{Charged gauge currents in flavor space;
  renormalized flavor states}
\label{subsub:charged}

It is now interesting to write back the renormalized Lagrangian in bare
flavor space (it is in this basis that we  uncovered empirical specific
breaking patterns). One starts from (\ref{eq:Cab2}) in bare mass space and
go to bare flavor space by the bare mixing matrices ${\cal C}_{u0}$ and
${\cal C}_{d0}$; this yields

\vbox{
\begin{eqnarray}
&&\left(\begin{array}{cc} \bar u_{mL} & \bar c_{mL}\end{array}\right)
{\mathfrak C} \gamma^\mu
\left(\begin{array}{c} d_{mL} \cr s_{mL}\end{array}\right)\cr
&&\hskip 1cm =\left(\begin{array}{cc} \bar u^0_{fL} & \bar c^0_{fL}\end{array}\right)
\frac12\left[ {\cal C}_{u0}\left(\begin{array}{cc} 1 & -A_u \cr -A_u & 1
\end{array}\right) {\cal C}_{u0}^\dagger
+ {\cal C}_{d0}\left(\begin{array}{cc} 1 & -A_d \cr -A_d & 1 \end{array}\right)
 {\cal C}_{d0}^\dagger
\right]\gamma^\mu \left(\begin{array}{c} d^0_{fL} \cr s^0_{fL}\end{array}\right)\cr
&& \hskip 1cm = \left(\begin{array}{cc} \bar u^0_{fL} & \bar c^0_{fL}\end{array}\right)
\left[ 1 +A_u {\cal T}_z(\theta_{uL}) +A_d
{\cal T}_z(\theta_{dL}) \right]\gamma^\mu
\left(\begin{array}{c} d^0_{fL} \cr s^0_{fL}\end{array}\right)\label{eq:ch1}\\
&&\hskip 1cm = \left(\begin{array}{cc} \bar u^0_{fL} & \bar c^0_{fL}\end{array}\right)
\left[ 1 +A_u {\cal T}_z(\theta_{uL})][ 1 +A_d
{\cal T}_z(\theta_{dL}) \right]\gamma^\mu
\left(\begin{array}{c} d^0_{fL} \cr s^0_{fL}\end{array}\right)\cr
&&\hskip 1cm \approx \overline{e^{A_u{\cal
T}_z(\theta_{uL})}\left(\begin{array}{c} u^0_{fL} \cr
c^0_{fL} \end{array}\right)}\gamma^\mu
e^{A_d{\cal T}_z(\theta_{dL})}\left(\begin{array}{c} d^0_{fL} \cr
s^0_{fL} \end{array}\right),
\label{eq:LC}
\end{eqnarray}
}

where we have used the expression for ${\cal T}_z(\theta)$ given in
(\ref{eq:Tzy})
 and  the relations
${\cal C}_{u0,d0}
\left(\begin{array}{cc} & 1 \cr 1 & \end{array}\right) {\cal C}_{u0,d0}^\dagger
= -2{\cal T}_z(\theta_{uL,dL})$.
It is therefore possible to define as
 ``renormalized flavor states'' the ones that appear in the last
line of (\ref{eq:LC})
\begin{equation}
\left(\begin{array}{c} d_{fL} \cr s_{fL}\end{array}\right)
 =e^{A_d {\cal T}_z(\theta_{dL})}\left(\begin{array}{cc} d^0_{fL} \cr
s^0_{fL}
\end{array}\right)\ \text{and}\ 
\left(\begin{array}{c} u_{fL} \cr c_{fL}\end{array}\right)
 = e^{A_u {\cal T}_z(\theta_{uL})}\left(\begin{array}{cc} u^0_{fL} \cr c^0_{fL}
\end{array}\right).
\label{eq:renflav}
\end{equation}
They are deduced from the bare flavor states by the non-unitary
transformations
$e^{A_{u,d}{\cal T}_z(\theta_{uL,dL})}$, and do not form anymore,
accordingly, an orthonormal basis.  In the renormalized flavor basis,
the $SU(2)_L$ generators write in their simplest form
\begin{equation}
T^3 = \frac12 
\left(\begin{array}{ccc}
1   & \vline &   \cr  
\hline
& \vline & -1    \end{array}\right),
T^+ = \left(\begin{array}{ccc}
 &   \vline &  1 \cr 
\hline
& \vline & \end{array}\right),   
T^- = \left(\begin{array}{ccc}
 &   \vline   &  \cr 
\hline
1 & \vline &  \end{array}\right);
\label{eq:TTTflav}
\end{equation}
universality is thus  achieved together with the absence of
FCNC's, like in the basis or renormalized mass states.
The two points of view describe of course the same physics: in the
non-orthonormal renormalized flavor basis, neutral currents are
controlled by the unit
matrix (seemingly absence of non-diagonal transitions, but they still occur
through the non-orthogonality of the states), and, in the bare flavor
basis, neutral currents are controlled by a matrix slightly different from
the unit matrix (non-diagonal transitions between orthogonal states are
then conspicuous).

The last step is to calculate the mixing matrices $\mathfrak C_{u,d}$
linking the renormalized mass states  (see (\ref{eq:xi}))
 to the renormalized flavor
states $f_{uL,dL}$ defined in (\ref{eq:renflav}).
From (\ref{eq:Cd1}), it is straightforward to deduce
\begin{eqnarray}
{\mathfrak C_d} &=&  e^{A_d T_z(\theta_{dL})}{\cal C}_d
\approx   \left( 1 + A_d T_z(\theta_{dL}) \right) {\cal C}_d\cr
&\stackrel{(\ref{eq:Cd1})}{=}&
\left( 1 + A_d T_z(\theta_{dL}) \right)\left[ 1 - {A_d}\left({\cal T}_z(\theta_{dL})
-i\rho_d {\cal T}_y\right)\right]{\cal R}(-\theta_{dL})\cr
&=& \big( 1 + i\rho_d A_d {\cal T}_y \Big) {\cal R}(-\theta_{dL})
\approx {\cal R}(-\theta_{dL} - \frac{\rho_d A_d}{2}).
\label{eq:Cdrenorm}
\end{eqnarray}
which is unitary. The relation
\begin{equation}
{\mathfrak C} = {\mathfrak C}_u^\dagger {\mathfrak C}_d,
\label{eq:Cproduni}
\end{equation}
is seen to be now restored.
So, renormalized mass states are connected to renormalized flavor states
through unitary mixing matrices.
In the renormalized flavor basis, the sole effects of Shabalin's 
counterterms is a  renormalization of the mixing angles.

After all these steps have been gone through, the 1-loop renormalized
 Lagrangian writes identically to the bare Lagrangian with:\newline
$\ast$\ renormalized masses;\newline
$\ast$\  renormalized, non-orthonormal (mass and flavor)
 eigenstates;\newline
$\ast$\ unitary mixing matrices with renormalized mixing angles.

It has the same form as the bare Lagrangian of the Standard Model, except
that the notion of flavor has been redefined, such that it
   no longer appears as a strictly conserved quantity.

\section{Neutral currents of bare flavor eigenstates; general QFT argumentation}
\label{section:general}

After establishing by perturbative arguments the {\em a priori}
non-unitarity of mixing matrices for non-degenerate coupled systems, we
come back to the argumentation of \cite{DuretMachet1}
based on general principles of Quantum Field Theory, then generalize
it to the case of three generations. Unlike in the previous section, the
argumentation goes beyond perturbation theory. There, for example, the two
mixing angles which could be introduced {\em de facto} in $\cal V$ (see
(\ref{eq:calV})), arose through perturbative arguments and were
 perturbatively close to each other;   we call them Cabibbo-like.
At the opposite, ``maximal mixing'' solutions of the
``unitarization equations'' (see subsection \ref{section:unitariz}
below), which occur in addition to Cabibbo-like solutions, form a discrete
set of solutions superimposed to the former,
and arise  independently of perturbative arguments. The property of maximal
mixing to be non-perturbative is  in agreement with its common association
with  quasi-degenerate systems (the smaller the mixing angle, the bigger the
mass hierarchy \cite{MachetPetcov}),
for which small variations (for example in the mass spectrum)
can have large  effects on eigenstates, and
thus on the mixing angles themselves.

The only ``perturbative expansions'' that will be performed (in
sections \ref{section:quarks} and \ref{section:neutrinos})  concern small
deviations from the solutions of the ``unitarization equations''.

\subsection{Different basis of fermions}
\label{subsection:basis}

Three bases generally occur in the treatment of fermions:

$\ast$ flavor eigenstates: 
$(u_f, c_f, t_f)$ and $(d_f, s_f, b_f)$ for quarks, $(e_f, \mu_f, \tau_f)$
and $(\nu_{ef}, \nu_{\mu f}, \nu_{\tau f})$ for leptons;

$\ast$ mass eigenstates:  $(u_m, c_m, t_m)$ and
$(d_m, s_m, b_m)$ for quarks, $(e_m, \mu_m, \tau_m)$
and $(\nu_{em}, \nu_{\mu m}, \nu_{\tau m})$ for leptons. They include 
in particular the charged leptons detected experimentally, since 
their identification proceeds  through the measurement
of their $charge/mass$ ratio in a magnetic field; these eigenstates are the
ones of the full renormalized propagator at its poles; at 1-loop, they can
be identified with components of the renormalized mass states  of
(\ref{eq:xi}) in section \ref{section:pert}; 

$\ast$ for leptons, one often invokes a third type of basis, made with
 the neutrino states that couple to the mass
eigenstates of charged leptons in charged weak currents.
These are the so-called "electronic'', ``muonic" and "$\tau$" 
neutrinos $(\nu_e, \nu_\mu, \nu_\tau)$ considered in SM textbooks:
they are indeed identified by the outgoing charged
leptons that they produce through
charged weak currents, and the latter are precisely mass eigenstates
 (see above).  They read 

\begin{equation}
\left(\begin{array}{c} \nu_e \cr \nu_\mu \cr \nu_\tau \end{array}\right)
= K^\dagger_\ell 
\left(\begin{array}{c} \nu_{ef} \cr \nu_{\mu f} \cr \nu_{\tau f}
 \end{array}\right)
= (K^\dagger_\ell K_\nu)
\left(\begin{array}{c} \nu_{em} \cr \nu_{\mu m} \cr \nu_{\tau m}
 \end{array}\right),
\label{eq:elecnu}
\end{equation}
where $K_\ell$ and $K_\nu$ are the mixing matrices respectively of charged
leptons and of neutrinos ({\em i.e.} the matrices that connect their flavor
to their mass eigenstates).
These neutrinos are neither flavor nor mass eigenstates; they
 coincide with the latter when the
mixing matrix {\em of charged leptons} is taken equal to unity
$K_\ell =1$, {\em i.e.} when the mass and flavor eigenstates
of charged leptons are aligned, which is often assumed  in the literature.

\subsection{Mixing matrices. Notations}
\label{subsection:notations}

We start again with the case of two generations, and use the
notations of \cite{DuretMachet1}. The situation is depicted on Fig.~2
\footnote{This figure was already published in \cite{DuretMachet1}. 
Its inclusion in the present work makes it more easily understandable and
self-contained.}.
The bare flavor states, independent of $q^2 =z$, are $\psi_1$ and
$\psi_2$ (they can be for example the $d^0_{fL}$ and $s^0_{fL}$ of section
\ref{section:pert}) and we suppose that they  are orthonormal.
Three orthonormal bases, respectively made of a pair of eigenvectors of the
(hermitian) renormalized quadratic Lagrangian at three different values of
$z$, can be seen.
The first corresponds to the  physical mass $z=z_1=m_1^2$;
the second, made of $\psi^1(z)$ and $\psi^2(z)$,
corresponds to  an arbitrary $z$; the last corresponds to the
second physical mass $z=z_2=m_2^2$.
Within the first basis one finds the first physical mass eigenstate,
$\phi_m^1$, and a second (non-physical) eigenstate, $\omega_1^2$;
the third basis is made
of the second physical mass eigenstate, $\phi_m^2$, and of a second (non-physical)
 eigenstate, $\omega_2^1$
\footnote{On Fig.~2, the $\lambda(z)$'s are the eigenvalues of the inverse
renormalized propagator at $z=q^2$.}.
For example, at 1-loop, $\phi_m^1$ and $\phi_m^2$ can be identified with
the two components of $\left(\begin{array}{c} d_{mL}\cr
s_{mL}\end{array}\right)$ (see (\ref{eq:xi})
 in section \ref{section:pert}).

\vbox{
\begin{center}
\includegraphics[height=7truecm,width=9truecm]{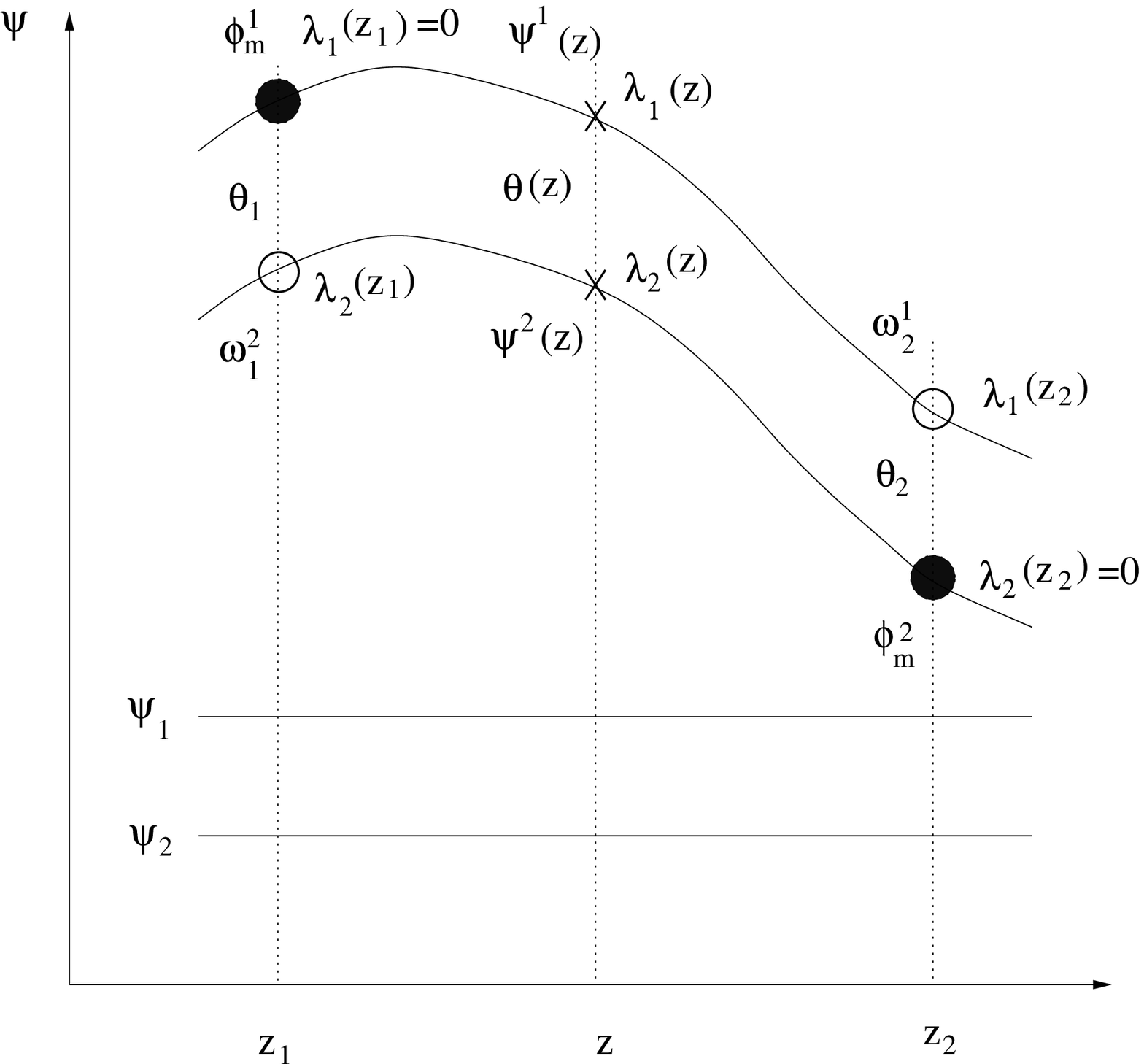}
\figskip
{\em Fig.~2: Eigenstates of a binary complex system}
\end{center}
}

The  flavor states $\psi_1$ and $\psi_2$ can be expressed
in both orthonormal  bases $(\phi_m^1, \omega_1^2)$ and
$(\phi_m^2,\omega_2^1)$ according to
\begin{eqnarray}
\psi_1 &=& c_1 \phi_m^1 - s_1 \omega_1^2 = c_2 \omega_2^1 - s_2
\phi_m^2,\cr
\psi_2 &=& s_1 \phi_m^1 + c_1 \omega_1^2 = s_2 \omega_2^1 + c_2
\phi_m^2,
\end{eqnarray}
which yields
\begin{subequations}
\begin{equation}\label{eq:f2m}
\left(\begin{array}{c} \phi_m^1 \cr \phi_m^2 \end{array}\right) =
\left(\begin{array}{rr} c_1 & s_1 \cr -s_2 & c_2 \end{array}\right)
\left(\begin{array}{c} \psi_1 \cr \psi_2 \end{array}\right) 
\end{equation}
\begin{equation}\label{eq:m2f}
\Leftrightarrow
\left(\begin{array}{c} \psi_1 \cr \psi_2 \end{array}\right) =
\frac{1}{c_1c_2 + s_1s_2}
\left(\begin{array}{rr} c_2 & -s_1 \cr s_2 & c_1 \end{array}\right)
\left(\begin{array}{c} \phi_m^1 \cr \phi_m^2 \end{array}\right).
\end{equation}
\end{subequations}
Since $\psi_1$ and $\psi_2$ have been assumed to form an orthonormal basis,
 eq.(\ref{eq:f2m}) entails
\begin{equation}
|\phi_m^1| = 1 = |\phi_m^2|,\ <\phi_m^2\;|\;\phi_m^1>= s_1c_2 -c_1s_2
\stackrel{\theta_2 \not=\theta_1}{\not =} 0.
\end{equation}
(\ref{eq:f2m}) shows that, for two generations, the mixing matrix ${\cal C}$
 satisfies
\footnote{This corresponds to  $\rho_d=1$ in formula (\ref{eq:Cd1}).}
\begin{equation}
{\cal C}^{-1} = \left(\begin{array}{rr} c_1 & s_1 \cr -s_2 & c_2
\end{array}\right).
\label{eq:invC}
\end{equation}
We generalize this, in the following, to the case of three generations by writing
the corresponding mixing matrix  $K^{-1}$ as a product of three 
matrices, which reduce,
in the unitarity limit, to the basic rotations by $-\theta_{12}$,
$-\theta_{23}$ and $-\theta_{13}$ (we are not concerned with $CP$ violation)
\begin{equation}
K^{-1} = \left(\begin{array}{rrr} 1 & 0 & 0 \cr
                             0 & c_{23} & s_{23} \cr
                             0 & - \tilde{s}_{23} & \tilde{c}_{23}
\end{array}\right) \times 
\left(\begin{array}{rrr} c_{13} & 0 & s_{13} \cr
                         0 & 1 & 0 \cr
                         - \tilde{s}_{13}& 0 & \tilde{c}_{13}
\end{array}\right) \times
\left(\begin{array}{rrr} c_{12} & s_{12} & 0 \cr
                         - \tilde{s}_{12} & \tilde{c}_{12} & 0 \cr
                         0 & 0 & 1
\end{array}\right).
\label{eq:CKM}
\end{equation}
We parametrize each basic matrix, which is {\em a priori} non-unitary, 
with two angles, respectively
 $(\theta_{12}, \tilde{\theta}_{12})$, $(\theta_{23},
\tilde{\theta}_{23})$ and $(\theta_{13}, \tilde{\theta}_{13})$
\footnote{So doing, we do not consider the most general non-unitary mixing
matrices. All possible phases were included in
\cite{DuretMachet1}\cite{DuretMachet2}, where they have been shown to
finally, in the case of two generations, drop out of the final results.
There is another reason to ignore them here, specially in the case of three
generations (in addition to the point that they would make the equations to
solve extremely difficult to handle analytically): such phases can be
expected to trigger $CP$ violation, even in the case of two generations.
We consider that the corresponding extensive study should be the subject of
a separate work. $CP$ violation is not our concern here.}.
We deal accordingly with six mixing
angles, instead of three in the unitary case (where
$\tilde{\theta}_{ij} = \theta_{ij}$). 
We will use throughout the paper the notations 
$s_{ij}= \sin(\theta_{ij}), \tilde{s}_{ij} = \sin(\tilde{\theta}_{ij})$,
and likewise, for the cosines, 
$c_{ij} = \cos(\theta_{ij}), \tilde{c}_{ij} = \cos(\tilde{\theta}_{ij})$. 

To lighten the text, the elements of $(K^{-1})^\dagger {K^{-1}}$ will
be abbreviated
 by $[ij], i,j=1\ldots 3$ instead of $((K^{-1})^\dagger K^{-1})_{[ij]}$,
and the
corresponding neutral current will be noted $\{ij\}$. So, in the quark case,
$\{12\}$ stands for $\bar u_f \gamma^\mu_L c_f$ or
$ \bar d_f \gamma^\mu_L s_f$, and, in
the neutrino case, for $\bar\nu_{ef} \gamma^\mu_L \nu_{\mu f}$ or $\bar
e_f \gamma^\mu_L \mu_f$.

\section{The unitary approximation}
\label{section:unitariz}

In a first approximation, mixing matrices are unitary, such that
 neutral currents are very close to being controlled in 
the renormalized mass basis, too,  by the unit matrix.
The corresponding equations (unitarization conditions)
 will determine the equivalent of
``classical solutions'', away from which we shall then consider
small deviations which exist because of  mass splittings:
 non-degeneracy generates a tiny
departure from unitarity of the corresponding mixing matrices and,
 accordingly, a tiny 
departure from unity of the matrix controlling neutral currents in bare
flavor space.

The unitarization conditions simply express the absence of
non-diagonal neutral  currents in flavor space, and universality for their
diagonal counterparts, assuming that the gauge Lagrangian of neutral currents
is controlled in mass space by the unit matrix;
they accordingly summarize into
\begin{equation}
(K^{-1})^\dagger K^{-1} = 1.
\end{equation}
There are five equations: three arise from the
absence of non-diagonal neutral currents, and two from
the universality of diagonal currents.  Accordingly,
one degree of freedom is expected to be unconstrained.

\subsection{Absence of non-diagonal neutral currents of flavor eigenstates}

The three conditions read:\newline

\vbox{
$\ast$ for the absence of $\{13\}$ and $\{31\}$ currents:
\begin{equation}
[13]=0=[31] \Leftrightarrow
c_{12}\left[c_{13}s_{13} -\tilde c_{13} \tilde s_{13}(\tilde c_{23}^2 + s_{23}^2)\right] 
- \tilde c_{13} \tilde s_{12}(c_{23} s_{23} - \tilde c_{23} \tilde s_{23}) = 0;
\label{eq:nodb}
\end{equation}
$\ast$ for the absence of $\{23\}$ and $\{32\}$ currents:
\begin{equation}
[23]=0=[32] \Leftrightarrow
s_{12}\left[c_{13}s_{13} -\tilde c_{13} \tilde s_{13}(\tilde c_{23}^2 + s_{23}^2)\right] 
+ \tilde c_{13} \tilde c_{12}(c_{23} s_{23} - \tilde c_{23} \tilde s_{23}) = 0;
\label{eq:nosb}
\end{equation}
$\ast$ for the absence of $\{12\}$ and $\{21\}$ currents:
\begin{eqnarray}
&& [12]=0=[21] \Leftrightarrow\cr
&&s_{12}c_{12} c_{13}^2 - \tilde s_{12} \tilde c_{12}(c_{23}^2 + \tilde s_{23}^2) + s_{12} c_{12} \tilde
s_{13}^2 ( s_{23}^2 + \tilde c_{23}^2) + \tilde s_{13} (s_{12} \tilde s_{12} - c_{12} \tilde
c_{12})(c_{23} s_{23} - \tilde c_{23} \tilde s_{23})=0.\cr
&&
\label{eq:nods}
\end{eqnarray}
}

\subsection{Universality of diagonal neutral currents of flavor eigenstates}

The two independent conditions read:

$\ast$ equality of $\{11\}$ and $\{22\}$ currents:
\begin{eqnarray}
&&[11]-[22]=0 \Leftrightarrow \cr
&&(c_{12}^2 - s_{12}^2)\left[ c_{13}^2 + \tilde s_{13}^2(s_{23}^2 + \tilde c_{23}^2)\right]
-(\tilde c_{12}^2 - \tilde s_{12}^2)(c_{23}^2+ \tilde s_{23}^2)\cr
&& \hskip 3cm + 2\tilde s_{13}(c_{23}s_{23}
- \tilde c_{23} \tilde s_{23})(c_{12} \tilde s_{12} + s_{12} \tilde
c_{12})=0;
\label{eq:ddss}
\end{eqnarray}

$\ast$ equality of $\{22\}$ and $\{33\}$ currents:
\begin{eqnarray}
&&[22]-[33]=0 \Leftrightarrow \cr
&&s_{12}^2 + \tilde c_{12}^2(c_{23}^2 + \tilde s_{23}^2) - (s_{23}^2 + \tilde c_{23}^2)
+(1+s_{12}^2)\left[ \tilde s_{13}^2(s_{23}^2 + \tilde c_{23}^2) -s_{13}^2
\right]\cr
&&\hskip 6cm
+ 2s_{12} \tilde s_{13} \tilde c_{12}(\tilde c_{23} \tilde s_{23} - c_{23} s_{23}) =0.
\label{eq:ssbb}
\end{eqnarray}
The equality  of $\{11\}$ and $\{33\}$ currents is of course
 not an independent condition.

\subsection{Solutions for $\boldsymbol{\theta_{13} = 0 =
\tilde{\theta}_{13}}$}
\label{subsection:vanish}

In a first step, to ease solving the system of trigonometric equations,
we shall study the configuration in which one of the two angles
parametrizing the 1-3 mixing vanishes  
\footnote{By doing so, we exploit the possibility to fix one degree of
freedom left {\em a priori} unconstrained by the five equations; see
subsection \ref{section:NCMM}.},
 which is very close to what is
observed experimentally in the quark sector, and likely in the neutrino
sector. It turns out, as demonstrated in Appendix \ref{section:theta3},
that the second mixing angle vanishes simultaneously.
We accordingly work in the approximation (the sensitivity of the
solutions to a small variation of $\theta_{13}, \tilde{\theta}_{13}$
will be studied afterwards)
\begin{equation}
\theta_{13} = 0 =\tilde{\theta}_{13}.
\label{eq:cond1}
\end{equation}
Eqs. (\ref{eq:nodb}), (\ref{eq:nosb}), (\ref{eq:nods}), (\ref{eq:ddss}) and
(\ref{eq:ssbb}),  reduce in this limit to

\vbox{
\begin{subequations}\label{subeq:geneq0}
\begin{equation}
 - \tilde s_{12}(c_{23} s_{23} - \tilde c_{23} \tilde s_{23}) = 0,
\label{eq:nodb0}
\end{equation}
\begin{equation}
 \tilde c_{12}(c_{23} s_{23} - \tilde c_{23} \tilde s_{23}) = 0,
\label{eq:nosb0}
\end{equation}
\begin{equation}
s_{12} c_{12}  - \tilde s_{12} \tilde c_{12}(c_{23}^2 + \tilde s_{23}^2) = 0,
\label{eq:nods0}
\end{equation}
\begin{equation}
(c_{12}^2 -s_{12}^2) -(\tilde c_{12}^2 - \tilde s_{12}^2)(c_{23}^2 + \tilde s_{23}^2) = 0,
\label{eq:ddss0}
\end{equation}
\begin{equation}
s_{12}^2 + \tilde c_{12}^2(c_{23}^2 + \tilde s_{23}^2) - (s_{23}^2 + \tilde c_{23}^2)  = 0.
\label{eq:ssbb0}
 \end{equation}
\end{subequations}
}

It is shown in Appendix \ref{section:sol0} that the only solutions
are:\newline
\qquad $\ast$ $\tilde{\theta}_{23} = \theta_{23} +
k\pi$ Cabibbo-like, associated with either $\theta_{12} =
\tilde{\theta}_{12} + m\pi$ Cabibbo-like or $\theta_{12}$ and
$\tilde\theta_{12}$ maximal;\newline
\qquad $\ast$ $\tilde\theta_{12} = \theta_{12} + r\pi$ Cabibbo-like,
 associated with $\theta_{23}$ and $\tilde\theta_{23}$ maximal.

Accordingly, the two following sections will respectively start from:

$\ast$ $\theta_{12}$ and $\theta_{23}$ Cabibbo-like (and, in a first step,
 vanishing
$\theta_{13}$), which finally leads to a mixing pattern similar to what is
observed for quarks;

$\ast$  $\theta_{23}$ maximal and $\theta_{12}$ Cabibbo like (and, in a
first step, vanishing
$\theta_{13}$), which finally leads to a mixing pattern similar to the one
observed for neutrinos.

\section{Beyond unitarity. The quark sector; constraining the  CKM angles}
\label{section:quarks}

Because of mass splittings, the ``unitarization equations'' of
subsection \ref{section:unitariz}  cannot be exactly satisfied.
This is why, in the following, 
mixing matrices connecting bare flavor states to (renormalized) mass states
are  considered to only belong to the vicinity of the (unitary)
solutions of these equations.
Characterizing this departure from unitarity is the subject of
this section and of the next one dealing with leptons.
We show that all their mixing angles satisfy
the straightforward generalization to three generations
of the empirical criterion satisfied to a high precision,
for two generations of quarks, by the Cabibbo angle \cite{DuretMachet2}:
for each pair of fermions of the same type, 
universality in the space of bare flavor states is verified
with the same accuracy as the absence of FCNC's.
It cannot be deduced, up to now,
from general principles and stays an empirical property
the origin of which should presumably be looked for
``beyond the Standard Model''
\footnote{Notice that it is satisfied a mixing matrix equal to the unit
matrix (alignment of mass and flavor states) since
universality  and absence of FCNC's are both fulfilled; accordingly
they both undergo identical (vanishing) violations.}.

We accordingly investigate, in the following, the possibility that,
in agreement with the reported criterion,
the  product  $(K^{-1}_{u,d})^\dagger K^{-1}_{u,d}$, with $K$ given by
(\ref{eq:CKM}), be  of the form
\begin{equation}
(K^{-1}_{u,d})^\dagger K^{-1}_{u,d} -1 = \left(\begin{array}{ccc}
            \alpha_{u,d}  & \pm(\alpha_{u,d} -\beta_{u,d})    &
\pm(\alpha_{u,d} - \gamma_{u,d})  \cr
  \pm(\alpha_{u,d} -\beta_{u,d})&      \beta_{u,d} &
\pm(\beta_{u,d}-\gamma_{u,d})        \cr
  \pm(\alpha_{u,d} - \gamma_{u,d}) & \pm(\beta_{u,d}-\gamma_{u,d})  &
\gamma_{u,d}   \end{array}\right);
\label{eq:KK3}
\end{equation}
so, as conspicuous on (\ref{eq:KK3}), any difference between two diagonal
elements (for example [11] - [33]) is identical to $\pm$ the corresponding
non-diagonal ones (in this case [13] and [31]).
The resulting conditions  yields a system of trigonometric equations for
the six mixing angles $\theta_{12},\tilde\theta_{12},
\theta_{23},\tilde\theta_{23},\theta_{13}$ and $\tilde\theta_{13}$.
 Without  exhibiting the whole set of its  solutions,
we are able to show that it includes all measured values of fermionic
mixing angles up to a precision smaller than the experimental uncertainty.

For the case of quarks, all mixing angles will be considered to belong to
the neighborhood of Cabibbo-like solutions of the unitarization equations
(this will be different for the case of leptons in section
\ref{section:neutrinos}, where $\theta_{23}$ will
be considered to belong to the neighborhood of the maximal mixing,
also solution of these equations).

\subsection{The simplified case $\boldsymbol{\theta_{13} = 0
=\tilde{\theta}_{13}}$}
\label{subsec:qapp}

In the neighborhood of the  solution with both $\theta_{12}$ and
$\theta_{23}$ Cabibbo-like, we write
\begin{eqnarray}
\tilde{\theta}_{12} &=& \theta_{12} + \epsilon,\cr
\tilde{\theta}_{23} &=& \theta_{23} + \eta.
\label{eq:q0}
\end{eqnarray}
The pattern $(\theta_{13} = 0 = \tilde{\theta}_{13})$ can be reasonably
considered to be  close to the experimental situation,
at least close enough for
trusting not only the relations involving the first and second generation,
but also the third one.

Like in \cite{DuretMachet2}, we impose that
the absence of $\{12\}, \{21\}$ neutral currents is violated with the
same strength as the universality of $\{11\}$ and $\{22\}$ currents.
(\ref{eq:nods0}) and (\ref{eq:ddss0}) yield
\begin{equation}
|2\eta s_{12} c_{12} s_{23} c_{23} + \epsilon (c_{12}^2 -s_{12}^2)|
= |-2\eta s_{23} c_{23}(c_{12}^2 -s_{12}^2) + 4\epsilon s_{12} c_{12}|.
\label{eq:q1}
\end{equation}
We choose the ``$+$'' sign for both sides, such that, for two generations only,
the Cabibbo angle satisfies $\tan(2\theta_{12}) = + 1/2$. (\ref{eq:q1}) yields
the ratio $\eta/\epsilon$, that we then plug into the condition
 equivalent to (\ref{eq:q1}) for the $(2,3)$ channel, coming from
(\ref{eq:nosb0})(\ref{eq:ssbb0})
\begin{equation}
|\eta c_{12}(c_{23}^2 - s_{23}^2)| = | 2\eta s_{23}c_{23}(1+c_{12}^2) 
-2\epsilon s_{12} c_{12}|.
\label{eq:q2}
\end{equation}
(\ref{eq:q1}) and (\ref{eq:q2}) yield
\begin{equation}
\tan(2\theta_{23}) = \displaystyle\frac{c_{12}}{1+c_{12}^2 -
2s_{12}c_{12}\displaystyle\frac{(s_{12}c_{12} + c_{12}^2 -
s_{12}^2)}{4s_{12}c_{12} -(c_{12}^2 -s_{12}^2)}}
\approx \displaystyle\frac{c_{12}}{2 -\displaystyle
\frac54\displaystyle\frac{s_{12}c_{12}}{\tan(2\theta_{12})
-\displaystyle\frac12}}.
\label{eq:q3}
\end{equation}
In the r.h.s. of (\ref{eq:q3}), we have assumed that $\theta_{12}$ is close to
its Cabibbo value $\tan(2\theta_{12}) \approx 1/2$. $\theta_{23}$ is  seen
to vanish with $[\tan(2\theta_{23}) -1/2]$.
The  value obtained for $\theta_{23}$ is plotted in Fig.~3  as a
function of $\theta_{12}$, together with the experimental intervals for
$\theta_{23}$ and $\theta_{12}$. There are two \cite{PDG}
 for $\theta_{12}$; the first comes
from the measures of $V_{ud}$ (black (dark) vertical lines on Fig.~3)
\begin{equation}
V_{ud} \in [0.9735,0.9740] \Rightarrow \theta_{12} \in [0.2285,0.2307],
\label{eq:Vud}
\end{equation}
and the second from the measures of $V_{us}$ (purple (light) vertical lines
 on Fig.~3)
\begin{equation}
V_{us} \in [0.2236,0.2278] \Rightarrow \theta_{12} \in [0.2255, 0.2298].
\label{eq:Vus}
\end{equation}
%
\begin{center}
\includegraphics[height=7truecm,width=9truecm]{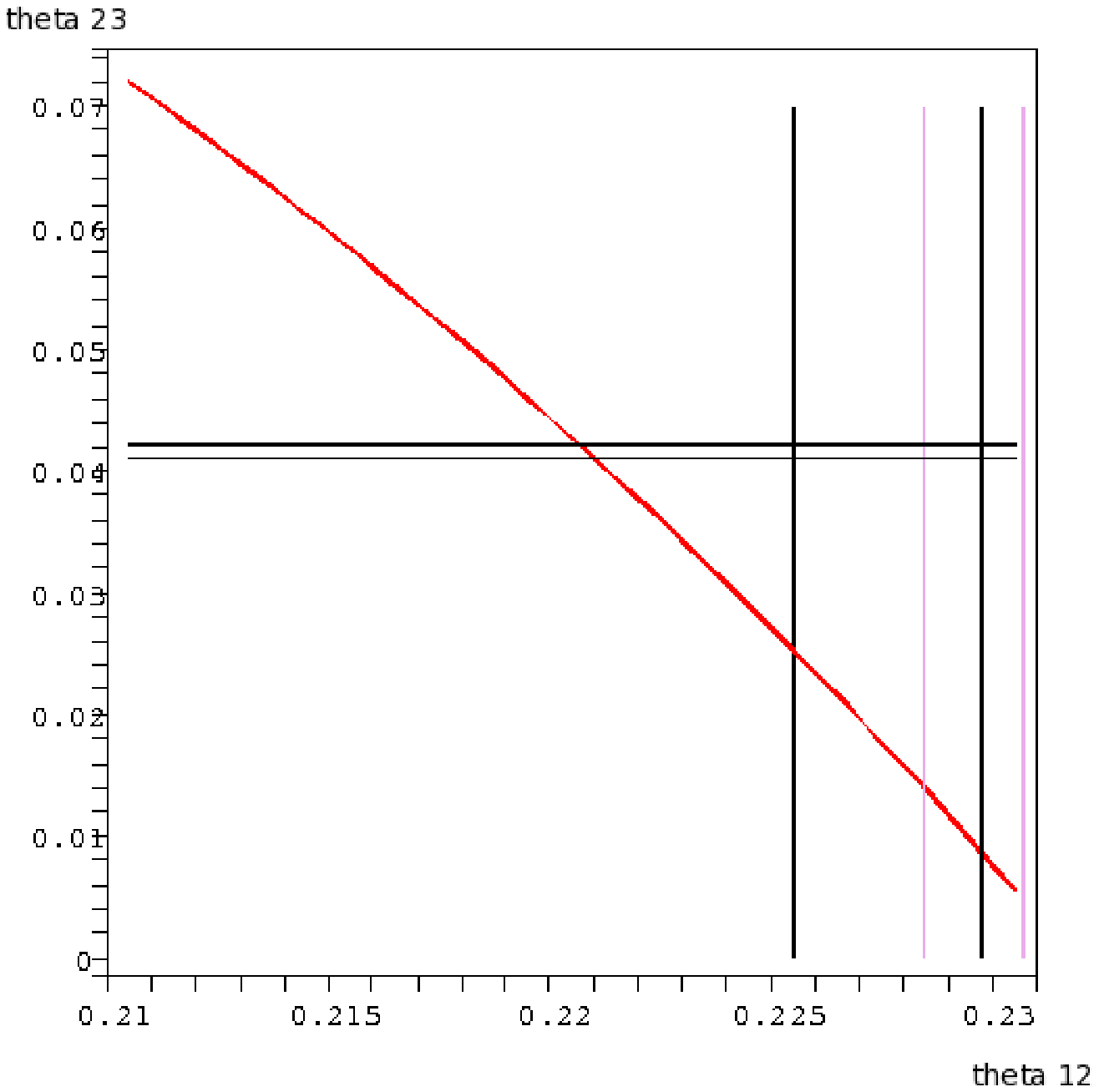}
\figskip
{\em Fig.~3: $\theta_{23}$ for quarks as a function of $\theta_{12}$;
simplified case $\theta_{13}=0=\tilde\theta_{13}$ }
\end{center}
\figskip

The measured value for $\theta_{23}$ is seen on Fig.~3
 to correspond to $\theta_{12}
\approx 0.221$, that is $\cos (\theta_{12}) \approx 0.9757$.
The value that we get for $\cos(\theta_{12})$ is accordingly $1.7\,10^{-3}$
 away from
the upper limit of the present upper bound for  $V_{ud} \equiv c_{12}c_{13}$
\cite{Vud} \cite{PDG}; it
corresponds to twice the experimental uncertainty.
It also  corresponds to $\sin(\theta_{12}) = 0.2192$,
while $V_{us} \equiv s_{12} c_{13}$ is measured to be $0.2247(19)$
 \cite{KLOE} \cite{PDG};
there, the discrepancy is $2/100$, only slightly above the $1.8/100$
relative width of the experimental interval.

The approximation which sets $\theta_{13} = 0 =\tilde{\theta}_{13}$
is accordingly reasonable,
though it yields results slightly away from experimental bounds.
We show in the next subsection that relaxing this approximation gives
results in very good agreement with present experiments.

\subsection{Going to $\boldsymbol{(\theta_{13}\not=0,
\tilde{\theta}_{13}\not=0)}$}

Considering all angles to be Cabibbo-like with, in addition to (\ref{eq:q0})

\vbox{
\begin{equation}
\tilde{\theta}_{13} = \theta_{13} + \rho,
\label{eq:q4}
\end{equation}
the l.h.s.'s of eqs. (\ref{eq:nodb}),(\ref{eq:nosb}),(\ref{eq:nods}),
(\ref{eq:ddss}), (\ref{eq:ssbb}) and the sum (\ref{eq:ddss} +
\ref{eq:ssbb}) depart respectively from zero by
\begin{subequations}
\label{subeq:general}
\begin{equation}
\eta c_{13} \left[s_{12}(c_{23}^2 - s_{23}^2) +
         2s_{13} c_{12} c_{23} s_{23}\right]
- \rho c_{12} (c_{13}^2 - s_{13}^2);
\label{eq:nodb3}
\end{equation}
\begin{equation}
\eta c_{13}\left[ - c_{12} (c_{23}^2 - s_{23}^2)
        + 2 s_{13} s_{12} c_{23} s_{23} \right]
- \rho s_{12} (c_{13}^2 - s_{13}^2);
\label{eq:nosb3}
\end{equation}
\begin{equation}
-\epsilon(c_{12}^2-s_{12}^2) + \eta\left[
s_{13}(c_{23}^2- s_{23}^2)(c_{12}^2
- s_{12}^2) -2 c_{23} s_{23} c_{12} s_{12} (1 + s_{13}^2) \right]
+ 2\rho c_{13} s_{13} c_{12} s_{12};
\label{eq:nods3}
\end{equation}
\begin{equation}
4\epsilon c_{12}s_{12}
+\eta \left[-4 s_{13} s_{12}c_{12}(c_{23}^2- s_{23}^2)
-2 c_{23} s_{23} (c_{12}^2 - s_{12}^2)(1 + s_{13}^2) \right]
+ 2\rho c_{13} s_{13} (c_{12}^2 - s_{12}^2 );
\label{eq:ddss3}
\end{equation}
\begin{equation}
-2\epsilon s_{12}c_{12}
+ \eta\left[ 2s_{13} c_{12} s_{12} (c_{23}^2 - s_{23}^2) + 2c_{23}s_{23}
\left((c_{12}^2 - s_{12}^2) + c_{13}^2 ( 1+s_{12}^2) \right)\right]
+ 2\rho c_{13} s_{13} (1+s_{12}^2); 
\label{eq:ssbb3}
\end{equation}
\begin{equation}
2\epsilon s_{12}c_{12}
+ \eta\left[ -2 s_{13} c_{12} s_{12} (c_{23}^2 - s_{23}^2) 
+ 2 c_{23} s_{23}
\left( c_{13}^2 (1+c_{12}^2) -(c_{12}^2 - s_{12}^2 ) \right) \right]
+ 2 \rho c_{13} s_{13} (1+ c_{12}^2).
\label{eq:ddbb3}
\end{equation}
\end{subequations}
}

We have added (\ref{eq:ddbb3}), which is not an independent relation, but
the sum of (\ref{eq:ddss3}) and (\ref{eq:ssbb3}); it expresses the
violation in the universality of diagonal $\{11\}$ and $\{33\}$
currents.

\subsubsection{A guiding calculation}
\label{subsub:guide}

Before doing the calculation in full generality, and to make a clearer
difference with the neutrino case, we first do it in the limit where one
neglects terms which are
quadratic in the small quantities $\theta_{13}$ and $\rho$.
By providing simple intermediate formul\ae, it enables in particular to
suitably choose the signs which occur in equating the moduli of two
quantities.  Eqs.(\ref{subeq:general}) become
\begin{subequations}
\label{subeq:simple}
\begin{equation}
\eta \left[s_{12}(c_{23}^2 - s_{23}^2)
+ 2s_{13} c_{12} c_{23} s_{23}\right] - \rho c_{12};
\label{eq:nodb3s}
\end{equation}
\begin{equation}
\eta \left[ - c_{12} (c_{23}^2 - s_{23}^2)
+ 2 s_{13} s_{12} c_{23} s_{23} \right] - \rho s_{12};
\label{eq:nosb3s}
\end{equation}
\begin{equation}
-\epsilon(c_{12}^2-s_{12}^2)
+ \eta\left[ s_{13}(c_{23}^2- s_{23}^2)(c_{12}^2 - s_{12}^2)
-2 c_{23} s_{23} c_{12} s_{12} \right];
\label{eq:nods3s}
\end{equation}
\begin{equation}
4\epsilon c_{12}s_{12} - 2\eta \left[2 s_{13} s_{12}c_{12}(c_{23}^2
- s_{23}^2) + c_{23} s_{23} (c_{12}^2 - s_{12}^2) \right];
\label{eq:ddss3s}
\end{equation}
\begin{equation}
-2\epsilon s_{12}c_{12} + 2\eta\left[ s_{13} c_{12} s_{12} (c_{23}^2
- s_{23}^2) + c_{23}s_{23} (1 +c_{12}^2 )\right]; 
\label{eq:ssbb3s}
\end{equation}
\begin{equation}
2\epsilon s_{12}c_{12} + 2\eta\left[ - s_{13} c_{12} s_{12} (c_{23}^2 
- s_{23}^2)  +  c_{23} s_{23} ( 1+ s_{12}^2)  \right].
\label{eq:ddbb3s}
\end{equation}
\end{subequations}
The principle of the method is the same as before.
From the relation (\ref{eq:nods3s}) = (-)(\ref{eq:ddss3s})
\footnote{The (-) signs ensures that $\tan(2\theta_{12}) \approx (+) 1/2$.}
, which
expresses that the absence of non-diagonal $\{12\}$ current is violated
with the same strength as the universality of $\{11\}$ and $\{22\}$
currents, one gets $\epsilon/\eta$ as a
function of $\theta_{12}, \theta_{23}, \theta_{13}$
\footnote{
\begin{equation}
\frac{\epsilon}{\eta} = s_{13}(c_{23}^2-s_{23}^2) + 2s_{23}c_{23}
\frac{s_{12}c_{12}+c_{12}^2-s_{12}^2}{4c_{12}s_{12} -(c_{12}^2-s_{12}^2)};
\label{eq:epsq}
\end{equation}
$\epsilon /\eta$ has a pole at $\tan(2\theta_{12}) = 1/2$, the suggested value
of the Cabibbo angle for two generations.
}.
 This expression is plugged in the relation 
(\ref{eq:nosb3s}) = (-)(\ref{eq:ssbb3s})\footnote{There, again, the (-)
sign has to be chosen so as to recover approximately (\ref{eq:q3}).},
which expresses the same condition for the $(2,3)$ channel;
from this, one 
extracts $\rho/\eta$ as a function of $\theta_{12}, \theta_{23},
\theta_{13}$
\footnote{
\begin{equation}
\displaystyle\frac{\rho}{\eta} = 2  c_{23} s_{23}\left[ s_{13}
-c_{12}
\left( 2\displaystyle\frac
{(c_{12}s_{12}+c_{12}^2-s_{12}^2)}{4s_{12}c_{12} -(c_{12}^2-s_{12}^2)} -
\displaystyle\frac{1+c_{12}^2}{c_{12}s_{12}}
+\displaystyle\frac{1}{s_{12}}
\displaystyle\frac{c_{23}^2-s_{23}^2}{2s_{23}c_{23}}\right)\right].
\label{eq:rhoq}
\end{equation}
$\rho/\eta$ has a pole at $\tan(2\theta_{12}) = 1/2$ 
and, for $\theta_{13}=0$, it vanishes, as expected, when
$\theta_{12}$ and $\theta_{23}$ satisfy the relation (\ref{eq:q3}),
which has been deduced for $\tilde{\theta}_{13}
(\equiv \theta_{13}+\rho)=0=\theta_{13}$.}.
The expressions that have been obtained
 for $\epsilon/\eta$ and $\rho/\eta$ are then
inserted into the third relation, \vline\ (\ref{eq:nodb3s})\ \vline\ =
\vline\ (\ref{eq:ddbb3s})\ \vline\ , which now corresponds to the $(1,3)$
channel.  This last step
yields a relation $F_0(\theta_{12},\theta_{23},\theta_{13})=1$
 between the three angles $\theta_{12},\theta_{23}, \theta_{13}$.

It turns out that $\frac{\partial F_0(\theta_{12},\theta_{23},\theta_{13})}{\partial
\theta_{13}}=0$, such that, in this case, a  condition between
  $\theta_{12}$ and $\theta_{23}$ alone eventually fulfills the three relations
under concern
\begin{equation}
1= \left|\frac{\text{viol}([11] = [22])}{\text{viol}([12] =0 = [21])}\right|
=\left|\frac{\text{viol}([22] = [33])}{\text{viol}([23] =0=[32])}\right|
=\left|\frac{\text{viol}([11] = [33])}{\text{viol}([13] =0=[31])}\right|
\Leftrightarrow \tilde F_0(\theta_{12},\theta_{23}) =1.
\label{eq:cond0}
\end{equation}
%

\vbox{
\begin{center}
\includegraphics[height=7truecm,width=9truecm]{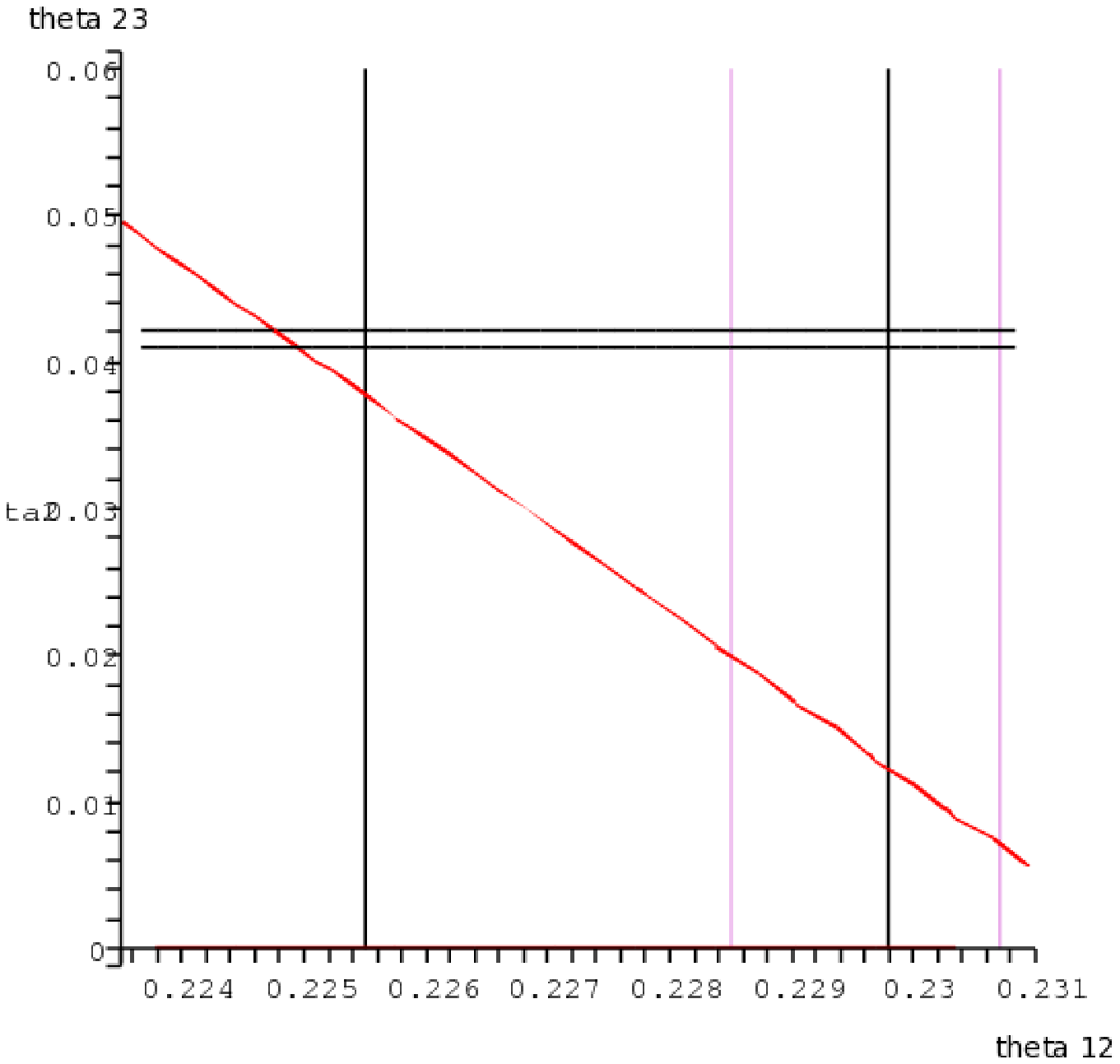}

{\em Fig.~4: $\theta_{23}$ for quarks
 as a function of $\theta_{12}$; neglecting terms
quadratic in $\theta_{13}$}
\end{center}
}
\figskip

$\theta_{23}$ is plotted on Fig.~4 as a function of $\theta_{12}$, together with the
experimental intervals for $\theta_{23}$ (black horizontal lines)
and $\theta_{12}$ (the intervals
for $\theta_{12}$  come respectively from $V_{ud}$ (eq.~(\ref{eq:Vud}),
black (dark) vertical lines)
and $V_{us}$ (eq.~(\ref{eq:Vus})), purple (light) vertical lines).

The precision obtained is much better than  in Fig.~3 since, in
particular, for $\theta_{23}$ within its experimental range,
the discrepancy
between the value that we get for $\theta_{12}$ and its lower experimental limit
coming from $V_{us}$ is smaller than the two experimental intervals, and
even smaller than their intersection.

\subsubsection{The general solution}
\label{subsub:general}

The principle for solving the general equations (\ref{subeq:general})
is the same as above.
One first uses the relation  (\ref{eq:nods3}) = (-) (\ref{eq:ddss3})
 to determine $\rho/\epsilon$ in terms of $\eta/\epsilon$.
The result is plugged in the relation (\ref{eq:nosb3}) = (-)
(\ref{eq:ssbb3}), which fixes $\eta/\epsilon$, and thus
$\rho/\epsilon$ as  functions of $(\theta_{12},\theta_{23},\theta_{13})$. These
expressions for $\eta/\epsilon$ and $\rho/\epsilon$ are finally plugged in
the relation \vline\  (\ref{eq:nodb3})\ \vline\ = \vline\ 
(\ref{eq:ddbb3})\ \vline\  , which provides a condition
$F(\theta_{12},\theta_{23},\theta_{13})=1$.
When it is fulfilled, the universality of
each pair of diagonal neutral currents of mass eigenstates
and the absence of the corresponding non-diagonal currents are violated
with the same strength, in the three channels
$(1,2)$, $(2,3)$ and $(1,3)$. 

The results are displayed in Fig.~5; $\theta_{23}$ is plotted
as a function of $\theta_{12}$ for $\theta_{13} =0, 0.004$ and $0.01$. Like
in Figs. 3 and 4, the experimental bounds on $\theta_{12}$ are depicted by
vertical black (dark) lines for the ones coming from $V_{ud}$ and purple
(light) for the ones coming from $V_{us}$; the experimental interval for
$\theta_{23}$ corresponds to the black horizontal lines.
The present experimental interval for $\theta_{13}$  is \cite{PDG}
\begin{equation}
V_{ub} = \sin(\theta_{13}) \approx \theta_{13} \in [4\,10^{-3},4.6\,10^{-3}].
\label{eq:theta3}
\end{equation}
%

\vbox{
\begin{center}
\includegraphics[height=7truecm,width=9truecm]{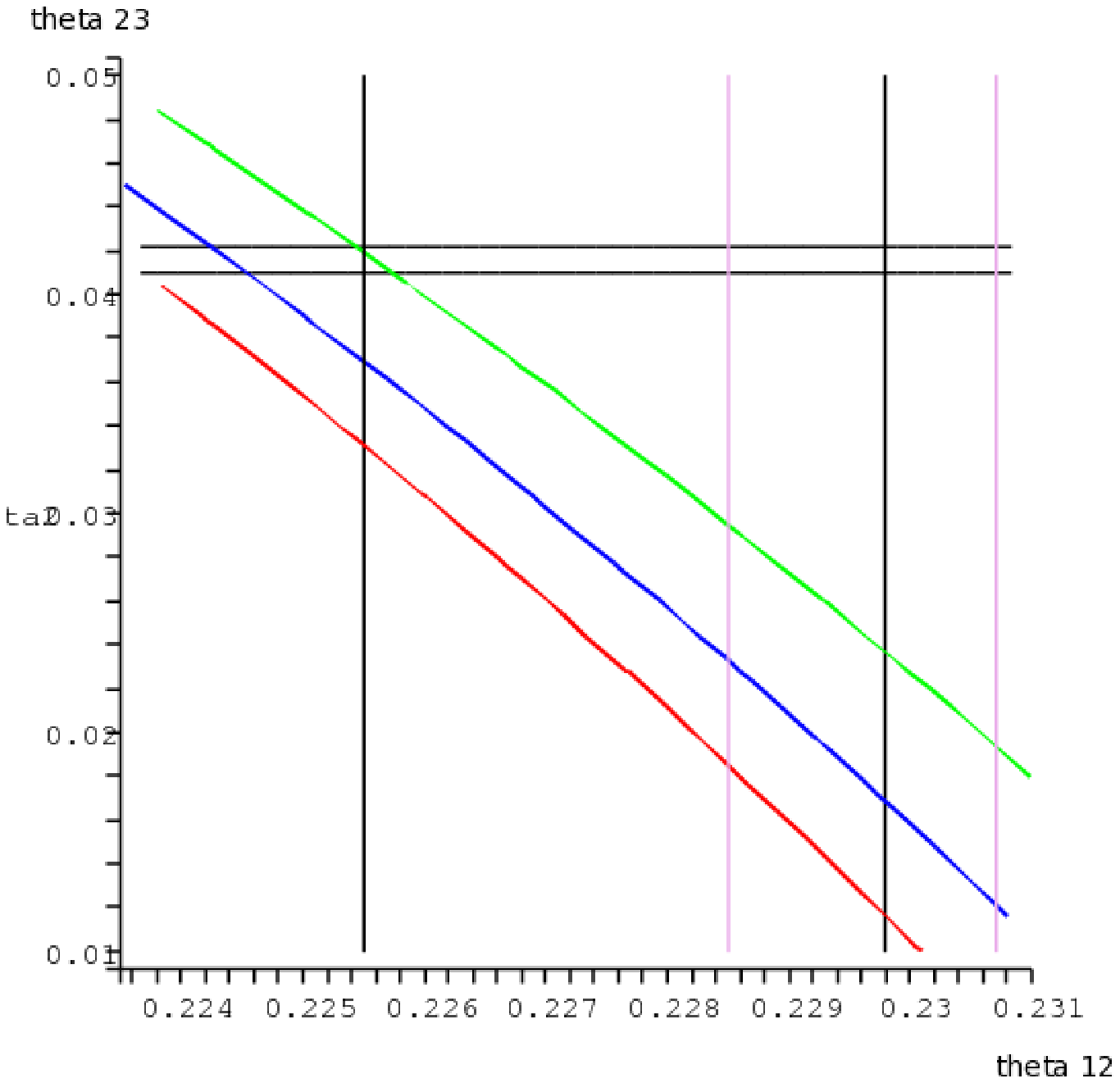}

{\em Fig.~5: $\theta_{23}$ for quarks as a function of $\theta_{12}$,
general case. 
$\theta_{13}=0$ (red, bottom), $0.004$ (blue, middle) and $0.01$ (green,
top)}
\end{center}}
\figskip
We conclude that:

$\ast$ The discrepancy between our results and experiments is
smaller than the experimental uncertainty;

$\ast$ a slightly larger value of $\theta_{13}$ and/or
 slightly smaller values of $\theta_{23}$ and/or $\theta_{12}$
still increase the agreement between our
results and experimental measurements;

$\ast$ the determination of $\theta_{12}$ from $V_{us}$ seems
preferred to that  from $V_{ud}$.

Another  confirmation of the relevance of our criterion
 is given in the next section concerning neutrino mixing angles.

\section{Beyond unitarity. A neutrino-like pattern; quark-lepton complementarity}
\label{section:neutrinos}

In the ``quark case'', we dealt with three ``Cabibbo-like'' angles.
The configuration that we investigate here is the one  in which
$\theta_{23}$ is, as observed experimentally \cite{PDG}, (close to)
 maximal, and $\theta_{12}$ and $\theta_{13}$ are
Cabibbo-like (see subsection \ref{subsection:vanish}). 
The two cases only differ accordingly from the ``classical solutions'' of the
unitarization equations away from which one makes small variations.
The criterion to fix the mixing angles stays otherwise the same.

\subsection{The  case $\boldsymbol{\theta_{13} = 0 = \tilde{\theta}_{13}}$}
\label{subsec:nu1}

We explore the vicinity of this solution,
slightly departing from the corresponding unitary mixing matrix,
by considering that $\tilde{\theta}_{12}$ now slightly differs
from $\theta_{12}$, and $\tilde{\theta}_{23}$
from its maximal value
\begin{eqnarray}
\tilde{\theta}_{12} &=& \theta_{12} + \epsilon,\cr
\theta_{23} = \pi/4 &,&  \tilde{\theta}_{23} = \theta_{23} + \eta.
\label{eq:eta}
\end{eqnarray}
The l.h.s.'s of eqs. (\ref{eq:nodb}) (\ref{eq:nosb}) (\ref{eq:nods})
(\ref{eq:ddss}) and (\ref{eq:ssbb}) no longer vanish, and become
respectively 

\vbox{
\begin{subequations}
\label{subeq:nueqs}
\begin{equation}
-\frac 12 \eta^2 (s_{12} + \epsilon c_{12}),
\label{eq:neut1}
\end{equation}
\begin{equation}
\frac 12 \eta^2 (c_{12} - \epsilon s_{12}),
\label{eq:neut2}
\end{equation}
\begin{equation}
(\ast)\ -\eta s_{12}c_{12} + \epsilon (s_{12}^2 -c_{12}^2)(1+\eta),
\label{eq:neut3}
\end{equation}
\begin{equation}
(\ast)\ -\eta (c_{12}^2 -s_{12}^2) + 4\epsilon s_{12}c_{12}(1+\eta),
\label{eq:neut4}
\end{equation}
\begin{equation}
\eta (1+c_{12}^2) -2\epsilon s_{12}c_{12}(1+\eta),
\label{eq:neut5}
\end{equation}
\end{subequations}
}

showing by which amount the five conditions under scrutiny are now 
violated. Some care has to be taken concerning the accurateness of
equations (\ref{subeq:nueqs}).
Indeed, we imposed a value of $\theta_{13}$ which is 
probably not the physical one (even if close to). It is then reasonable to
consider that channel $(1,2)$ is the less sensitive to this approximation
and that, accordingly,
 of the five equations above, (\ref{eq:neut3}) and (\ref{eq:neut4}),
marked with an ``$\ast$'', are  the most accurate
\footnote{The limitation of this approximation also appears in the fact
that  (\ref{eq:neut2}), of second order in $\eta$, is
not compatible with (\ref{eq:neut5}), which is of first order.}
.

The question: is there a special value of $\theta_{12} = \tilde{\theta}_{12}$
Cabibbo-like for which  small deviations $(\epsilon, \eta)$ from 
unitarity  entail equal strength violations of \newline
$\ast$ the absence of $\{12\}, \{21\}$ non-diagonal neutral
currents;\newline
$\ast$ the universality of $\{11\}$ and $\{22\}$ neutral currents ?

gets then a simple answer
\begin{equation}
s_{12}c_{12} = c_{12}^2 - s_{12}^2 \Rightarrow \tan(2\theta_{12}) = 2.
\label{eq:qlc2}
\end{equation}
We did not take into account the terms proportional to 
$\epsilon$ because we assumed that the mass splittings
between the first and second generations
(from which the lack of unitarity originates)
are much smaller that the ones  between
the second and the third generation
.

In the case of two generations, only $\epsilon$ appears, and
one immediately recovers from (\ref{eq:neut3}) and (\ref{eq:neut4}) the
condition fixing $\tan(2\theta_c) = 1/2$ for the Cabibbo angle.

Accordingly, the same type of requirement that led to a value of the
Cabibbo angle for two generations very close to the observed value leads,
for three generations, to a value of the first mixing angle satisfying
the quark-lepton complementarity relation (\ref{eq:qlc}) \cite{QLC}.

The values of $\theta_{12}$ and $\theta_{23}$  determined through this
procedure are very close to
the observed neutrino mixing angles \cite{PDG} \cite{GGM}. 

Though we only considered the two equations that are {\em a priori}
the least sensitive to our choice of a vanishing third mixing angle
(which is not yet confirmed experimentally),
it is instructive to investigate the sensitivity of
our solution to a small non-vanishing value of $\theta_{13}$.
This is done in Appendix \ref{section:maxsens} in which, for this purpose,
we  made the simplification $\tilde{\theta}_{13} \approx \theta_{13}$.
It turns out
that the terms proportional to $s_{13}$ in the two equations
$[12]=0=[21]$ and $\vline\ [11]\ \vline =\ \vline[22]\ \vline$ are also
proportional to $(c_{23}^2 - s_{23}^2)$, such that our solution
with $\theta_{23}$ maximal is very stable with
respect to a variation of $\theta_{13}$ around zero.
This may of course not be the case
for the other three equations, which are expected to be more sensitive to
the value of $\theta_{13}$.

\subsection{Solutions for the angle $\boldsymbol{\theta_{13}}$}
\label{subsec:nugen}

We now consider, like we did for quarks,
the general case $\theta_{13} \not= 0
\not = \tilde{\theta}_{13} (\rho\not=0)$, $\tilde{\theta}_{12} \not= \theta_{12} (\epsilon \not=
0)$, $\tilde{\theta}_{23} \not= \theta_{23} (\eta\not=0)$,  while
assigning to $\theta_{12}$ and $\theta_{23}$
their values obtained in subsection \ref{subsec:nu1}.

We investigate the eight different relations between
$\theta_{12}$, $\theta_{23}$
and $\theta_{13}$ which originate from the $2 \times 2 \times 2$ possible
 sign combinations in the conditions (\ref{eq:cond0}) (the r.h.s. is now
replaced by a condition $F(\theta_{12}, \theta_{23}, \theta_{13})=1$
involving the three mixing angles),
where each  modulus can be alternatively replaced by ``$+$'' or ``$-$''.

Among the solutions found for $\theta_{13}$,  only two (up to a
sign) satisfy the very loose experimental bound
\begin{equation}
\sin^2 (\theta_{13}) \leq 0.1.
\end{equation}
They correspond respectively to the sign combinations
$(+/-/-)$, $(+/+ /+)$, $(-/+/+)$ and $(-/-/-)$
\begin{eqnarray}
\theta_{13} = \pm 0.2717 &,& \sin^2(\theta_{13}) = 0.072,\cr
\theta_{13} = \pm 5.7\,10^{-3} &,& \sin^2(\theta_{13}) = 3.3\,10^{-5}.
\label{eq:nupred}
\end{eqnarray}
The most recent experimental bounds can be found in \cite{GGM}. They read
\begin{equation}
\sin^2 (\theta_{13}) \leq 0.05,
\end{equation}
which only leaves the smallest solution in (\ref{eq:nupred})
\footnote{These values substantially differ from the ones in
\cite{Picariello}, which mainly focuses on special textures for 
the product of the quark and neutrino mixing matrices \cite{Xing}.}.
Future experiments will confirm, or infirm, for neutrinos, the properties
that we have shown to be satisfied with an impressive accuracy
by quark mixing angles.

\section{Flavor transformations}
\label{section:NCMM}

Up to now, the observed ``pattern'' of flavor mixing has been disconnected
from flavor symmetries.  It has instead been connected with a precise scheme
of departure from unitarity of the matrix controlling gauge neutral
currents in bare flavor space.
This contrasts with most approaches which, first, focus on mass matrices
rather than gauge currents, secondly try to induce precise forms of the
latter from horizontal symmetries \cite{Ma}.
The goal of this section is to (start to) span a bridge between the two.
We shall investigate unitary flavor transformations, while restricting, for
the sake of simplicity, to the case of two flavors, in which symmetry
patterns are more
conspicuous (the presence of a third generation has been seen, for example,
 to only lightly affect the Cabibbo angle).

The most natural unitary flavor group  to be investigated is then $U(2)_f$, or 
$U(1)_f \times SU(2)_f$. For degenerate systems, this is a symmetry group of
the Lagrangian. As soon as the degeneracy is lifted, it is no longer so,
though an arbitrary unitary transformation on fermions should not change
``physics'' {\em i.e.} the physical masses and mixing angles. This last
property means that unitary flavor transformations have
 to be considered from two points of view: on one side, we will check
that physical mixing angles  do not change when fermions are
transformed, in particular that the process of renormalization by the
counterterms of Shabalin goes also unaltered in the transformation,
and, on the other side,  we will investigate which changes they induce on the
(different parts of) the Lagrangian, and how their  breaking 
can eventually be associated with the pattern of neutral currents
that seemingly controls  mixing angles observed in nature.

For non-degenerate masses, the explicit form  for the matrix
$({\cal C}^{-1})^\dagger {\cal C}^{-1}$ controlling neutral currents in the 
bare flavor basis, that has been obtained in section
\ref{subsection:1loop}) (see \ref{eq:Cd2})) provides an
``orientation'' of the relevant $SU(2)_f$ with respect to the trivial
one (the generators of which are the Pauli matrices),
which depends on the mixing angle $\theta$: 
there arises the generator ${\cal T}_z(\theta)$.
A trivial invariance of the effective Lagrangian of gauge neutral currents by 
transformation $e^{i(\alpha + \beta_z{\cal T}_z(\theta))}$ follows, which is
broken for charged currents (unless the $(d,s)$ and $(u,c)$ sectors undergo
identical transformations).

We then study possible connections between mass matrices and gauge neutral
currents.
We start by the simple case of  a constant (symmetric) mass matrix.
A link with neutral currents rapidly appears because,
apart from  trivial  terms proportional to the unit matrix,
the mass matrix is deduced from 
$({\cal C}^{-1})^\dagger {\cal C}^{-1}$ by a translation 
$\theta \to \theta -\pi/4$ of the mixing angle.
The departure of the mass matrix from (a term proportional to) unity
 is then represented
by the ${\cal T}_x$ generator of the rotated $SU(2)_f(\theta)$ mentioned
above. The commutator $ {\cal T}_y =[{\cal T}_z(\theta), {\cal T}_x(\theta)]$ is
the standard Pauli matrix,  independent of the mixing angle,

In a short paragraph, we single out  a special invariance of both
(non-trivial parts of) neutral currents and mass terms by the orthogonal,
hermitian but non unitary transformations $e^{\alpha {\cal T}_y}$ (it is
not a symmetry of the whole Lagrangian).

Then, we study general $2 \times 2$ unitary transformations on fermions.
We demonstrate, through various steps, that these transformations go across
Shabalin's renormalization and finally leave unchanged the renormalized
mixing angles.
Flavor rotations, equivalent to $e^{i\alpha {\cal T}_y}$ transformations
appear of special relevance. They are shown to continuously rotate neutral
currents into mass terms and to preserve their group structure.
As far as charged currents are concerned, their group structure
only stays unchanged when
the same rotation is performed in the $(u,c)$ ans $(d,s)$ sectors.
It then occurs that mass and flavor eigenstates can only be
aligned in  one of the two sectors.
The mixing angle of the non-aligned sector becomes then identical to the
Cabibbo angle occurring in charged currents.  
In this framework, as commented upon more at length in subsection
\ref{subsection:cc}, flavor rotations appear  as a very mildly broken flavor
subgroup of the electroweak Standard Model.

Last, we generalize this result to the renormalized mass
matrix (fermionic self-energy). Its dependence on $q^2$ leads, like 
in the general argumentation of QFT used in \cite{DuretMachet1}, 
to the presence of an  orthonormal basis of eigenstates
 for each value of $q^2$,  containing one at most among the physical
mass eigenstates. Accordingly, one reaches  the same conclusion
concerning the non-unitarity {\em a priori} of mixing matrices.
The $U(1)_{em}$ Ward identity that connects the photon-fermion-antifermion
 vertex function at zero external
momentum  to the
derivative of the inverse propagator requires that the two sides of the
identity be invariant by the same transformation
 $e^{i(\alpha + \beta_z{\cal T}_z(\theta))}$ mentioned above.
This yields a constraint that we propose to adopt because it  guarantees that
($q^2$ dependent) neutral currents and the fermionic self energy 
keep the same structure as that  encountered in the case of a constant
mass matrix;  they are in particular, again,
continuously transformed into each other by flavor rotations.
The resulting expressions are in particular, unlike textures,
stable by these transformations.

\subsection{A first type of horizontal symmetries}
\label{subsection:firstunit}

We exhibit below specific  flavor transformations transformations that
leave parts  of the gauge electroweak Lagrangian invariant.
We deal with the case of two generations, which makes  an easy link with
\cite{DuretMachet2}, and  consider for example the $(d,s)$ channel.
The corresponding neutral currents in the basis of bare flavor eigenstates
are controlled by the product $({\cal C}_d^{-1})^\dagger {\cal C}_d^{-1}$.

When ${\cal C}_d$ departs from unitarity, we parametrize it like in
(\ref{eq:Cd1})
in which the role of $A_d$ is now played by $\epsilon_d$,
such that (\ref{eq:Cd2}) becomes
\begin{equation}
({\cal C}_d^{-1})^\dagger(\theta_{dL}) {\cal C}_d^{-1}(\theta_{dL})
= 1 + 2\epsilon_d\, {\cal T}_z(\theta_{dL}),
\label{eq:CC}
\end{equation}
where the expression for ${\cal T}_z$ has been given in (\ref{eq:Tzy}).
Whatever be $\theta_{dL}$, the  unitary transformation
\begin{equation}
\Omega_z(\alpha_d,\beta_d,\theta_{dL}) = e^{i(\alpha_d +\beta_d {\cal
T}_z(\theta_{dL}))}
\label{eq:Omega}
\end{equation}
with  arbitrary $\alpha_d$ and  $\beta_d$, acting on
$\left(\begin{array}{c} d^0_{fL} \cr s^0_{fL} \end{array}\right)$,
satisfies
\begin{equation}
\Omega_z^\dagger(\alpha_d,\beta_d,\theta_{dL})\ \left[({\cal
C}_d^{-1})^\dagger(\theta_{dL})
{\cal C}_d^{-1}(\theta_{dL})\right]\ 
\Omega_z(\alpha_d,\beta_d,\theta_{dL}) =
({\cal C}_d^{-1})^\dagger(\theta_{dL}) {\cal C}_d^{-1}(\theta_{dL}),
\end{equation}
and, thus, leaves invariant Lagrangian for gauge neutral currents
\begin{equation}
\left(\begin{array}{cc} \bar d^0_{fL} & \bar s^0_{fL} \end{array}\right)
 W_\mu^3  \gamma^\mu_L\,
\left[({\cal C}_d^{-1})^\dagger(\theta_{dL}){\cal C}_d^{-1}(\theta_{dL})\right]\,
\left( \begin{array}{c} d^0_{fL} \cr s^0_{fL} \end{array}\right).
\end{equation}
It is accordingly a horizontal
group of invariance of the gauge Lagrangian for neutral currents
in the space of bare flavor states
\footnote{
Things become clearer in the proper basis of $\Omega_z$, which is also the
one of ${\cal T}_z(\theta_{dL})$. Its eigenvectors are
\begin{equation}
        \frac{1}{\sqrt{2}}\left(\begin{array}{r} -\sqrt{1+\sin 2\theta_{dL}} \cr
  \sqrt{1-\sin 2\theta_{dL}}   \end{array}\right),\quad
      \frac{1}{\sqrt{2}}\left(\begin{array}{r}   \sqrt{1-\sin 2\theta_{dL}} \cr
  \sqrt{1+\sin 2\theta_{dL}} \end{array}\right)
\label{eq:proper}
\end{equation}
and the diagonalized ${\cal T}_z (\theta_{dL})$ is
\begin{equation}
D_{T_z} =\frac12\left(\begin{array}{cc}  1  &  \cr  & -1
\end{array}\right) = T^3.\label{eq:diagC}
\end{equation}
Accordingly, in the proper ($\theta_{dL}$ dependent) basis, the horizontal group
of invariance is
\begin{equation}
\Omega(\alpha_d,\beta_d) = e^{i(\alpha_d + \beta_d T^3)},
\end{equation}
that is, up to an arbitrary phase, a $U(1)$ transformation by the $T^3$
subgroup of the horizontal $SU(2)_f$ symmetry associated to the triplet of
neutral currents in the $(d,s)$ channel.
In the proper basis, the neutral currents become controlled by $1 + \epsilon_d
D_{T_z} = 1 + \epsilon_d T^3$, close to unity like in the mass basis
 and in the flavor basis.

The proper basis $\left(\begin{array}{cc} d_p \cr s_p\end{array}\right)$
 of $({\cal C}_d^{-1})^\dagger {\cal C}_d^{-1}, {\cal T}_z(\theta_{dL})$,
and $\Omega_z$ can be easily expressed in terms of the  renormalized
mass basis (\ref{eq:xi})
$\left(\begin{array}{cc} d_p \cr s_p\end{array}\right)
= P_d^\dagger 
\left(\begin{array}{cc} d^0_{fL} \cr s^0_{fL}\end{array}\right)
=
P_d^\dagger {\cal C}_d \left(\begin{array}{c} d_{mL}\cr s_{mL}\end{array}\right)$,
where $P_d$ is the unitary matrix the columns of which are the eigenvectors
(\ref{eq:proper}): $P_d = P_d^\dagger = \left(\begin{array}{rr}
-\cos\zeta_d & \sin\zeta_d \cr \sin\zeta_d & \cos \zeta_d
\end{array}\right)$ with $\tan \zeta_d =
\frac{\sqrt{ 1 - \sin 2\theta_{dL}}}{ \sqrt{1+\sin 2\theta_{dL}}} \Rightarrow \tan
2\zeta_d = \frac{1}{\tan 2\theta_{dL}} \Rightarrow \zeta_d = -\theta_{dL} +
\frac{\pi}{4} + k\frac{\pi}{2}$. ${\cal C}_d$ being given by
(\ref{eq:Cud0}) and (\ref{eq:Cd1}), $P^\dagger_d {\cal C}_d$ writes
$\left(\begin{array}{rr} -\cos(\theta_{dL}+\zeta_d)
& \sin(\theta_{dL}+\zeta_d) \cr \sin(\theta_{dL}+\zeta_d) & \cos(\theta_{dL}+\zeta_d)
\end{array}\right) \equiv \left(\begin{array}{rr}
-\cos(\frac{\pi}{4}+k\frac{\pi}{2}) & \sin(\frac{\pi}{4}+k\frac{\pi}{2})
\cr \sin(\frac{\pi}{4}+k\frac{\pi}{2}) &
\cos(\frac{\pi}{4}+k\frac{\pi}{2}) \end{array}\right)$ up to corrections in
$\epsilon_d$.
}.

As can be seen on (\ref{eq:LC}), such transformations
$\left(\begin{array}{cc} d^0_{fL} \cr s^0_{fL} \end{array}\right) 
\to e^{i(\alpha_d +\beta_d {\cal T}_z(\theta_d))}\left(\begin{array}{cc}
d^0_{fL} \cr s^0_{fL}
\end{array}\right)$,
$\left(\begin{array}{cc} u^0_{fL} \cr c^0_{fL} \end{array}\right) 
\to e^{i(\alpha_u +\beta_u {\cal T}_z(\theta_{uL}))}\left(\begin{array}{cc}
u^0_{fL} \cr c^0_{fL}
\end{array}\right)$,
 acting independently in
the $(u,c)$ and $(d,s)$ sectors (with different parameters), do not leave
the gauge charged currents invariant.

A special invariance of the non-trivial parts of both neutral gauge
currents and mass terms by a non-unitary transformation will also be
exhibited in subsection \ref{subsub:special}.

\subsubsection{The example of the Cabibbo angle $\boldsymbol{\tan 2\theta_c
= \frac12}$}

The value of the Cabibbo angle $\tan 2\theta_c =\frac12$
\cite{DuretMachet2} corresponds to
$\sin 2\theta_c = \frac{1}{\sqrt{5}}$ and, so,
${\cal T}_z(\theta_c) =\displaystyle\frac{1}{2\sqrt{5}}\left(\begin{array}{rr} 1 & -2
\cr -2 & -1 \end{array}\right) = \frac12 t_c, t_c^2 =1$. In this case, the
horizontal group of invariance of the corresponding neutral currents
$({\cal C}^{-1})^\dagger {\cal C}^{-1}$ in bare flavor space is,
$(\alpha_d, \beta_d)$ being arbitrary parameters
\begin{equation}
\Omega_z(\alpha_d, \beta_d, \theta_c) =  e^{i(\alpha_d + \beta_d t_c)}
= e^{i\alpha_d}(\cos \beta_d + it_c \sin\beta_d) 
=e^{i\alpha_d}
\left(\begin{array}{cc}
\cos\beta_d + \frac{i}{\sqrt{5}}\sin\beta_d &
-\frac{2i}{\sqrt{5}}\sin\beta_d \cr
-\frac{2i}{\sqrt{5}}\sin\beta_d &
\cos\beta_d - \frac{i}{\sqrt{5}}\sin\beta_d \end{array}\right).
\end{equation}
Note that it is of the form $\left(\begin{array}{cc} \alpha  & -2(\alpha
-\beta) \cr -2(\alpha-\beta) & \beta \end{array}\right)$ like $({\cal
C}^{-1})^\dagger {\cal C}^{-1}$, belonging to the same group of matrices
(see subsection \ref{subsub:consM}).

\subsection{Gauge currents versus mass matrices}

In this work, the determination of mixing angles has been disconnected from
the knowledge and / or assumptions concerning mass matrices, {\em e.g.}
textures.
In addition to the fact, already mentioned, that a single constant mass
matrix cannot account for the properties of coupled systems in QFT
 \cite{MaNoVy}\cite{Novikov}, the
limitations of putting the emphasis on mass matrices have already often been
stressed. Textures are unstable by unitary transformations on fermions
 and cannot
represent genuine physical properties of the system under consideration.
In \cite{JarlskogQLC} it was explicitly shown how one can obtain, for
example, bi-maximal mixing matrices without Dirac mass matrices playing any
role. On these grounds, this last work casts serious doubts on the
relevance of the Quark-Lepton Complementarity relation, which does not rely
on ``invariant'' relations and quantities.
That the Golden ratio value for $\tan \theta_c$
 can be recovered from special textures (see for
example \cite{Strumia}) can thus only be considered as a special case of
some more general properties.  

It is noticeable that the way we obtained these two properties stays
independent of any assumption concerning mass matrices, since the
remarkable properties at work concern gauge currents.

The problem that comes to mind is clearly whether a bridge can be
spanned between gauge currents and some class of mass matrices.
We just make a few remarks below; in a first step we shall consider a
(abusively) single constant mass matrix; then, we will consider
renormalized, $q^2$ dependent, mass matrices.

\subsubsection{The case of a constant mass matrix}
\label{subsub:consM}

We have demonstrated in subsection \ref{subsection:1loop} that 
non-unitary mixing matrices arise in the diagonalization of renormalized
kinetic terms; this does not depend on the form of the classical mass matrix
$M_0$.
We consider, in a first step, the simple case of a binary system endowed
with a real symmetric  mass matrix 
\begin{equation}
M_0 = \left(\begin{array}{cc} a & c \cr c & b \end{array}\right).
\end{equation}
Calling $m_1$ and $m_2$ its eigenvalues, one can re-parametrize
\begin{eqnarray}
{M_0}  &=& m + \Delta m\  {\cal T}_x(\theta),\ 
{\cal T}_x(\theta)=\frac12
\left(\begin{array}{rr} \cos 2\theta & \sin 2\theta \cr
\sin 2\theta & -\cos 2\theta \end{array}\right),\cr 
&& m=\frac{m_1 + m_2}{2},\ \Delta m = {m_1 - m_2},
\label{eq:Mbin}
\end{eqnarray}
where $\theta$ is the (classical) mixing angle arising from the diagonalization of
$M_0$.  It satisfies
\begin{equation}
\tan 2\theta = \frac{2c}{a-b}.
\label{eq:tan2}
\end{equation}

That a given mixing angle can be related to infinitely many 
different mass patterns clearly appears since, for example, shifting $M$
by $\kappa\ \times$ the unit matrix does not change
the mixing angle, does not change $\Delta m$ either and shifts each individual
eigenvalue by $\kappa$.  In particular, a value of $\kappa$ much larger than
$m_{1,2}$ leads to a quasi-degenerate binary system, the mixing angle of
which stays nevertheless the same since
$\tan 2\theta = 2c/\sqrt{(\Delta m)^2 -4c^2}$ is unchanged. Also, two mass
matrices proportional to each other have the same mixing angle though their
eigenvalues have the same proportionality factor (mass ratios keep the same
in this case)
\footnote{Any homographic transformation on a mass matrix $M$:
 $M \to \frac{\alpha M +
\beta}{\delta M + \gamma}$ preserves the eigenvectors of $M$ and thus the
mixing angles.}. Trying to
explain a given mixing pattern by a specific mass matrix is thus illusory
because it cannot tackle the problem in its generality.

Shifting $M_0$ by a constant 
is a particular one among the transformations that leave the r.h.s. of
 (\ref{eq:tan2}) unchanged, {\em i.e.} the ones such that
$\frac{2c}{a-b} =$ cst.  The set $\{\Theta(u)\}$ of such matrices
\footnote{This set is of interest to us because, as we recalled in section
\ref{section:quarks}, the Cabibbo angle empirically corresponds to $u\equiv
\tan 2\theta_c =1/2$, and, as we showed in sections \ref{section:quarks} and
\ref{section:neutrinos}, the same structure underlies,
 for three generations, quark and leptonic mixing angles. The empirical
criterion equating the violation of universality and that of the absence of
FCNC's corresponds to the same structure, in which the difference of
diagonal elements of a symmetric $2\times 2$ matrix is identical, up to a
sign, to its off-diagonal one.} 
\begin{equation}
\Theta(u) =\left(\begin{array}{cc}
a & \displaystyle\frac u2 (a-b) \cr \displaystyle\frac u2(a-b) & b \end{array}\right)
= \frac{a+b}{2} + \frac{a-b}{2}
\left(\begin{array}{rr} 1 & u \cr u & -1 \end{array}\right)
\label{eq:group}
\end{equation}
form,  for any given $u$, a real abelian group, spanned by the two
matrices $1$ and $\displaystyle\frac12\frac{1}{\sqrt{1+u^2}}
\left(\begin{array}{rr} 1 & u \cr u & -1 \end{array}\right)$.

Interesting connections can  be obtained as follows.
Comparing (\ref{eq:CC}) and  (\ref{eq:Mbin}), one gets:
\begin{equation}
\frac{({\cal C}^{-1})^\dagger {\cal C}^{-1} -1}{2\epsilon}(\theta)
=\frac{{M_0}-m}{\Delta m}(\theta-\pi/4).
\end{equation}
It then appears natural to consider the three
 $SU(2)_f(\theta)$ generators (anticommuting matrices with eigenvalues
$\pm 1/2$)
\begin{equation}
{\cal T}_x(\theta) = \frac12 \frac{1}{\sqrt{1+u^2}} \left(\begin{array}{rr} 1 & u \cr u & -1
\end{array}\right),\ 
{\cal T}_y = \frac12 \left(\begin{array}{rr} & -i \cr i &
\end{array}\right),\ 
{\cal T}_z(\theta) = \frac12  \frac{1}{\sqrt{1+u^2}}\left(\begin{array}{rr} u & -1
\cr -1 & -u
\end{array}\right),
\label{eq:SU2}
\end{equation}
such that, parametrizing
\footnote{For $u\equiv \tan 2\theta$ to be continuous, we have to restrict, for
example, $\theta$ to the interval $]-\pi/4, +\pi/4[$.}
\begin{equation}
\cos 2\theta = \displaystyle\frac{1}{\sqrt{1+u^2}},
\  \sin 2\theta = \displaystyle\frac{u}{\sqrt{1+u^2}},
\end{equation}
one has, like in (\ref{eq:Mbin}) and (\ref{eq:CC})
\begin{eqnarray}
&&{M_0} = m + \Delta m\, {\cal T}_x(\theta), \cr
 &&({\cal C}^{-1})^\dagger {\cal C}^{-1} = 1 + 2\epsilon\, {\cal T}_z(\theta).
\end{eqnarray}
The $\vec{\cal T}(\theta)$'s are related to the standard $SU(2)$ generators $\vec
T$ defined in (\ref{eq:TTTflav}) by 
\begin{equation}
\left(\begin{array}{r} {\cal T}_x(\theta) \cr {\cal T}_z(\theta) \end{array}\right) =
 R(u) \left(\begin{array}{cc} {T}_x \cr {T}_z \end{array}\right),\ 
 R(u) = \frac{1}{\sqrt{1+u^2}}
\left(\begin{array}{rr}  u & 1 \cr -1 & u \end{array}\right);\
R^T(u)R(u)=1.
\label{eq:rotTT}
\end{equation}
``Mass terms'' and neutral currents 
are transformed into one another by the action of (see (\ref{eq:rot}) for the
definition of the rotation $\cal R$)
\begin{equation}
e^{i\gamma {\cal T}_y} = \cos \frac{\gamma}{2} + 2i{\cal T}_y
\sin\frac{\gamma}{2} = {\cal R}(\frac{\gamma}{2}).
\label{eq:Ty1}
\end{equation}
Indeed,
\begin{equation}
\left(\begin{array}{c} 
e^{-i\gamma {\cal T}_y}  {\cal T}_x(\theta) e^{i\gamma {\cal T}_y}\cr
e^{-i\gamma {\cal T}_y}  {\cal T}_z(\theta) e^{i\gamma {\cal T}_y}
\end{array}\right) = {\cal R}(-\gamma)
\left( \begin{array}{c} 
{\cal T}_x(\theta) \cr {\cal T}_z(\theta)
\end{array}\right),
\end{equation}
which we rewrite
\footnote{In terms of neutral currents and mass terms, one has
\begin{eqnarray}
 e^{-i\gamma {\cal T}_y} {M_0} e^{i\gamma {\cal T}_y}&=&
m + \Delta m ( {\cal T}_x(\theta) \cos \gamma -{\cal T}_z(\theta) \sin\gamma)
\approx (m + \Delta m {\cal T}_x(\theta)) -\gamma \Delta m {\cal T}_z(\theta)\cr
&& = {M_0} -\gamma \frac{\Delta m}{2\epsilon} \left( ({\cal
C}^{-1})^\dagger {\cal C}^{-1}-1
\right),\cr
 e^{-i\gamma {\cal T}_y} ({\cal C}^{-1})^\dagger {\cal C}^{-1} e^{i\gamma {\cal T}_y}&=&
1  + 2\epsilon ( {\cal T}_x(\theta) \sin \gamma + {\cal T}_z(\theta) \cos\gamma)
\approx (1 +2\epsilon {\cal T}_z(\theta)) + 2\epsilon \gamma{\cal T}_x(\theta)\cr
&& = ({\cal C}^{-1})^\dagger {\cal C}^{-1} +  \gamma \frac{2\epsilon}{\Delta m} ({M_0}-m).
\label{eq:MCC}
\end{eqnarray}
}

\begin{equation}
\left(\begin{array}{r} \hat{\cal T}_x(\theta) \cr \hat{\cal T}_z(\theta) \end{array}\right)
= {\cal R}(-\gamma)
\left(\begin{array}{r} {\cal T}_x(\theta) \cr {\cal T}_z(\theta) \end{array}\right),
\quad
\hat{\cal T}_{x,z}(\theta)=e^{-i\gamma{\cal T}_y} {\cal T}_{x,z}(\theta) e^{i\gamma{\cal T}_y}.
\label{eq:rotTbis}
\end{equation}
Comparing (\ref{eq:Ty1}) and (\ref{eq:rotTbis}) shows that  $e^{i\gamma T_y}$,
 which shifts $\theta$ by $\gamma/2$,
rotates fermions by  $\gamma/2$, but rotates
the ${\cal T}_x(\theta)$ and ${\cal T}_z(\theta)$ generators by $(-\gamma)$. In
particular, when rotating the fermions by $\pi/4$, {\em i.e.} taking
$\gamma = \pi/2$, ${\cal T}_x(\theta) \to 
-{\cal T}_z(\theta), \ {\cal T}_z(\theta) \to {\cal T}_x(\theta)$.

Combining with (\ref{eq:rotTT}), one finds

\vbox{
\begin{eqnarray}
\left(\begin{array}{r} \hat{\cal T}_x(\theta) \cr \hat{\cal T}_z(\theta) \end{array}\right)
&=& \left(\begin{array}{rr} \sin(\gamma + 2\theta) & \cos(\gamma + 2\theta) 
\cr -\cos(\gamma + 2\theta)  & \sin(\gamma + 2\theta)  
\end{array}\right)
\left(\begin{array}{r} T_x \cr T_z \end{array}\right)\cr
&=&  {\cal R}\left(-(\gamma + 2\theta - \frac{\pi}{2})\right)
\left(\begin{array}{r} T_x \cr T_z \end{array}\right)\cr
&=& \left(\begin{array}{c}
e^{-i(2\theta + \gamma -\frac{\pi}{2})T_y}\; T_x\; e^{i(2\theta + \gamma
-\frac{\pi}{2})T_y} \cr
e^{-i(2\theta + \gamma -\frac{\pi}{2})T_y}\; T_z\; e^{i(2\theta + \gamma
-\frac{\pi}{2})T_y} \end{array}\right).
\end{eqnarray}
}

The rotation matrix occurring is exactly of the same type as $R(u)$
occurring in (\ref{eq:rotTT}), with
its argument shifted from $2\theta$ to $2\theta + \gamma$.
(\ref{eq:rotTT}) rewrites in particular
\begin{equation}
\left(\begin{array}{c} {\cal T}_x(\theta) \cr {\cal T}_y \cr {\cal T}_z(\theta)
\end{array}\right)
=e^{-2i(\theta-\frac{\pi}{4})T_y}\left(\begin{array}{c}
 T_x \cr T_y \cr T_z
\end{array}\right)e^{2i(\theta-\frac{\pi}{4})T_y}.
\end{equation}
(\ref{eq:SU2}) shows that one recovers the standard $SU(2)$ generators
$T_x, T_y, T_z$ at the limit $u \to +\infty$ ($\theta \to\pi/4$);
when $u \to -\infty$ ($\theta \to -\pi/4$), ${\cal T}_{x,z}(\theta) \to
-T_{x,z}$;
at the limit $u \to 0$ ($\theta=0$), ${\cal T}_x(\theta) \to T_z,
{\cal T}_z(\theta) \to -T_x$.

By the transformation (isomorphic to $Z_2$)  $u \rightarrow -1/u$,
${\cal T}_x(\theta) \to {\cal T}_z(\theta),\ {\cal T}_z(\theta) \to -{\cal
T}_x(\theta)$. 
It corresponds to the transformation $\tan 2\theta \to -1/\tan 2\theta$,
which is an outer automorphism
of the $SU(2) \times U(1)$ (or $U(2)$) algebra under scrutiny.
One can also speak of an infinite set of $SU(2)_f$, depending of the
continuous parameter $u$. This set is divided by the transformation $u \to
-1/u$ into two  subsets, respectively with generators
$\{{\cal T}_x(\theta),{\cal T}_y,{\cal T}_z(\theta)\}$ and
$\{{\cal T}_z(\theta),{\cal T}_y,-{\cal T}_x(\theta)\}$. They intersect
along the $U(1)$ group with generator ${\cal T}_y$, which is independent of
$\theta$.

\bigskip

\subsubsection{A special invariance}
\label{subsub:special}

In subsection \ref{subsection:firstunit}, we encountered the unitary
transformations $\Omega_z$ which leave invariant the Lagrangian of neutral
currents. Likewise, we can define  transformations
$\Omega_x = e^{i(\alpha + \beta {\cal T}_x(\theta))}$, which, due to
(\ref{eq:Mbin}), leave mass terms invariant. None is a symmetry of both
terms: neutral currents are not
invariant by $\Omega_x$, nor are mass terms by $\Omega_z$.

There exist a special invariance satisfied by the non-trivial parts of the
mass matrix and of neutral currents, which results from the anticommutation
of ${\cal T}_y$ with ${\cal T}_x$ and ${\cal T}_z$.
Both $\displaystyle\frac{{M_0}-m}{\Delta m}$
and $\displaystyle\frac{({\cal C}^{-1})^\dagger {\cal C}^{-1} -1}{2\epsilon}$
satisfy
\begin{equation}
O^\dagger \frac{{M_0}-m}{\Delta m} O = \frac{{M_0}-m}{\Delta m},\ 
O^\dagger \frac{({\cal C}^{-1})^\dagger {\cal C}^{-1} -1}{2\epsilon} O =
\frac{({\cal C}^{-1})^\dagger {\cal C}^{-1} -1}{2\epsilon},
\end{equation}
where $O$ is the orthogonal matrix depending on an arbitrary real parameter
$\alpha$
\begin{equation}
O= \left(\begin{array}{rr} \cosh \alpha & -i \sinh\alpha \cr
i \sinh \alpha & \cosh\alpha \end{array}\right)
\equiv   e^{i\alpha\fontsize{7}{7}\selectfont \left(\begin{array}{rr}  & -1 \cr  1 &
\end{array}\right)} = e^{2\alpha{\cal T}_y},\ 
 O O^T =1,\ O^\dagger O  \not = 1.
\end{equation}
This transformation is non-unitary, such that the trivial parts of the
matrices for mass and neutral currents (the ones proportional to the unit
matrix) are not invariant.

It is also noticeable that the corresponding parts of the Lagrangian are
{\em not}
invariant by the unitary $U(1)$ rotation $\tilde O$  obtained by going to
imaginary $\alpha$. At the opposite, the trivial parts of the
corresponding mass terms and
gauge neutral currents, which are not invariant by $O$,
are  invariant by $\tilde O$.

\subsubsection{Unitary transformations on fermions}
\label{subsub:unitfer}

$\Delta m$ (mass splitting) and $2\epsilon$ (lack of unitarity of the
mixing matrix) cannot  be but tightly connected; they are in particular
 expected to
vanish simultaneously. When both vanish, the mixing angle is undetermined:
mass terms, proportional to the unit matrix, are trivially invariant by
the $SU(2)_f(\theta) \times U(1)_f$ flavor symmetry;  so are the terms
corresponding to neutral currents in the gauge Lagrangian.

As soon as the degeneracy is lifted, this symmetry is broken:
mass terms and neutral currents  are no longer invariant.
However, it is the
common belief that ``physics'' should not depend on arbitrary unitary
flavor transformations on the fermion fields.
So, on one side, we will check this point and, on the other side, we will
study how different parts of the Lagrangian transform,
 putting a special emphasis on flavor rotations.

\paragraph{Flavor rotations}
\label{par:flarot}
\hfill\newline

According to (\ref{eq:Ty1}), they are strictly equivalent (up to a phase) to
transformations $\Omega_y = e^{i(\alpha + \beta {\cal T}_y(\theta))}$.
We consider
\begin{equation}
 \left(\begin{array}{c} d^0_{fL} \cr s^0_{fL}\end{array}\right) \to
{\cal R}(\varphi) \left(\begin{array}{c} d^0_{fL} \cr s^0_{fL}\end{array}\right),
\label{eq:unitr}
\end{equation}
which is  equivalent to a transformation $e^{2i\varphi {\cal T}_y}$.

Concerning mass terms and neutral currents (in the original
(bare) flavor basis), they respectively transform according to
(see (\ref{eq:CC}) and (\ref{eq:Mbin}))
\begin{eqnarray}
{\cal R}^\dagger(\varphi) {\cal T}_x(\theta) {\cal R}(\varphi) =
{\cal T}_x(\theta+\varphi),\quad
{\cal R}^\dagger(\varphi) {\cal T}_{z}(\theta) {\cal R}(\varphi) =
{\cal T}_z(\theta+\varphi),
\label{eq:neutran}
\end{eqnarray}
which consistently shifts the  angle $\theta \to \theta+\varphi$.
Such transformations, in particular, rotate continuously mass terms
into neutral currents (see also (\ref{eq:rotTbis})).

To ascertain that they ``do not change physics'' (given that
they are obviously not symmetries of the Lagrangian), we must check that
physical mixing angles are not changed by such transformations, in
particular the Cabibbo angle occurring in charged currents of renormalized
mass states.
We accordingly consider (\ref{eq:unitr}) acting on $(d^0_{fL}, s^0_{fL})$,
together with
\begin{equation}
\left(\begin{array}{c} u^0_{fL} \cr c^0_{fL}\end{array}\right)\to
{\cal R}(\vartheta) \left(\begin{array}{c} u^0_{fL} \cr c^0_{fL}\end{array}\right).
\label{eq:unitr2}
\end{equation}

By the unitary ${\cal R}(\vartheta,\varphi)$, the $(u,c)$ classical
 mass matrix is 
left-multiplied by ${\cal R}^\dagger(\vartheta)$ and the $(d,s)$ one by
${\cal R}^\dagger(\varphi)$.  In the diagonalization process by a bi-unitary
transformation, the unitary matrices
${\cal C}_{d0}$ and ${\cal C}_{u0}$ (see subsection \ref{subsection:1loop})
have simply to be changed into
 $\hat{\cal C}_{u0}= {\cal R}^\dagger(\vartheta) {\cal C}_{u0}$ and  $\hat{\cal
C}_{d0}= {\cal R}^\dagger(\varphi) {\cal C}_{d0}$ (in this simple case of
rotations, the classical angles linking the original flavor states to
the new mass states have become
$\hat\theta_{uL} = \theta_{uL} +\vartheta,\ \hat\theta_{dL} = \theta_{dL} +
\varphi$
\footnote{Since $\varphi$ and  $\vartheta$ are both free, by tuning the
former to $-\theta_{dL}$ and the latter to  $-\theta_{uL}$,
 one can tune  both $\hat{\cal C}_{d0} = {\cal R}^\dagger(\varphi) {\cal C}_{d0}$
and $\hat{\cal C}_{u0}= {\cal R}^\dagger(\vartheta){\cal C}_{u0}$ to the unit
matrix: the mixing angles connecting, in both sectors, the new classical 
mass states to the initial flavor states,  can thus be cast to zero
(the mixing angles connecting the rotated bare flavor states to the
new classical mass eigenstates are left unchanged by the rotation).
However, if one considers charged currents in flavor space, we show after
(\ref{eq:ch2})  that their group structure stays unaltered only if the
two arbitrary flavor rotations become identical; this  accordingly
favors a common arbitrary flavor rotation in the two sectors.  See also
appendix \ref{section:align} for the reverse statement that, by a flavor
rotation, one can always align the new flavor states to the classical mass
states.  \label{footnote:newmix}}).
So doing, the bare masses stay the same.
The new classical mass eigenstates are 
$\left(\begin{array}{c} \hat u^0_{mL} \cr \hat c^0_{mL} \end{array}\right)
 =  {\cal C}^\dagger_{u0} {\cal R}(\vartheta) \left(\begin{array}{c} u^0_{fL} \cr
c^0_{fL} \end{array}\right), \ 
\left(\begin{array}{c} \hat d^0_{mL} \cr \hat
s^0_{mL} \end{array}\right) ={\cal C}^\dagger_{d0} {\cal R}(\varphi) 
\left(\begin{array}{c} d^0_{fL} \cr s^0_{fL} \end{array}\right)$; they are
deduced from the original ones by the transformations ${\cal C}^\dagger_{u0}
{\cal R}(\vartheta) {\cal C}_{u0} \equiv {\cal R}(\vartheta)$ and
${\cal C}^\dagger_{d0}{\cal R}(\varphi) {\cal C}_{d0} \equiv {\cal
R}(\varphi)$.
By the action of  ${\cal R}(\vartheta)$ and ${\cal R}(\varphi)$,
the classical charged currents Lagrangian  becomes
$\left(\begin{array}{cc} \bar u^0_{fL} & \bar c^0_{fL}\end{array}\right)
 {\cal R}^\dagger(\vartheta) W\hskip -3mm/ \; {\cal R}(\varphi)
\left(\begin{array}{c} d^0_{fL} \cr s^0_{fL} \end{array}\right)$, which
writes in  terms of the new classical mass eigenstates as (using the
unitarity of ${\cal R}(\vartheta)$ and ${\cal R}(\varphi)$)
$ \left(\begin{array}{cc} \overline{\hat u^0_{mL}} & \overline{\hat
c^0_{mL}}\end{array}\right) {\cal C}^\dagger_{u0} W\hskip -3mm/\; {\cal
C}_{d0} \left(\begin{array}{c} \hat d^0_{mL}\cr \hat s^0_{mL}
\end{array}\right)$.
So, at the classical level, the mixing (Cabibbo) matrix occurring in
charged currents is unchanged.
This means that, in the equivalent of (\ref{eq:TTT}), involving the new
classical mass eigenstates defined above, ${\cal C}_0$ is formally
unchanged and so are the $SU(2)_L$ generators.
This is precisely the ingredients which are used to calculate Shabalin's
counterterms. So, in the new classical mass basis, the
$A_{u,d},B_{u,d},D_{u,d}$'s are unchanged. This entails that the
renormalized  matrix $\cal C$ expressed by (\ref{eq:Cab2}) is also unchanged.
The last step is to go to the basis of the new renormalized mass eigenstates
$\left(\begin{array}{c}\hat d_{mL} \cr \hat s_{mL}\end{array}\right)$
 (see (\ref{eq:xi})).
 Since the $A_{u,d}$ Shabalin's counterterms are unchanged, so are, 
formally, the matrices ${\cal V}_{u,d}$ (see
(\ref{eq:kincon},\ref{eq:calV})), which still
depend on arbitrary angles $\varphi_{Lu}$ and $\varphi_{Ld}$ and parameters
$\rho_u$ and $\rho_d$.
Since Shabalin's counterterms $B_{u,d}$ and $D_{u,d}$ are 
unchanged, so are the unitary matrices $V_{u,d}$ and $U_{u,d}$.
Since ${\cal C}_{d0}$ and ${\cal C}_{u0}$ have been changed (see above), so
have ${\cal C}_d$ and ${\cal C}_u$ (see (\ref{eq:Cd1})), in which
$\theta_{dL}$ and $\theta_{uL}$ are now also respectively shifted by $\varphi$ and
$\vartheta$. Let us keep as before
$\varphi_{Lu} + \theta_{2Lu} =0 = \varphi_{Ld}
+\theta_{2Ld}$;  ${\cal V}_d V_d$, which does not depend on
$\theta_{dL}$ (see (\ref{eq:VV})), stays unchanged (and so does ${\cal V}_u V_u$).
Since $\cal C$ has been seen to be unchanged, too, the Cabibbo matrix $\mathfrak
C$, expressed by the first line of (\ref{eq:C2}), is unchanged.

${\cal C}_d$, which connects original flavor states to renormalized mass
states, becomes $\hat{\cal C}_d \equiv \hat{\cal C}_{d0} {\cal V}_d V_d
= {\cal R}^\dagger(\varphi) {\cal C}_d$
and one gets a similar expression for $\hat{\cal C}_u$.

We introduce, like before, renormalized flavor states (see
(\ref{eq:renflav})) and
 the renormalized mixing matrices $\hat{\mathfrak C}_u$ and $\hat{\mathfrak
C}_d$ connecting the latter to renormalized mass states.
Redoing the manipulations that led from (\ref{eq:Cab2}) to (\ref{eq:Cfin}),
one finds that (\ref{eq:Cfin}) stays unchanged. So do the three first terms
of (\ref{eq:Cdrenorm}), as well as  $({\mathfrak C}_u)^{-1}$.
The rotation
angles $\varphi$ and $\vartheta$  can be absorbed in the definition
of the new renormalized flavor states 
 which are, as expected, deduced from the initial
ones (see (\ref{eq:renflav})) by ${\cal R}(\varphi)$ and ${\cal
R}(\vartheta)$.
Finally, $\hat{\mathfrak C}_d = {\mathfrak C}_d$, $\hat{\mathfrak C}_u =
{\mathfrak C}_u$: each renormalized mixing
matrix stays unchanged and the relation
 ${\mathfrak C} \equiv \hat{\mathfrak C} = \hat{\mathfrak
C}^\dagger_u\hat{\mathfrak C}_d$ still holds.
The mixing angles are renormalized as before according to
\begin{equation}
\tilde \theta_{uL} = \theta_{uL} + \frac{\rho_u\epsilon_u}{2},\ 
\tilde \theta_{dL} = \theta_{dL} + \frac{\rho_d\epsilon_d}{2},\ 
\tilde \theta_c = \tilde \theta_{dL} - \tilde\theta_{uL}.
\end{equation}

Let us also write what happens in
there for charged currents (the transformations of mass terms and neutral
currents are given in (\ref{eq:neutran})).
It is convenient for this to use the second line of (\ref{eq:ch1}): 
\begin{eqnarray}
\hskip -1cm\left(\begin{array}{cc} \bar u_{mL} & \bar c_{mL}\end{array}\right)
 {\mathfrak C} \gamma^\mu \left(\begin{array}{c} d_{mL}\cr s_{mL}
\end{array}\right)=
 \left(\begin{array}{cc} \bar u^0_{fL} & \bar c^0_{fL}\end{array}\right)
\left[ 1 +\epsilon_u {\cal T}_z(\theta_{uL}) +\epsilon_d
{\cal T}_z(\theta_{dL})\right]\gamma^\mu
\left(\begin{array}{c} d^0_{fL} \cr s^0_{fL}\end{array}\right).
\label{eq:ch2}
\end{eqnarray}
We recall (see subsection \ref{subsection:1loop})
 that $\epsilon_d$ and $\epsilon_u$ are
proportional to $\sin(\theta_{dL} - \theta_{uL})\cos(\theta_{dL} -
\theta_{uL})$.
By the transformations (\ref{eq:unitr}) and (\ref{eq:unitr2})
the arguments of ${\cal T}_z(\theta_{uL})$ and ${\cal T}_z(\theta_{dL})$ in (\ref{eq:ch2}) are
both shifted by $(\varphi + \vartheta)$:
$2\theta_{uL} \to 2\theta_{uL} + \varphi+\vartheta$, and
$2\theta_{dL} \to 2\theta_{dL} + \varphi+\vartheta$, such that their difference
stays the same. The structure of (\ref{eq:ch2}) stays unchanged,
but for the $1$, which becomes ${\cal R}(\varphi - \vartheta)$
\footnote{One can check directly
 this statement by starting again from (\ref{eq:C2}),
which we have shown to be unchanged (though
${\cal C}_{u0}$ and ${\cal C}_{d0}$ have changed, ${\cal C}_0$ stays
unchanged). We just have to make the transformation from the new classical
mass eigenstates to the original flavor states. This is the role of the
transformations $\hat{\cal C}_{d0}$ and $\hat{\cal C}_{u0}$ such that, in the
original flavor basis the charged currents write (omitting the $W_\mu
\gamma^\mu$)
\begin{eqnarray}
&& \left(\begin{array}{cc} \bar u^0_{fL} & \bar c^0_{fL}\end{array}\right)
\hat{\cal C}_{u0}
 \frac12 
\left[ \left(\begin{array}{cc} 1 & -A_u \cr -A_u & 1
\end{array}\right) {\cal C}_{0}
+{\cal C}_{0} \left(\begin{array}{cc} 1 & -A_d \cr -A_d & 1
\end{array}\right)\right]
\hat{\cal C}^\dagger_{d0}\left(\begin{array}{c} d^0_{fL} \cr
s^0_{fL}\end{array}\right)\cr
&=&  \left(\begin{array}{cc} \bar u^0_{fL} & \bar c^0_{fL}\end{array}\right)
{\cal R}^\dagger(\vartheta)\, 
\left\{\frac12\;{\cal C}_{u0} 
\left[\left(\begin{array}{cc} 1 & -A_u \cr -A_u & 1
\end{array}\right) {\cal C}_{0}
+{\cal C}_{0} \left(\begin{array}{cc} 1 & -A_d \cr -A_d & 1
\end{array}\right)\right]
{\cal C}^\dagger_{d0}\right\}{\cal R}(\varphi)\left(\begin{array}{c} d^0_{fL} \cr
s^0_{fL}\end{array}\right),\cr
&&
\label{eq:ch3}
\end{eqnarray}
which yields the same conclusion as operating with ${\cal R}(\vartheta)$ and
${\cal R}(\varphi)$ directly on (\ref{eq:ch2}).}. Because of this term, the group
structure of charged currents is modified, since
it no longer projects only  on ${\cal T}_z$,
 unless $\varphi =\vartheta$; accordingly, if one wants to preserve it,
the same  flavor rotation should be performed in both sectors.

So, while independent $e^{i\alpha {\cal T}_y^u} \times e^{i\beta {\cal T}_y^d}$
 flavor rotations  do not change, in the new bases,
the mixing angles (see also footnote
\ref{footnote:newmix}), they modify the different parts 
of the Lagrangian in different ways. The tightly connected structure of neutral
currents and mass terms stay unchanged  and they are continuously rotated
into one another. The modification of charged currents is more important
unless the two rotations are identical.
Accordingly, requesting that neutral and charged gauge currents
exhibit the
 same flavor structure provides a constraint on the arbitrary flavor
rotations that can be performed and thus a connection between sectors of
different electric charge. This is one of the consequences of the fact that
the angles of the two sectors get entangled by radiative corrections. 
It has also  consequences for the alignment (up to small radiative
corrections) of mass
and flavor eigenstates in {\em one} of the two sectors $(u,c)$ or $(d,s)$
(see subsection \ref{subsection:cc}).

\paragraph{Arbitrary unitary transformations}
\hfill\newline

We now consider
 arbitrary $2 \times 2$ unitary transformations $\Omega^u$ and $\Omega^d$
on fermions.

Like for rotations, it is straightforward to show that ${\cal C}_0$,
Shabalin's counterterms, $\cal C$, ${\cal V}_{u,d}$, $V_{u,d}$ and
$U_{u,d}$ stay unchanged. So do ${\cal V}_d V_d$ and ${\cal V}_u V_u$ and,
finally, the Cabibbo matrix $\mathfrak C$ between the new renormalized mass
states.
${\cal C}_d$ becomes $\hat{\cal C}_d = \Omega^{d\dagger} {\cal
C}_d$ and ${\cal C}_u$ becomes $\hat{\cal C}_u = \Omega^{u\dagger} {\cal
C}_u$.

We parametrize, with the appropriate $u$ or $d$ index for $\alpha$ and
$\vec\beta$
\begin{equation}
\Omega = e^{i(\alpha + \beta_x {\cal T}_x(\theta) + \beta_y {\cal
T}_y + \beta_z{\cal T}_z(\theta))}.
\end{equation}
Concerning mass terms and neutral currents, in the original flavor basis
one gets: 
\begin{eqnarray}
\Omega^\dagger {\cal T}_x(\theta) \Omega \approx
{\cal T}_x(\theta + \frac{\beta_y}{2})  +\beta_z {\cal T}_y,\quad
\Omega^\dagger {\cal T}_z(\theta) \Omega \approx
{\cal T}_z(\theta + \frac{\beta_y}{2}) - \beta_x {\cal T}_y,
\end{eqnarray}

while, for charged currents (\ref{eq:ch2}) becomes:

\vbox{
\begin{eqnarray}
&&\hskip -2cm \Omega^{u\dagger} \left[ 1 +\epsilon_u {\cal T}_z(\theta_{uL})
+\epsilon_d {\cal T}_z(\theta_{dL})\right]\Omega^d \cr
\approx\  \Omega^{u\dagger} \Omega^d 
&+&  
\epsilon_u \Bigg[ 
{\cal T}_z\left(\theta_{uL} +\frac{\beta_y^u + \beta_y^d}{4}\right)
+ i(\alpha_d -\alpha_u) {\cal T}_z(\theta_{uL})
-\frac12 \beta_x^u\, {\cal T}_y -\frac{i}{4} \beta_z^u\cr
&& \hskip 4cm
+ \frac{i}{4}\beta_x^d\; F\big((\theta_{uL}-\theta_{dL})\big)
+ \frac{i}{4}\beta_z^d\; G\big((\theta_{uL}-\theta_{dL})\big)
\Bigg]\cr
&+&
\epsilon_d \Bigg[ 
{\cal T}_z\left(\theta_{dL} +\frac{\beta_y^u + \beta_y^d}{4}\right)
+ i(\alpha_d -\alpha_u) {\cal T}_z(\theta_{dL})
-\frac12 \beta_x^d\, {\cal T}_y + \frac{i}{4}\beta_z^d\cr
&& \hskip 4cm
+ \frac{i}{4}\beta_x^u\; F\big((\theta_{uL}-\theta_{dL})\big)
-\frac{i}{4} \beta_z^u\; G\big((\theta_{uL}-\theta_{dL})\big)
\Bigg],\cr
&&\hskip -2cm \text{with}\   F(\tau) = \left(\begin{array}{rr}
\sin 2\tau         & \cos 2\tau  \cr
-\cos 2\tau        &  \sin 2\tau \end{array}\right), \quad
G(\tau) = \left(\begin{array}{rr}
\cos 2\tau         & \sin 2\tau  \cr
-\sin 2\tau        &  \cos 2\tau \end{array}\right).
\end{eqnarray}
}

So,  mass terms, neutral currents and charged currents are all in general
deeply modified, which corresponds to a strong breaking of the $SU(2)_f
\times U(1)_f$ flavor symmetry.

\subsubsection{Self energy, electromagnetic current and Ward identity}

Departure from the inappropriate Wigner-Weisskopf approximation
\cite{MaNoVy}\cite{Novikov} can also be done by  working
with an effective renormalized $q^2$-dependent mass matrix (self-energy)
$M(q^2)$.

The eigenvalues of $M(q^2)$  are now $q^2$-dependent,
 and are determined by the equation $\det[M(q^2) -\lambda(q^2)]=0$
\footnote{This is the simple  case of a normal mass
matrix,  which can be diagonalized by a single ($q^2$-dependent) unitary
matrix. When it is non-normal, the standard procedure uses a bi-unitary
diagonalization.}.
 Let them be $\lambda_1(q^2) \ldots
\lambda_n(q^2)$. The physical masses satisfy the $n$ self-consistent
equations $q^2 = \lambda_{1\ldots n}(q^2)$, such that
$m_1^2 = \lambda_1(m_1^2) \ldots m_n^2 = \lambda_n(m_n^2)$. At each
$m_i^2$, $M(m_i^2)$ has $n$ eigenvectors, but only one corresponds to the
physical mass eigenstate; the others are ``spurious'' states \cite{MaNoVy}.
Even if the renormalized mass matrix is hermitian at any given $q^2$,
the physical mass eigenstates corresponding to different $q^2$ belong to as
many different orthonormal sets of eigenstates and  thus, in general, do
not form an orthonormal set. The discussion proceeds like in the core of
the paper, leading to similar conclusions.

We study below the role of the $U(1)_{em}$ Ward Identity connecting
the inverse fermionic propagator $S^{-1}(q)$ to the
 photon-fermion-antifermion vertex $\Gamma_\mu (q,q)$ at vanishing
incoming photon momentum.
In each sector of (bare) flavor space, the vertex function is (due to the
Gell-Mann-Nishijima relation  between neutral $SU(2)_L$ and $U(1)_{em}$
generators in the standard model,  and up to $\gamma^\mu \times$ the
electric charge in the given sector)
nothing more than
$({\cal C}^{-1})^\dagger (q^2) {\cal C}^{-1}(q^2)$ encountered before for neutral
currents (see also footnote {\ref{footnote:em}).
Requesting that the two sides of the identity be invariant by the same
flavor transformation (\ref{eq:Omega})
will induce constraints which do not suffer the major drawback
of textures, their instability  by such  transformations.

In each channel, for example $(d,s)$, the aforementioned Ward Identity
writes
\begin{equation}
\Gamma_\mu(q,q) = \frac{\partial}{\partial q_\mu} S^{-1}(q).
\label{eq:Ward}
\end{equation}
Accordingly, both sides of (\ref{eq:Ward}) should be invariant by
 the same group of symmetry.

We write the $(d,s)$ propagator $S(q^2)$ (we suppose that it is symmetric,
such that left and right eigenstates are obtained by the same rotation) as
\begin{equation}
S^{-1}(q^2) = /\!\!\!q - M(q^2),\ 
M(q^2) = \left(\begin{array}{cc} a(q^2) & c(q^2) \cr c(q^2) & b(q^2)
\end{array}\right).
\label{eq:Ward2}
\end{equation}
Defining $\theta(q^2)$ such that $\tan 2 \theta(q^2) = \displaystyle\frac
{2c(q^2)}{a(q^2) - b(q^2)}$,  $M(q^2)$ can then be rewritten
\begin{equation}
M(q^2) = \frac{a(q^2) + b(q^2)}{2} + \frac{a(q^2) - b(q^2)}{2 \cos
2\theta(q^2)}\; {\cal T}_x(\theta(q^2)).
\label{eq:W3}
\end{equation}
Differentiating both sides of (\ref{eq:Ward2}) with respect to $q_\mu$
and using (\ref{eq:W3}) yields

\vbox{
\begin{eqnarray}
\frac{\partial}{\partial q_\mu} S^{-1}(q) &=& \gamma^\mu
+ 2q_\mu \Bigg[
\displaystyle\frac{\partial (a(q^2)+b(q^2))}{2\,\partial q^2} 
-\displaystyle\frac{a(q^2)-b(q^2)}{2\cos 2\theta(q^2)}\;
{\cal T}_x(\theta(q^2))
 \frac{\partial \theta (q^2)}{\partial q^2}
\cr
 && + \left[\displaystyle\frac{\partial}{\partial q^2}\ \left(\frac{a(q^2)-b(q^2)}{2
\cos 2\theta(q^2)}\right)\right] \;
{\cal T}_z(\theta(q^2))
\Bigg],
\end{eqnarray}
}

in which only the first two terms, respectively proportional to the unit
matrix and to ${\cal T}_z$,  are invariant by the same transformation
$\Omega_z$ (\ref{eq:Omega}) as $({\cal C}^{-1})^\dagger {\cal C}^{-1}$ which
controls both neutral and electromagnetic gauge currents;
the last term, proportional to
$\displaystyle\frac{\partial \Delta m(q^2)}{2\,\partial q^2}, m(q^2) =
\frac{\lambda_+(q^2) + \lambda_-(q^2)}{2}$ (see footnote
\ref{foot:racines}), is not.
The invariance can be recovered if we constrain this derivative to vanish,
that is the
self-energy to satisfy the condition
\begin{equation}
a(q^2) - b(q^2) = 2\mu \cos 2\theta(q^2),\ \mu =\text{cst},
\label{eq:condM}
\end{equation}
(of course trivially satisfied for $a(q^2)=b(q^2)$, in
which case $\theta(q^2) = \pi/4$) or, equivalently
\footnote{
The eigenvalues of $M(q^2)$ are $\lambda_+(q^2) =
a(q^2) + 2\mu \sin^2\theta(q^2)$ and 
$\lambda_-(q^2) = a(q^2) - 2\mu \cos^2\theta(q^2)$ (thus $\mu =
\frac{\lambda_+(q^2) - \lambda_-(q^2)}{2}$),
such that the physical masses (poles of the propagator) satisfy
\begin{equation}
m_1 = a(m_1^2) + 2\mu \sin^2\theta(m_1^2),\ m_2 = a(m_2^2)
-2\mu\cos^2\theta(m_2^2).
\end{equation}
The degenerate case $m_1=m_2$ corresponds to $\mu=0$. By (\ref{eq:condM}),
this is equivalent to $a(q^2) = b(q^2)$ and to $\theta=\frac{\pi}{4}$. For
quasi-degenerate systems $m_1 \approx m_2 \approx m$,
 one has $\displaystyle\frac{m_1-m_2}{m_1+m_2} =
\displaystyle\frac{\mu}{a(m^2)-\mu\cos 2\theta(m^2)}\approx
\displaystyle\frac{\mu}{a(m^2)}$ and
$\mu \approx \displaystyle\frac{m_1-m_2}{2}$.\label{foot:racines}}
\begin{equation}
M(q^2) = a(q^2) -\mu\cos 2\theta(q^2) + \mu\; {\cal T}_x(\theta(q^2)).
\label{eq:Mq}
\end{equation}

Unlike textures, this form of the self-energy
is stable by flavor rotations\hfill\break
$\left(\begin{array}{c} d^0_{fL} \cr s^0_{fL}\end{array}\right) \to
{\cal R}(\varphi)\left(\begin{array}{c} d^0_{fL} \cr
s^0_{fL}\end{array}\right)$; $M(q^2)$ is transformed into
$ a(q^2) -\mu\cos 2\theta(q^2) + \mu \;{\cal T}_x(\theta(q^2) + \varphi)
$,
which shows that the mixing angle $\theta(q^2)$ has simply become, as
expected, $\theta(q^2) + \varphi$
while the spectrum is unchanged.
So is the form (\ref{eq:CC}) for the vertex function $\Gamma_\mu$.

Our conjecture is accordingly that any self-energy or vertex function
should be of the form
\begin{equation}
 \Xi(q^2) + \mu \left(\begin{array}{rr} \cos 2\theta(q^2) & \pm \sin
2\theta(q^2) \cr
\pm \sin 2\theta(q^2) & -\cos 2 \theta(q^2) \end{array}\right)
\ \text{or}\ 
 \Sigma(q^2) + \mu \left(\begin{array}{rr} \sin 2\theta(q^2) & \pm \cos
2\theta(q^2) \cr
\pm \cos 2\theta(q^2) & -\sin 2 \theta(q^2) \end{array}\right),
\label{eq:W4}
\end{equation}
which make them stable by flavor rotations.
They are in particular normal, and thus can always
be diagonalized by a unique unitary transformation, which can be used to
define both left and right eigenvectors.

\medskip

Eq.(\ref{eq:Mq}) trivially rewrites
\begin{equation}
M(q^2) = a(q^2)  +
\mu \left(\begin{array}{cc} 0  & \sin 2\theta(q^2)  \cr \sin
2\theta(q^2) & -2\cos 2\theta(q^2) \end{array}\right),
\label{eq:Mq2}
\end{equation}
reminiscent, up to $a(q^2)$ (which does not change $\theta(q^2)$) of the
triangular matrix suggested in \cite{Strumia} for $\tan 2\theta(q^2) = -2$;
however, while  the expressions (\ref{eq:W4}) are
stable by flavor rotations, this particular texture is not.
Indeed, rotating (\ref{eq:Mq}) and (\ref{eq:Mq2}), one gets respectively 

\vbox{
\begin{eqnarray}
\label{eq:R1}
&& R^\dagger(\varphi) \left[a(q^2) -\mu\cos 2\theta(q^2) +
\mu\; {\cal T}_x(\theta(q^2))
\right]R(\varphi)
= a(q^2) -\mu\cos 2\theta(q^2) +\mu\;{\cal T}_x(\theta(q^2)+\varphi)
,\cr && \\
\label{eq:R2}
&& R^\dagger(\varphi)\left[ a(q^2)  +
\mu \left(\begin{array}{cc} 0  & \sin 2\theta(q^2)  \cr \sin
2\theta(q^2) & -2\cos 2\theta(q^2) \end{array}\right)\right] R(\varphi)\cr
&& = a(q^2)  +
\mu \left(\begin{array}{cc}  -\sin 2\theta(q^2)\sin 2\varphi -2 \cos
2\theta \sin^2\varphi & \sin 2(\theta(q^2)+\varphi)  \cr
\sin 2(\theta(q^2) + \varphi)  &
\sin 2\theta(q^2) \sin 2\varphi -2 \cos 2\theta(q^2) \cos^2\varphi
 \end{array}\right).\cr
&&
\end{eqnarray}
}
By evaluating the ratio between twice the non-diagonal term and
the difference of diagonal ones, one finds, on both (\ref{eq:R1}) and
(\ref{eq:R2}), that, as expected, the mixing angle has become
$\theta(q^2)+\varphi$. However, while the ``structure'' of (\ref{eq:R1})
is manifestly preserved by the rotation, the $0$
texture in (\ref{eq:R2}) is not.

\section{Conclusion,  open issues and outlook}
\label{section:questions}

\subsection{Summary}

That mixing matrices connecting flavor to mass eigenstates of
non-degenerate coupled fermion systems should not be
considered {\em a priori} as unitary has been given in this work, in
addition to  general QFT arguments, a perturbative basis from the
calculation of radiative corrections at 1-loop to fermionic self-energies
and neutral currents.
The counterterms of Shabalin, in particular kinetic counterterms (wave
function renormalization), have been  shown to play an
important role, controlling  the departure from 1 of the matrix
of neutral currents in  bare flavor space.

We have shown that, in the renormalized mass basis, which,
unlike the bare one, is no longer
orthonormal, the renormalized mixing (Cabibbo) matrix stays unitary and, as
required by the closure of the $SU(2)_L$ gauge algebra, neutral currents
are, like in the bare mass basis, controlled by the unit matrix.

The peculiar feature that is satisfied for two
generations by the Cabibbo angle, that universality of neutral
currents is violated with the same strength as the absence of FCNC's, has
been shown to be  compatible with all mixing angles of quarks and
leptons for three generations, too. For neutrinos, we have shown that there
exists only one solution for $\theta_{13}$ to the corresponding equations
that rigorously falls within present experimental limits,
and we have obtained,
without any hypothesis (textures) concerning mass matrices,
the property of ``quark-lepton complementarity'' between the Cabibbo
angle and their $\theta_{12}$.

Flavor symmetries, and their entanglement with $SU(2)_L$ gauge symmetry,
have been shown to underlie the physics of mixing angles.
In particular, for two generations, the ways gauge currents and fermionic
mass terms (or self-energy) transform by flavor rotations  bear
common footprints left by a non-degenerate mass spectrum.

\subsection{Comparison with previous works}
\label{subsection:comp}

At this stage, it can be  useful to stress that, in this 
approach to the renormalization of mixing matrices,  both kinetic and
mass terms + counterterms have been simultaneously diagonalized.
Having dealt with self-mass as well as wave function renormalization, the
mixing matrices that we define connect bare mass states to renormalized
mass states which do not anymore undergo non-local non-diagonal transitions.

This is not the case of previous approaches, in particular of
\cite{KniehlSirlin}, in which the sole diagonalization of mass terms +
counterterms defines renormalized mass states; so there still
exist among them non-diagonal kinetic-like transitions
\footnote{In the renormalization scheme proposed in \cite{KniehlSirlin}, it
is furthermore impossible to cancel finite contributions to self-masses in all
channels. As a results, in some of them, finite non-diagonal fermionic mass
counterterms  stay present, which, when inserted on external legs of
$Wq_1\bar q_2$ vertex, can trigger right-handed currents at ${\cal O}(g^5)$
in the standard model.}.

We have shown that Shabalin's kinetic counterterms are the ones that drive
the non-unitarity of mixing matrices. Would we have left  them aside,
like in \cite{KniehlSirlin}, we would have reached the same
conclusion as theirs, that renormalized individual mixing matrices are unitary.

The mechanism that keeps unitary the CKM or PMNS matrices occurring in charged
current is thus  different from the one advocated in
\cite{KniehlSirlin}; it results from subtle cancellations between two
individually non-unitarity mixing matrices and the fact that, because of
$SU(2)_L$ gauge invariance which dictates the form of covariant
derivatives, the customary expression $K= K_u^\dagger K_d$ for the CKM
matrix in terms of individual mixing matrices for $u$- and $d$-type quarks
is no longer valid.

Another important feature of our work is that the general QFT argument
leading to non-unitary mixing matrices makes use of pole masses. These are
the only ones which are gauge independent. This choice goes along with the
existence of several $q^2$ scales. One can instead choose to consider
the renormalized mass matrix (self-energy) at a
given unique $q^2$, and to define the renormalized mass eigenstates through a
bi-unitary diagonalization of this mass matrix. This leads to unitary
mixing matrices. However, the renormalized masses that appear by
this procedure (which are not the eigenvalues of the mass
matrix) do not match the poles of the renormalized propagator
(which correspond to different values of $q^2$). Because of this,
they certainly cannot satisfy the criterion of gauge independence.

As for the fate of $SU(2)_L$ Ward Identities in a
multiscale renormalization approach, in addition to the fact that $SU(2)_L$
gauge invariance is compatible with the existence of different
mass scales, we  refer to \cite{BogShir}:
 the regularization method might violate some invariance (gauge, Lorentz ...);
 one has then to introduce counterterms which violate it, too.
 After the regularization has been taken away,  the S-matrix so
 obtained satisfies the requested invariance. There
 appears accordingly to be no fundamental obstacle (only technical
 difficulties) if the regularization procedure does not respect the Ward
 Identities corresponding to the invariance of the theory.

\subsection{Physically relevant mixing angles}
\label{subsection:cc}

The results that have been exposed are valid for fermions of both electric
charges. They concern the mixing angles which parametrize

$\ast$ for quarks, the mixing matrix $K_u$
of $u$-type quarks as well as $K_d$ of d-type quarks;

$\ast$ for leptons, the mixing matrix $K_\nu$ of neutrinos as well as
that of charged leptons $K_\ell$,

and we have shown that our approach accounts for the observed
 values of the mixing angles.

However, a question arises :  the measured values  of the mixing angles are
commonly attached, not to a single mixing matrix, {e.g.} $K_u$ or $K_d$,
but to the product  $K=K^\dagger_u K_d$  which occurs in charged
currents when both quark types are mass eigenstates.
Thus, in the standard approach, they are {\em a priori} related 
to an entanglement of the 
mixing angles of quarks (or leptons) of different charges.
Then, if mixing angles in each sector are expected to satisfy the same
criterion, their difference, which makes up, up to small approximation, the
Cabibbo angle, would be expected to vanish.

The same issue arises  in the leptonic sector.
Let us consider for example the case of solar
neutrinos: the flux of ``electron neutrinos'' detected on earth is (roughly)
half the one predicted by solar model to be emitted from the sun.
Would the flux predicted in solar models concern  flavor neutrinos, and
would also the detection process  counts flavor neutrinos, the sole mixing
matrix which controls their evolution and oscillations would be $K_\nu$; 
it is indeed the only matrix involved in the projection of flavor states
onto mass (propagating) states.
The  situation is different if the comparison is made between the
(emitted and detected) fluxes of states $\nu_e, \nu_\mu, \nu_\tau$
 defined in subsection \ref{subsection:basis}; since their
projections on the mass eigenstates now involve the product $K_\ell^\dagger
K_\nu$, their oscillations are, like for quarks,
 controlled by an entanglement
of the mixing angles of neutrinos and charged leptons.
The nature of the neutrino eigenstates that are
produced and detected is also sometimes questioned
 (see also for example \cite{Giunti}).
An often proposed  solution is that, for charged
leptons, their flavor is defined to coincide with their mass
\cite{Akhmedov}, which amounts  to setting $K_\ell = 1$. 

This is indeed the solution that comes naturally to the mind since, as
we stated in subsection \ref{subsub:unitfer} (flavor rotations)
(see also appendix \ref{section:align}):
 while arbitrary independent flavor rotations are a priori allowed in
each sector of different charge, with the corollary statement that  the
only physically relevant mixing angles are the ones occurring in the Cabibbo
matrix,  these two rotations are constrained to be identical if
one likes to preserve the group structure (breaking pattern) of both
neutral and charged currents  in bare flavor space.
For $\vartheta = -\theta_{uL}= \varphi$,
only the  mixing angles in the $(u,c)$ sector becomes vanishing
(alignment of bare mass and flavor states in this sector).
 The structure (\ref{eq:ch2})
 of charged gauge
currents in bare flavor space becomes $\big(1 +  \epsilon_u(\theta_{uL}=0)
 {\cal T}_z(0) + \epsilon_d(\theta_{uL}=0)
 {\cal T}_z(\theta_{dL}- \theta_{uL})\big)$.
Now, as discussed in subsection \ref{subsub:finite}, the value of the
parameter $\epsilon_u \equiv A_u$ depends on the renormalization scheme;
for example its values in $MS$ and $\overline{MS}$ differ by a
constant proportional to $\gamma g^2 \sin\theta_c \cos\theta_c
(m_s^2 - m_d^2)/M_W^2$, $\gamma$ being the Euler constant.
A ``physical'' renormalization scheme 
\footnote{Its existence is only a conjecture. One could simply
subtract from $A_u^{\overline{MS}}$  its value
at $\theta_{uL}=0$, that is, another constant like when going from $MS$ to
$\overline{MS}$. However it is not clear that such a scheme respects
the gauge Ward Identities, nor how to implement it in practice
at the level of individual Feynman diagram. Subtracting from each one
its value at $\theta_{uL}=0$ is the simplest choice  as a
``physical'' renormalization prescription suitable to the
alignment of flavor and mass states in the $(u,c)$ sector. When applied
to Fig.~1 or to its equivalent for $u^0_m \leftrightarrow c^0_m$ transitions
(hence, in practice, to the functions $f_{u,d}$
(see footnote \ref{footnote:fu})), it also modifies the values of the
$B_{u,d},E_{u,d},D_{u,d}$ counterterms and that of the
combination (\ref{eq:consA}), which keeps non-vanishing,
 because the subtracted constants have,
in this case, a  dependence on fermion masses and on $p^2$ more involved
than the sole difference of $(mass)^2$ that factorizes the Euler constant
$\gamma$ in (\ref{eq:f}).\label{footnote:scheme}}
 would correspond to the condition $\epsilon_u(\theta_{uL}=0) =0$.
In this scheme, a classical unit mixing matrix (vanishing classical
 mixing angle) would not be modified by (1-loop) radiative corrections:
mass and flavor eigenstates could keep aligned in the corresponding sector
 both at the classical level and at 1-loop
\footnote{The resulting mixing matrix, which is identical to the unit
matrix, trivially satisfies  the criterion under
 consideration, {\em i.e.} that the violation of
universality (presently non-existing) is equal to that of the
absence of FCNC's (also vanishing).}.
Then,
after aligning mass and flavor states in the $(u,c)$ sector, that is, in
practice, turning  $\theta_u$  to zero by a flavor rotation,
the formula (\ref{eq:ch2}) for
 charged current in bare flavor space  becomes
$\left(\begin{array}{cc} \bar u_{mL} & \bar c_{mL}\end{array}\right)
 {\mathfrak C} \gamma^\mu  \left(\begin{array}{c} d_{mL}\cr s_{mL}
\end{array}\right)=
 \left(\begin{array}{cc} \bar u^0_{fL} & \bar c^0_{fL}\end{array}\right)
\left[ 1  +\epsilon_d(\theta_{uL}=0)
{\cal T}_z(\theta_{dL} -\theta_{uL})\right]\gamma^\mu
\left(\begin{array}{c} d^0_{fL} \cr s^0_{fL}\end{array}\right)$, in which
the only argument of the ${\cal T}_z$ generator is the Cabibbo angle.
The criterion linking universality and
FCNC's could accordingly be
applied to charged gauge currents in the bare flavor basis, which controls
the observed Cabibbo angle.

\subsection{Shabalin's counterterms in the calculation of physical
transitions}
\label{subsection:phys}

As far as physics is concerned, some remarks are due concerning decays like
$K \to \pi \nu \bar\nu, \mu \to e \gamma, \mu \to e \nu\bar\nu$, for which 
1-loop flavor changing neutral currents play an important role.
One could indeed wonder what are the consequences on these transitions of the
introduction of Shabalin's counterterms.

The first way to proceed is the usual one: no counterterm  ``\`a la
Shabalin'' is introduced and calculations are done in the bare mass
basis, which is orthonormal as soon as the bare flavor basis is supposed to
be so.

However, $d_m^0 \leftrightarrow s_m^0$ transitions occur at 1-loop, which
can be considered to jeopardize the standard CKM phenomenology \cite{DMV}.
To remedy this, the $A_{u,d},B_{u,d},E_{u,d},D_{u,d}$ counterterms
are added, and should then be included  in any perturbative calculation.
This second possibility may be cumbersome,  due to their twofold nature 
(kinetic and mass) and the fact that they have both chiralities.
Furthermore,  for $d$ and $s$ off
mass-shell (which occurs at 2-loops and more),
 their action can be no longer reduced to the cancellation of
non-diagonal $d^0_m \leftrightarrow s^0_m$ transitions.
\newline
Note that the one-loop amplitude of, for example, $s_m^0 \to d_m^0Z$ or
$s_m^0 \to d_m^0\gamma$
transition does not change when these counterterms
are introduced since, on one side, they kill
$s_m^0 \to Z(\gamma)s_m^0 \to Z(\gamma)d_m^0$ and $s_m^0
\to d_m^0 \to Z(\gamma)d_m^0$
transitions but, on the other side, the covariant 
derivative associated with the $p\!\!/$ in Eq.~(\ref{eq:AC}) restores them
(see Appendix \ref{section:dsgamma}).
So, the standard (without counterterms) 1-loop calculation of FCNC's
stays valid when counterterms are introduced, and the latter do not
accordingly play, there, any physical role.

The third possibility is to diagonalize the bare Lagrangian + Shabalin's
counterterms and to perform calculations in the so-defined renormalized
mass basis, in which it has 
the standard canonical form except that, as usual when counterterms are
introduced, the parameters (masses, mixing angles) become the renormalized ones.

This form of the Lagrangian is extremely simple since
all effects of Shabalin's counterterms have
been re-absorbed in a renormalization of the masses and mixing angles, and
a change of status (orthogonality or not) of the mass basis.
However, again, non-diagonal $d_m \leftrightarrow s_m$
transitions can occur at 1-loop, between renormalized mass states.
They  are similar to the ones occurring in the bare Lagrangian without
counterterms, except that their amplitudes are now expressed in terms of
renormalized parameters.
Two attitudes are then possible:\newline
* either one applies standard Feynman rules to this Lagrangian without
worrying about the orthonormality of the basis of reference, which leads
back to  the usual way of performing calculations;
this is tantamount to considering that Shabalin's counterterms do not play any
physical role.  This can look a reasonable attitude
since one does not know {\em a priori} whether a set of vectors is orthonormal
or not, except on physical grounds; \newline
* or, before starting any perturbative calculation, one first worries
whether the reference basis  is orthonormal or not. This is now tantamount to 
considering that physical predictions could  depend on this property and
that any sensible Lagrangian should be written, before any perturbative
expansion is performed, in an orthonormal reference basis.
These considerations go however beyond the scope of the present work.

\subsection{Flavor rotations as a very softly broken symmetry
of the Standard Model}
\label{subsection:fr}

Performing a rotation by an angle $\varphi$
 in the two sectors $(u,c)$ and
$(d,s)$ (or $(e,\mu)$ and $(\nu_e, \nu_\mu)$):\newline
* shifts both arguments $\theta_{uL}$ and $\theta_{dL}$  in the $SU(2)_f$
generators ${\cal T}_z(\theta_{uL,dL})$ and ${\cal T}_x(\theta_{uL,dL})$
which occur respectively in neutral (and electromagnetic)
currents and mass matrices by $\varphi$ (see  (\ref{eq:neutran}));
\newline
* yields equivalent shifts in charged currents (see (\ref{eq:ch2}));\newline
* does not modify the physical Cabibbo angle (see ``Flavor rotations'' in
subsection \ref{subsub:unitfer});\newline
* leaves invariant the rest of the Standard Model Lagrangian and does
not change the physical masses.

The rotation angle $\varphi$ and the resulting modifications of the Lagrangian appear
unphysical. This is why flavor rotations (identical in the two sectors)
 can be considered to be a symmetry
of the Standard Model.

\subsection{$\boldsymbol{CP}$ violation}

In this work we have deliberately ignored $CP$ violating mixing angles
and all effects of $CP$ violation.
There are several reasons for this:\newline
* they are {\em a priori} small and should not quantitatively alter the
results that have been obtained  for the other type of mixing angles;\newline
* since the renormalized Cabibbo matrix is constrained by $SU(2)_L$ gauge
invariance to stay unitary, we do not expect strong deviations from the
customary results;\newline
* the introduction of $CP$ violating phases would
considerably complicate the trigonometric equations to solve, which are
already highly non-trivial.

There is however an interesting point:
 the most general non-unitary mixing matrix allows {\em a priori}
 $CP$ violation even
for two generations. But we consider this as another matter which deserves a
separate investigation.

\subsection{Open issues. Beyond the Standard Model}
\label{subsection:pert}

The present work raises several questions and challenges.

A first type of challenge concerns experimentally observable consequences
of the issues raised in this work, specially the {\em a priori}
non-unitarity of mixing matrices. Unlike the $CP$-violating parameters
$\epsilon_L$ and $\epsilon_S$ of neutral kaons the difference of which we
could estimate  in \cite{MaNoVy}, we are not yet able to exhibit and
estimate  observables which would be sensitive to this non-unitarity, or,
equivalently, to the energy dependence of eigenstates induced by radiative
corrections.  This is all the more challenging as we
have shown that the renormalized mixing  matrix occurring
in charged currents (Cabibbo, CKM, PMNS) keeps unitary as a consequence of
$SU(2)_L$ gauge invariance. So, no deviation from unitarity can be expected
in charged currents from this mechanism. 
The finite renormalization of mixing angles in charged currents by the
simple function $\frac{\rho_d\epsilon_d - \rho_u\epsilon_u}{2}$ of Shabalin's
counterterms is itself non-physical since the parameters $\rho_{u,d}$ are
arbitrary. 
Non-unitary is thus expected to only be at work in neutral currents in bare
flavor space, where only one fermionic sector gets involved.
However, it is a much debated issue whether individual mixing angles,
corresponding to a given sector, are observable, or whether the sole
observable angles are the ones occurring in charged currents. 

A connection should also be made with the non-unitary equivalence of mass
and flavor Fock spaces investigated in \cite{GiuntiFock}. We have shown
that renormalized mass states are {\em a priori} connected to bare flavor
space by non-unitary transformations, which preaches in favor of the
propositions in \cite{GiuntiFock}. However, we have also proved
in subsection \ref{subsub:charged} that one
can define renormalized flavor states which deduce from bare flavor states
by a non-unitary transformation and which, now, connect to renormalized mass
states by unitary transformations. The issue arises accordingly (see also
subsection \ref{subsection:phys}) of which
basi(e)s can be considered to be orthonormal. Renormalized mass states and
bare flavor states we have shown cannot be simultaneously orthonormal.
Since, and this is the point of view of \cite{GiuntiFock}, physical mass
states (that is renormalized mass states) are expected to have the standard
anticommutation relations and to form a Fock space of orthonormal states,
renormalized flavor states, which are unitarily connected to the latter,
would then form, too, an orthonormal basis (such that bare flavor states
should not be anymore considered to form an orthonormal set, nor bare mass
states).
 Then the two spaces of
renormalized flavor states and renormalized mass states would be two
 unitarily connected Fock spaces. This issue is  currently under
investigation.

It is to be mentioned that, often, mixing angles are not
defined, like we did, through fundamental parameters of the Lagrangian, but
as ratios of  amplitudes among physical bound states (mesons).
The connection between the two approaches is certainly to be investigated,
but it is clear that it faces the tedious problem of bound states, in
which any tentative calculation is doomed to uncertainties largely
exceeding the effects that need to be tested.
  
The last type of challenge concerns the criterion that
seemingly controls observed mixing angles: it connects in the simplest
possible way the violation of unitarity to FCNC's in bare flavor space.
We have no reason to believe that the Standard Model possesses, in itself,
even including the refinements of QFT that we have implemented, the
necessary ingredients to give birth to such a property. All 
it can tell is that, due to non-degeneracy, one expects
both violation of unitarity and the presence of FCNC's
for all gauge currents in bare flavor space.
So, it seems reasonable to think that the realm
of any possible connection between the two lies  ``beyond the Standard
Model'', and that only there can one hope to  ultimately
find a theoretical explanation to the observed pattern, and to
the relation between the $\tan$ of the Cabibbo angle and the Golden
ratio \cite{DuretMachet2}\cite{Strumia}.

\subsection{Conclusion and perspective}
\label{subsection:conclusion}

This work does not, obviously, belong to what is nowadays referred 
to as "Beyond the Standard Model", since it does not incorporate any 
``new physics'' such as supersymmetry, ``grand unified theories (GUT)''
 or extra-dimensions.
However it does not strictly lie within
 the SM either, even if it is very close to.  Of course,
 it shares with the latter its general framework
(mathematical background and physical content), 
 and also borrows from it the two physical conditions of universality for 
 diagonal neutral currents  
 and absence of FCNC's, which play a crucial role in the process.
But, on the basis of the most general arguments of 
QFT, we make a decisive use of 
the essential non-unitarity of the mixing matrices,
whereas only unitary matrices are present in the SM.
This property  may be considered, in the SM,
as an "accidental" characteristic of objects which are
intrinsically non-unitary.

The mixing angles experimentally observed get constrained in the vicinity
of this ``standard'' situation, a slight departure from which
being due to  mass splittings.
Hence our approach can be considered to explore the
"Neighborhood of the Standard Model", which 
is likely to exhibit low-energy manifestations
of physics "Beyond the Standard Model".

While common approaches limit themselves to guessing symmetries
for mass matrices (see for example \cite{Ma} and references therein),
we showed that relevant patterns reveal instead
themselves in the violation of properties attached to gauge currents:
in each given $(i,j)$ flavor channel, two dimensional flavor rotation
appears as  a  flavor subgroup softly broken by the presence of mass
splittings, which continuously connects neutral currents and the fermionic
self-energy.

When two generations are concerned, nature seems to exhibit a quantization
of the $\tan$ of twice the mixing angles as multiples of $1/2$. This
corresponds to the property that, in the original flavor basis, the effects
of lifting the mass degeneracy are such that  universality for
neutral currents is violated with the same strength as the absence of
FCNC's. The third generations appears as a small perturbation of this
property. Whether this quantization really exists and whether it can be
cast on a firm theoretical background, in particular through
perturbative calculations, stays unfortunately an open question.

It is remarkable that the same type of symmetry underlies both the
quark and leptonic sectors; they only differ through the $0th$ order
solution to the ``unitarization equations",
the twofold-ness of which was recently uncovered in \cite{DuretMachet1}.
In the neutrino case, the values that we obtain for the
 mixing angles (with the smallest one of $\theta_{13}$)
do not deviate by more than $10\%$ from the tri-bimaximal pattern
\cite{HPS}.

To conclude, this  work demonstrates that flavor physics offers to our
investigation very special and simple patterns  which had been, up to now,
unnoticed.  Strong arguments in favor of them have been given in both
the quark and leptonic sectors, and they will be further tested when
the third mixing angle of neutrinos is accurately determined.


\vskip .5cm 
\begin{em}
\underline {Acknowledgments}: Discussions, comments and critics with / from
 A.~Djouadi, M.B.~Gavela,
C.~Giunti, S.~Lavignac, V.A.~Novikov, L.B.~Okun, J.~Orloff, E.P.~Shabalin
 and J.B.~Zuber are gratefully acknowledged.
\end{em}


\newpage\null

\appendix

{\Large \bf Appendix}

\vskip 1cm

\section{Calculation of Shabalin's counterterms}
\label{section:Shab}

We derive here the expressions (\ref{eq:BD}) for Shabalin'
s counterterms $A_d,B_d,E_d,D_d$.

Requesting that the sum of (\ref{eq:kinren}) and (\ref{eq:AC}) vanishes for
$s_m^0$ one mass-shell gives the first equation
\begin{eqnarray}
&&\hskip -1.5cm f_d(p^2=m_s^2) \bar d_m^0 (1+\gamma^5) m_s\, s_m^0 = \cr
&&
 A_d\, \bar d_m^0 (1+\gamma^5) m_s\, s_m^0 + B_d\, \bar d_m^0 (1-\gamma^5) s_m^0
+E_d\, \bar d_m^0(1-\gamma^5) m_s\, s_m^0 + D_d\, \bar d_m^0 (1+\gamma^5)
s_m^0.\cr
&&
\label{eq:sha1}
\end{eqnarray}
Likewise, using $\bar d\, \overleftrightarrow{\partial_\mu} s \equiv
\bar d (\partial_\mu s) - (\partial_\mu \bar d) s$ and
$i\gamma^\mu\partial_\mu \bar s = -m_s \bar s$, 
 the equivalent request for $d_m^0$ on mass-shell yields
\begin{eqnarray}
&&\hskip -.5cm f_d(p^2=m_d^2) \bar d_m^0 m_d (1-\gamma^5) s_m^0 = \cr
&&
 A_d\, \bar d_m^0 m_d(1-\gamma^5) s_m^0 + B_d\, \bar d_m^0 (1-\gamma^5) s_m^0
+E_d\, \bar d_m^0 m_d(1+\gamma^5)  s_m^0 + D_d\, \bar d_m^0 (1+\gamma^5)
s_m^0.\cr
&&
\label{eq:sha2}
\end{eqnarray}
(\ref{eq:sha1}) yields the two conditions, respectively for $(1+\gamma^5)$
and $(1-\gamma^5)$ terms:
\begin{eqnarray}
m_s f_d(p^2 = m_s^2) &=& m_s A_d + D_d,\cr
B_d + m_s E_d &=& 0. 
\label{eq:sha3}
\end{eqnarray}
while (\ref{eq:sha2} yields the two other conditions
\begin{eqnarray}
m_d f_d(p^2 = m_d^2) &=& m_d A_d + B_d,\cr
D_d + m_d E_d &=& 0. 
\label{eq:sha4}
\end{eqnarray}
The solutions of the  four equations in (\ref{eq:sha3}) and (\ref{eq:sha4})
are given by (\ref{eq:BD}).

\section{The inclusion of Shabalin's counterterms does not modify
$\boldsymbol{s \to d \gamma}$ transition.}
\label{section:dsgamma}

Making use of formula (\ref{eq:kinren}) for the 1-loop $s_m^0 \to d_m^0$ transition
of Fig.~1, the left and center diagrams of Fig.~6
 write respectively (we omit the $\epsilon^\mu$
of the photon and use and abbreviated notation $f_d(m_d^2) = f_d(p^2=m_d^2,
m_u^2, m_c^2, m_W^2)$)
\begin{eqnarray}
\label{eq:diag1a}
&& \bar d_m^0(p) f_d(m_s^2) \psl(1-\gamma^5)\, \displaystyle\frac{1}{\psl - m_s}\gamma^\mu
s_m^0(p+q), \\
\label{eq:diag1b}
&& \bar d_m^0(p) \gamma_\mu\, \displaystyle\frac{1}{\psl-\qsl - m_d} f_d(m_s^2) (\psl +
\qsl)(1-\gamma^5) s_m^0(p+q).
\end{eqnarray}
%

\vbox{
\begin{center}
\includegraphics[height=2truecm,width=5truecm]{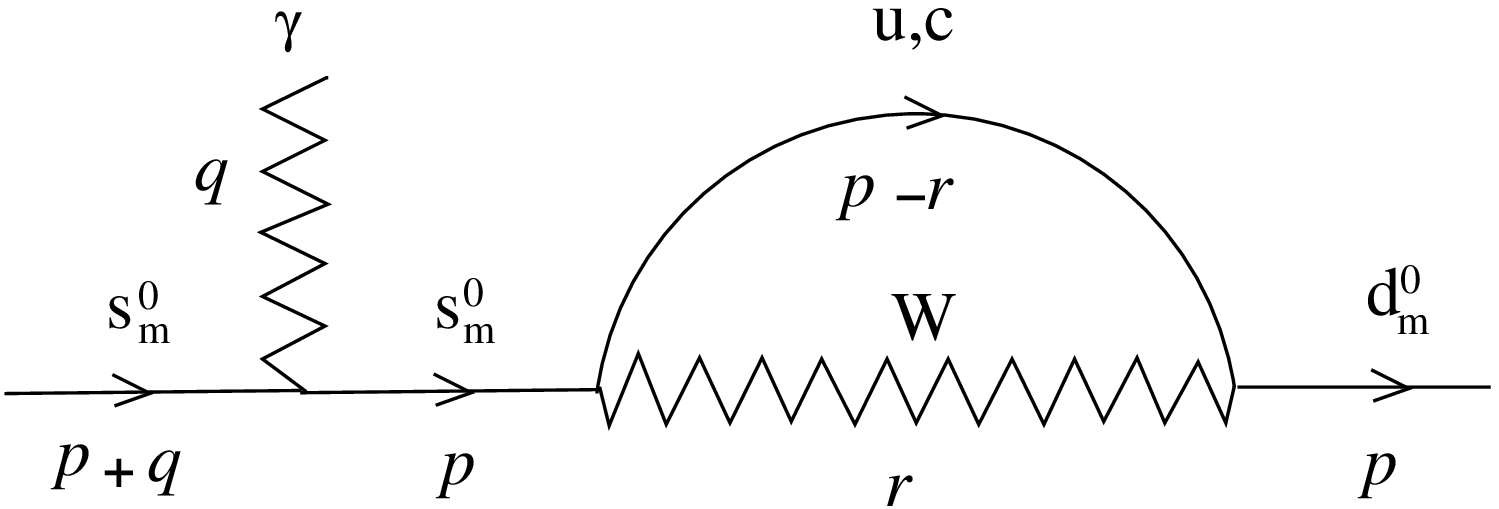}\quad
\includegraphics[height=2truecm,width=5truecm]{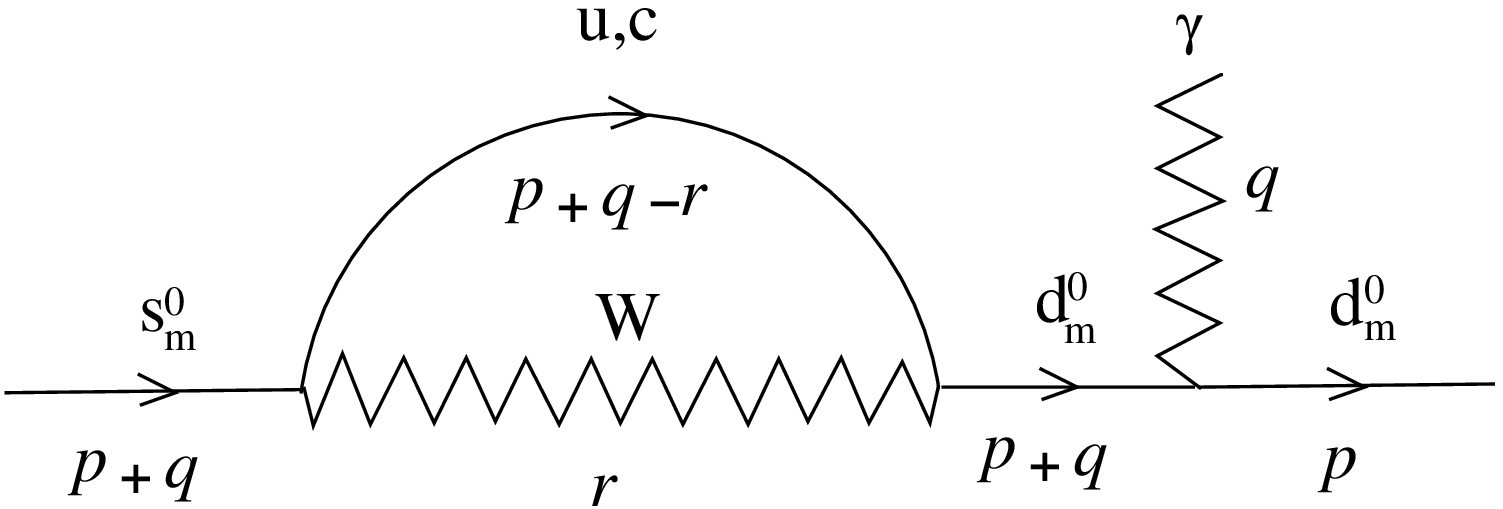}\quad
\includegraphics[height=2truecm,width=3truecm]{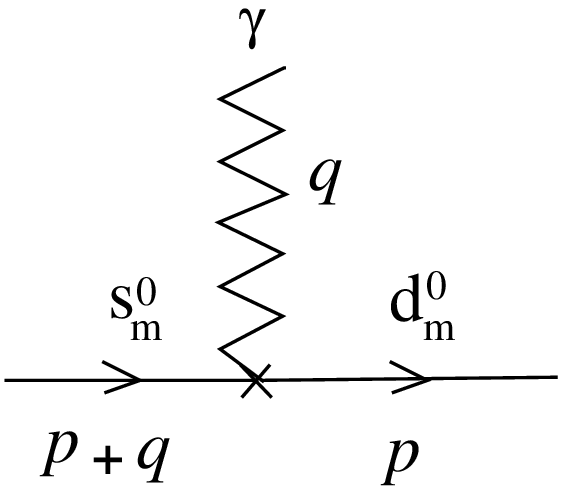}

{\em Fig.~6: diagrams contributing to $s_m^0\to d_m^0\gamma$ sensitive to
Shabalin's counterterms, which cancel  the left and center amplitudes;
the latter are re-created via the covariant
derivatives inside $A_d$ and $E_d$ which yield the diagram on the right.}
\end{center}}
\figskip
Using that the $d$ quark is on mass-shell in Fig.~6 left  and the $s$ quark is
on mass-shell in Fig.~6 center, straightforward manipulations transform
(\ref{eq:diag1a}) and (\ref{eq:diag1b}) respectively into
\begin{eqnarray}
& f_d(m_d^2)&\left[ \frac{m_d^2}{m_d^2 - m_s^2} \bar d_m^0 \gamma_\mu
(1-\gamma^5) s_m^0 + \frac{m_d m_s}{m_d^2 - m_s^2} \bar d_m^0\gamma_\mu(1+\gamma^5)
s_m^0\right],\\
 -&f_d(m_s^2)&\left[ \frac{m_s^2}{m_d^2 - m_s^2} \bar d_m^0 \gamma_\mu
(1-\gamma^5) s_m^0 + \frac{m_d m_s}{m_d^2 - m_s^2} \bar d_m^0\gamma_\mu(1+\gamma^5)
s_m^0\right],
\end{eqnarray}
the sum of which yields
$A_d\, \bar d_m^0\, \gamma_\mu(1-\gamma^5) s_m^0 + E_d\, \bar d_m^0\, \gamma_\mu(1-\gamma^5)
s_m^0$, where $A_d$ and $E_d$ are the Shabalin's counterterms given in
(\ref{eq:BD}).

So, while the two corresponding amplitudes  are
canceled by Shabalin's counterterms (since, in both diagrams, a $s_m^0 \to
d_m^0$
transition occurs with either $s$ or $d$ on mass-shell),
the photonic parts in the covariant derivatives which should be used inside
 $A_d$ and $E_d$ re-create the same transition amplitude (Fig.~6 right).
$s_m^0 \to d_m^0\gamma$ is thus left unchanged by the introduction of these
counterterms.

The same demonstration holds for $s_m^0 \to d_m^0 Z$  transitions.

\section{$\boldsymbol{\tilde{\theta}_{13}=0 \Rightarrow \theta_{13}=0}$}
\label{section:theta3}

Using the notations of section \ref{section:general}, we start with
the following system of equations:
\begin{subequations}
\label{subeqs:A}
\begin{equation} \label{eq:I}
  \frac{[11]+[22]}{2}=[33] \Leftrightarrow
s_{13}^2+s_{23}^2+\tilde{c}_{23}^2=1;
\end{equation}
\begin{equation} \label{eq:IIA}
 [11]=[22] \Leftrightarrow  c_{13}^2\cos(2\theta_{12})=(c_{23}^2+
\tilde{s}_{23}^2)\cos(2\tilde{\theta}_{12});
\end{equation}
\begin{equation}\label{eq:IIB}
 [12]=0=[21] \Leftrightarrow 
c_{13}^2\sin(2\theta_{12})=(c_{23}^2+\tilde{s}_{23}^2)
\sin(2\tilde{\theta}_{12});
\end{equation}
\begin{equation}\label{eq:IIIA}
 [13]=0=[31] \Leftrightarrow
 \tilde{s}_{12}\left(\sin(2\theta_{23})-\sin(2\tilde{\theta}_{23})\right)
=c_{12}\sin(2\theta_{13});
\end{equation}
\begin{equation}\label{eq:IIIB}
 [23]=0=[32] \Leftrightarrow
\tilde{c}_{12}\left(\sin(2\tilde{\theta}_{23})-\sin(2\theta_{23})\right)
=s_{12}\sin(2\theta_{13}).
\end{equation}
\end{subequations}
From equation (\ref{eq:I}), we have 
$c_{23}^2+\tilde{s}_{23}^2 \neq 0$, 
which entails $c_{13}^2 \neq 0
$\footnote{Indeed, let us suppose that $c_{13}$ vanishes. 
Then $\cos(2\tilde{\theta}_{12})$ and $\sin(2\tilde{\theta}_{12})$ 
must vanish simultaneously, which is impossible.}.
Let us study the consequence on the two equations
(\ref{eq:IIA}) and (\ref{eq:IIB}).

$\bullet$ the two sides of (\ref{eq:IIA}) vanish for $\cos(2\theta_{12})=
0=\cos(2\tilde{\theta}_{12})$, {\em i.e.} 
$\theta_{12}=\frac{\pi}{4} [\frac{\pi}{2}]=\tilde{\theta}_{12}$.\\
(\ref{eq:IIB}) then gives  $c_{13}^2=c_{23}^2+\tilde{s}_{23}^2$,
which, associated with (\ref{eq:I}), yields the following solution
\footnote{$ \left\{\begin{array}{l}
c_{13}^2=c_{23}^2+\tilde{s}_{23}^2 \\ s_{13}^2+s_{23}^2+\tilde{c}_{23}^2=1 
\end{array}\right.
\qquad
 \Longrightarrow \qquad \left\{\begin{array}{l} s_{23}^2+\tilde{c}_{23}^2=1 \\
s_{13}^2=0
 \end{array}\right.$}: 
$\theta_{13}=0 [\pi]$ and $\tilde{\theta}_{23}=\pm \theta_{23} [\pi]$.

$\bullet$ the two sides of (\ref{eq:IIB}) vanish for $\sin(2\theta_{12})=
0=\sin(2\tilde{\theta}_{12})=0$, i.e. 
$\theta_{12}=0 [\frac{\pi}{2}]=\tilde{\theta}_{12}$.\\
 (\ref{eq:IIA}) gives then $c_{13}^2=c_{23}^2+\tilde{s}_{23}^2$, 
hence, like previously, $\theta_{13}=0 [\pi]$ and 
$\tilde{\theta}_{23}=\pm \theta_{23} [\pi]$.

$\bullet$ in the other cases we can calculate
 the ratio (\ref{eq:IIA}) / (\ref{eq:IIB}), which gives 
$\tan(2\theta_{12})=\tan(2\tilde{\theta}_{12})$, 
hence $\theta_{12}=\tilde{\theta}_{12} [\pi]$ or $\theta_{12}=
\frac{\pi}{2}+\tilde{\theta}_{12} [\pi]$:
 
\quad$\ast$ $\theta_{12}=\frac{\pi}{2}+\tilde{\theta}_{12} [\pi]$ implies 
for (\ref{eq:IIA})(\ref{eq:IIB}) $c_{13}^2=-c_{23}^2-\tilde{s}_{23}^2$, 
which, together with (\ref{eq:I}) ($c_{13}^2=s_{23}^2+\tilde{c}_{23}^2$),
gives a contradiction : $2=0$:

\quad$\ast$ $\theta_{12}=\tilde{\theta}_{12} (\neq 0)[\pi]$ implies, like 
previously, $c_{13}^2=c_{23}^2+\tilde{s}_{23}^2$, 
which gives, when combined with (\ref{eq:I}):
$\theta_{13}=0 [\pi]$ and $\tilde{\theta}_{23}=\pm \theta_{23} [\pi]$.

Hence, it appears that whatever the case, the solution gives rise
to $\theta_{13}=0 [\pi]$.

Let us now look at (\ref{eq:IIIA}) and (\ref{eq:IIIB}).
Since $\theta_{13}=0$, the two r.h.s.'s vanish, 
and we obtain 
the twin equations $\tilde{s}_{12}(\sin(2\theta_{23})-\sin(2\tilde{\theta}_{23}))=0$ and 
 $\tilde{c}_{12}(\sin(2\theta_{23})-\sin(2\tilde{\theta}_{23}))=0$,
which, together, imply 
 $\sin(2\theta_{23})=\sin(2\tilde{\theta}_{23})$. It follows that, either 
 $\theta_{23}=\tilde{\theta}_{23} [\pi]$ or 
 $\theta_{23}=\frac{\pi}{2}-\tilde{\theta}_{23} [\pi]$;
 
\quad$\ast$ $\theta_{23}=\tilde{\theta}_{23} [\pi]$ matches the result of the 
previous discussion in the ``+" case, whereas, in the ``-" case,
 the matching leads to 
$\theta_{23}=\tilde{\theta}_{23}=0$, which is to be absorbed
as a particular case in the ``+" configuration;
 
\quad$\ast$ $\theta_{23}=\frac{\pi}{2}-\tilde{\theta}_{23} [\pi]$ matches 
the result of the previous discussion in the ``+" configuration,
in which case it leads to 
$\theta_{23}=\tilde{\theta}_{23}=\frac{\pi}{4} [\frac{\pi}{2}]$, {\em i.e.}
maximal mixing 
between the fermions of the second and third generations.

\section{$\boldsymbol{(\theta_{12},\theta_{23})}$ solutions of eqs.
(\ref{eq:nodb}) (\ref{eq:nosb}) (\ref{eq:nods}) (\ref{eq:ddss})
(\ref{eq:ssbb}) for
$\boldsymbol{\theta_{13} = 0 = \tilde{\theta}_{13}}$}
\label{section:sol0}

Excluding $\tilde{\theta}_{12}=0$, (\ref{eq:nodb0}) and (\ref{eq:nosb0}) require
$\sin(2\theta_{23}) = \sin(2\tilde{\theta}_{23}) \Rightarrow \tilde{\theta}_{23} = \theta_{23} + k\pi$
 or $\tilde{\theta}_{23} = \pi/2 - \theta_{23} + k\pi$.

$\bullet$ for $\tilde{\theta}_{23} = \theta_{23} + k\pi$ Cabibbo-like,

(\ref{eq:nods0}) requires
$\sin(2\theta_{12}) = \sin(2\tilde{\theta}_{12})  \Rightarrow
\tilde{\theta}_{12} = \theta_{12} + n\pi$ or $\tilde{\theta}_{12} = \pi/2 - \theta_{12} + n\pi$;

(\ref{eq:ddss0}) requires
$\cos(2\theta_{12}) = \cos(2\tilde{\theta}_{12}) \Rightarrow \tilde{\theta}_{12} = \pm \theta_{12} +
p\pi$;

(\ref{eq:ssbb0}) requires $s_{12}^2 + \tilde c_{12}^2 -1 =0 \Rightarrow
\tilde{\theta}_{12} = \pm \theta_{12} + r\pi$.

The solutions of these three equations are
$\theta_{12} = \tilde{\theta}_{12} + m\pi$ Cabibbo-like or
$\theta_{12} = \pi/4 + q\pi/2$ maximal ($\tilde\theta_{12} = \pm
\theta_{12} + r\pi$ is then also maximal).
They are associated with  $\tilde{\theta}_{23} =
\theta_{23} + k\pi$, condition heading this paragraph.

$\bullet$ for $\tilde{\theta}_{23} = \pi/2 - \theta_{23} + k\pi$,

(\ref{eq:nods0}) requires $s_{12}c_{12} = 2c_{23}^2 \tilde s_{12} \tilde
c_{12}$;

(\ref{eq:ddss0}) requires $c_{12}^2 -s_{12}^2 = 2 c_{23}^2 (\tilde c_{12}^2 - \tilde
s_{12}^2)$;

(\ref{eq:ssbb0}) requires $s_{12}^2 + 2c_{23}^2 \tilde c_{12}^2 - 2s_{23}^2
=0$.

Taking the ratio of the first two conditions yields
 $\tan(2\theta_{12}) = \tan(2\tilde{\theta}_{12}) = 2c_{23}^2
\Rightarrow \tilde{\theta}_{12} = \theta_{12} + k\pi/2 + n\pi$, which entails
$2c_{23}^2 = 1 \Rightarrow \theta_{23} = \pm \pi/4 + p\pi/2$ maximal; by
the condition $\tilde{\theta}_{23} = \pi/2 - \theta_{23} + k\pi$ heading
this paragraph, $\tilde{\theta}_{23}$ is then maximal, to. The third
condition becomes $s_{12}^2 + \tilde c_{12}^2 -1=0$, which requires
$\tilde\theta_{12} = \pm\theta_{12} + r\pi$. Then, the second condition is
automatically satisfied, but the first requires that the $``+''$ sign be
chosen; so, $\tilde\theta_{12} = \theta_{12} + r\pi$ is Cabibbo-like.

$\bullet$ Summary: the solutions are:\newline
\qquad $\ast$ $\tilde{\theta}_{23} = \theta_{23} +
k\pi$ Cabibbo-like, associated with either $\theta_{12} =
\tilde{\theta}_{12} + m\pi$ Cabibbo-like or $\theta_{12}$ and
$\tilde\theta_{12}$ maximal;\newline
\qquad $\ast$ $\tilde\theta_{12} = \theta_{12} + r\pi$ Cabibbo-like,
 associated with $\theta_{23}$ and $\tilde\theta_{23}$ maximal.

\section{Sensitivity of the neutrino solution to a small variation of
$\boldsymbol{\theta_{13}}$}
\label{section:maxsens}
 
If one allows for a small $\theta_{13} \approx \tilde{\theta}_{13}$, 
(\ref{eq:nods}) and
(\ref{eq:ddss}) become respectively
\begin{equation}
-2\eta  s_{12} c_{12} s_{23} c_{23} + \epsilon (s_{12}^2 - c_{12}^2) 
            +\eta s_{13} (c_{23}^2 -s_{23}^2)(c_{12}^2 - s_{12}^2)=0
\end{equation}
and
\begin{equation}
-2\eta s_{23}c_{23} (c_{12}^2 - s_{12}^2) + 4 \epsilon s_{12} c_{12} 
     -2\eta s_{13} (c_{23}^2 - s_{23}^2)(2s_{12}c_{12} + \epsilon(c_{12}^2 - s_{12}^2))
=0.
\end{equation}
For $\theta_{23}, \tilde{\theta}_{23}$ maximal, the dependence on $\theta_{13}$ drops out.

\section{Aligning classical flavor states and classical mass states}
\label{section:align}

We show below that, at the classical level of mass matrices, one can always
perform, in each sector, a flavor rotation such that the classical mass
eigenstates and the rotated flavor states get aligned.
Since the logic is slightly different from the one in paragraph
\ref{par:flarot}
\footnote{The change in flavor states was defined, there,
by (\ref{eq:unitr}) and the transformed Lagrangian was expressed in terms
of the original bare flavor fields.  Classical mass eigenstates got changed
such that the  new ones are deduced from the starting ones by the
rotation ${\cal R}(\varphi)$: $\left(\begin{array}{c}\hat d_m^0 \cr \hat
s_m^0\end{array}\right) = {\cal R}(\varphi)
\left(\begin{array}{c} d_m^0 \cr s_m^0\end{array}\right)$. The new
classical mass eigenstates could then be aligned with the starting bare
flavor states. In the present approach, it is the new flavor states which
can get aligned with the bare mass eigenstates, the latter staying
 unchanged.},
we chose to explain things in detail here.

Let us now consider the {\em change of variables} in flavor space
$\left(\begin{array}{c} d'_{fL} \cr s'_{fL}\end{array}\right) =
{\cal R}(\varphi)
\left(\begin{array}{c} d^0_{fL} \cr s^0_{fL}\end{array}\right)$.
In terms of the primed fields, the mass terms in the Lagrangian rewrite
$\left(\begin{array}{c} d'_{fL} \cr s'_{fL}\end{array}\right)^\dagger
{\cal R}(\varphi) M_0 {\cal S}^\dagger(\varphi) 
\left(\begin{array}{c} d'_{fR} \cr s'_{fR}\end{array}\right)$, in which
${\cal S}(\varphi)$ is the equivalent of ${\cal R}(\varphi)$ for right-handed fields.
Since $M_0$ was diagonalized according to
${\cal C}_{d0}^\dagger M_0 {\cal H}_{d0} = diag(m^0_d, m^0_s)$,
${\cal R}(\varphi) M_0 S^\dagger_\varphi$ is now diagonalized according to
${\cal C}_{d0}^\dagger  R^\dagger_\varphi ({\cal R}(\varphi) M_0 {\cal
S}^\dagger(\varphi))
 {\cal S}(\varphi) {\cal H}_{d0} = diag(m^0_d, m^0_s)$.
Accordingly, the new classical mass eigenstates are
$\left(\begin{array}{c} d'_{mL} \cr s'_{mL}\end{array}\right) =
{\cal C}_{d0}^\dagger  {\cal R}^\dagger(\varphi)
\left(\begin{array}{c} d'_{fL} \cr s'_{fL}\end{array}\right)
={\cal C}_{d0}^\dagger 
\left(\begin{array}{c} d^0_{fL} \cr s^0_{fL}\end{array}\right)
\equiv \left(\begin{array}{c} d^0_{mL} \cr s^0_{mL}\end{array}\right)$.
So, the classical mass eigenstates are unchanged, but are now deduced from
the new classical flavor states by the product
 ${\cal C}_{d0}^\dagger  {\cal R}^\dagger(\varphi)$. The angle $\varphi$ can
accordingly be tuned such that this product is the unit matrix. When it is
so, the new classical flavor states are aligned with the bare mass states.

The same demonstration holds in the $(u,c)$ sector. This shows that, at
the classical level of mass matrices, mixing angles in each sector, when
defined as the one connecting bare flavor states to original bare mass
states have no
physical meaning and can always be tuned to zero. So, 
the only physical mixing angles are the ones occurring in charged currents.
Indeed, since mass states are
unchanged, it is even more trivial than in subsection \ref{par:flarot} to
show that these angles stay unchanged by arbitrary flavor rotations.

We recall however that, as emphasized in footnote \ref{footnote:newmix}, a
common flavor rotation of both sectors is required as soon as one wants
to preserve the group structure of charged currents in bare flavor space.

\newpage\null
\begin{em}

\end{em}

\end{document}